\documentclass[a4paper,11pt]{article}
\usepackage[english]{babel}
\usepackage{jheppub}
\usepackage{amsmath,amssymb,graphicx,textcomp,enumerate,alltt,xspace,multirow}

\newcommand\POWHEGBOX{{\tt POWHEG BOX}}
\newcommand\POWHEG{{\tt POWHEG}}
\newcommand\MCatNLO{{\tt MC@NLO}}

\newcommand\PythiaEight{{\tt Pythia8}}

\newcommand\HerwigSevenPone{{\tt Herwig7.1}}

\newcommand\PythiaEightPtwo{{\tt Pythia8.2}}

\newcommand\RES{{\tt POWHEG BOX RES}}
\newcommand\VTWO{{\tt POWHEG BOX V2}}

\newcommand\Openloops{{\tt OpenLoops}}

\newcommand{\mathd}{\mathrm{d}}

\newcommand\sss{\mathchoice%
{\displaystyle}%
{\scriptstyle}%
{\scriptscriptstyle}%
{\scriptscriptstyle}%
}
\newcommand{\pt}{\ensuremath{p_{\sss T}}\xspace}

\newcommand{\muF}{\ensuremath{\mu_{\sss \mathrm{F}}}}
\newcommand{\muR}{\ensuremath{\mu_{\sss \mathrm{R}}}}
\newcommand{\KF}{\ensuremath{K_{\sss \mathrm{F}}}}
\newcommand{\KR}{\ensuremath{K_{\sss \mathrm{R}}}}

\def\beq{\begin{equation}}
\def\beqn{\begin{eqnarray}}
\def\eeq{\end{equation}}
\def\eeqn{\end{eqnarray}}

\def\lq{\left[} 
\def\rq{\right]}

\def\({\left(} 
\def\){\right)} 
 
\newcommand     \MSB            {\ifmmode {\overline{\rm MS}} \else
                                 $\overline{\rm MS}$\fi}

\newcommand\as{\alpha_{\sss\rm S}}

\newcommand\mZ{m_{\sss Z}}

\newcommand\mt{m_{t}}
\newcommand\mtc{m_{t,\, c}}

\newcommand\fourl{e^+\nu_{\sss e}\, \mu^-\bar{\nu}_{\sss \mu}}

\newcommand\pwgopt[1]{{\tt #1}}
\newcommand\bjet{\ensuremath{b}-jet}

\newcommand\mwbj{\ensuremath{m_{Wb_j}}}
\newcommand\mwbjmax{\ensuremath{m_{Wb_j}^{\max}}}
\newcommand\Ebj{\ensuremath{E_{b_j}}}
\newcommand\Ebjmax{\ensuremath{E_{b_j}^{\max}}}
\newcommand\pT{\ensuremath{p_{\sss\rm  T}}}

\newcount\minutes 
\newcount\scratch 
\def\timestamp{%
\scratch=\time 
\divide\scratch by 60 
\edef\hours{\the\scratch} 
\multiply\scratch by 60 
\minutes=\time 
\advance\minutes by -\scratch 
---$\,$\hours:\null 
\ifnum\minutes< 10 0\fi 
\the\minutes}

\definecolor{mygray}{gray}{0.5}

\newcommand\aprod{\ensuremath{\alpha_{\rm \scriptscriptstyle ISR}}}

\newcommand\PythiaEightPlot{{\tt Py8.2}}
\newcommand\HerwigSevenPlot{{\tt Hw7.1}}
\newcommand\ttbnlodecPlot{$t\bar{t}dec$}
\newcommand\bbllllPlot{$b\bar{b}4\ell$}
\newcommand\hvqPlot{$hvq$}

\newcommand\hvq{\hvqPlot}
\newcommand\bbfourl{\bbllllPlot}
\newcommand\ttNLOdec{\ttbnlodecPlot}
\newcommand\ttbnlodec{\ttNLOdec}

\newcommand{\subtopic}[1]{\vspace{4mm}}

\preprint{
\begin{flushright}
  CERN-TH-2018-001 \\
  ZU-TH 01/18
\end{flushright}
}

\title{A theoretical study of top-mass measurements at the LHC using NLO+PS
  generators of increasing accuracy}

\author[a]{Silvia Ferrario Ravasio,}
\author[b]{Tom\'a\v{s} Je\v{z}o,} 
\author[c]{Paolo Nason,}
\author[a]{Carlo Oleari}

\emailAdd{silvia.ferrario@mib.infn.it}
\emailAdd{tomas.jezo@physik.uzh.ch}
\emailAdd{paolo.nason@mib.infn.it}
\emailAdd{carlo.oleari@mib.infn.it}

\affiliation[a] {Universit\`a di Milano-Bicocca and INFN, Sezione di
  Milano-Bicocca, Piazza della Scienza 3, 20126 Milano, Italy}

\affiliation[b] {Physics Institute, Universit\"at Z\"urich, Z\"urich,
  Switzerland}

\affiliation[c]{CERN, CH-1211 Geneve 23, Switzerland, and INFN, Sezione di Milano-Bicocca,
  Piazza della Scienza 3, 20126 Milano, Italy}

\abstract{In this paper we study the theoretical uncertainties in the determination of
the top-quark mass using next-to-leading-order~(NLO) generators interfaced to
parton showers~(PS) that have different levels of accuracy. Specifically we
consider three generators: one that implements NLO corrections in the
production dynamics, one that includes also NLO corrections in top decay in
the narrow width approximation, and one that implements NLO corrections for
both production and decay including finite-width and interference effects.
Since our aim is to provide an assessment of the uncertainties of purely
theoretical origin, we consider simplified top-mass related observables that
are broadly related to those effectively used by experiments, eventually
modelling experimental resolution effects with simple smearing procedures.
We estimate the differences in the value of the extracted top mass that would
occur due to the use of the three different NLO generators, to the variation
of scales, to the choice of parton distribution functions and to the matching
procedure.  Furthermore, we also consider differences due to the shower and
to the modelling of non-perturbative effects by interfacing our NLO
generators to both \PythiaEightPtwo{} and \HerwigSevenPone{}, with various
settings. We find very different results depending upon the adopted shower
model. While with \PythiaEightPtwo{} we find moderate differences between the
different NLO+PS generators, with \HerwigSevenPone{} we find very large
ones. Furthermore, the differences between \PythiaEightPtwo{} and
\HerwigSevenPone{} generators are also remarkably large.}


\newcommand\diffdiffRttdec{$   55$}
\newcommand\diffdiffRhvq{$   34$}
\newcommand\pyminushwRfour{$  830$}
\newcommand\pyminushwRsix{$ 1267$}
\newcommand\pyminushwdeltaRfoursix{$  437$}
\newcommand\Bfrommwbjbbfourl{     1.008}
\newcommand\Berrfrommwbjbbfourl{     0.002}
\newcommand\Bfrommwbjsmearbbfourl{     0.958}
\newcommand\Berrfrommwbjsmearbbfourl{     0.001}
\newcommand\Bfrommwbjttdec{     1.000}
\newcommand\Berrfrommwbjttdec{     0.002}
\newcommand\Bfrommwbjsmearttdec{     0.957}
\newcommand\Berrfrommwbjsmearttdec{     0.001}
\newcommand\Bfrommwbjhvq{     1.002}
\newcommand\Berrfrommwbjhvq{     0.002}
\newcommand\Bfrommwbjsmearhvq{     0.949}
\newcommand\Berrfrommwbjsmearhvq{     0.001}
\newcommand\diffttdecbbfourl{$ 140$}
\newcommand\diffhvqbbfourl{$-147$}
\newcommand\diffaverage{$ 140$}
\newcommand\varPDF{$   9$}
\newcommand\varscalemax{  86}
\newcommand\varscalemin{  53}
\newcommand\varscaleothers{$   7$}
\newcommand\deltaalphashvq{$  18$}
\newcommand\deltaalphasttdec{$ 108$}
\newcommand\deltaalphasbbfourl{$ 128$}
\newcommand\deltaalphasunsmear{$   8$}
\newcommand\pdferrorhvqsmear{$   5$}
\newcommand\pdferrorhvqnosmear{$   3$}
\newcommand\pyminushwhvq{$ 240$}

\newcommand\hwTSsmearbbfourlttdec{$ 130$}

\newcommand\BfromEbjbbfourl{   0.54}
\newcommand\BerrfromEbjbbfourl{   0.07}
\newcommand\BfromEbjttdec{   0.50}
\newcommand\BerrfromEbjttdec{   0.03}
\newcommand\BfromEbjhvq{   0.50}
\newcommand\BerrfromEbjhvq{   0.03}
\newcommand\diffEbjcml{    3.4}
\newcommand\diffEbjumc{    3.3}
\newcommand\Ebjbbfourlmhvq{$  460$}
\newcommand\Ebjbbfourlmhvqerr{$  100$}

\newlength{\wfigsing}
\newlength{\wfigsingmulti}
\newlength{\wfigdoub}
\newlength{\wtablarge}

\def\whichjournal{1}
\ifnum \whichjournal=0 
  \wtablarge=\textwidth
  \newcommand\fignewline{\\}
  \newcommand\writeApp{}
\else
  \newcommand\fignewline{}
  \newcommand\writeApp{Appendix~}
\fi

\begin{document}
\maketitle

\flushbottom

\section{Introduction}
The abundant production of top pairs at the Large Hadron Collider~(LHC)
provides an opportunity for detailed studies of top-quark properties, for
tests of the Standard Model~(SM) in the top sector, and for measurements of fundamental
parameters such as the top-quark mass.  With the Higgs boson mass now known
with high precision, the $W$-boson and top-quark masses have become strongly
correlated, and an accurate determination of both would lead to a SM test of
unprecedented precision~\cite{Patrignani:2016xqp-EWreview, Baak:2014ora}.
The present value of the indirect top-mass determination from electroweak
precision data~($176.7\pm 2.1$~GeV, see~\cite{Patrignani:2016xqp-EWreview})
is in slight tension, at the $1.6\,\sigma$ level, with the direct
measurements. The latest combination of the Tevatron and the LHC
results~\cite{ATLAS:2014wva} yields~$173.34\pm 0.76$~GeV, but more recent
results favour even smaller values, close to $172.5$~GeV,
see~\cite{Aaboud:2016igd, Khachatryan:2015hba, CMS-PAS-TOP-17-007,
  ATLAS-CONF-2017-071}. Recent reviews of top-mass measurements by the ATLAS
and CMS collaborations can be found in Refs.~\cite{Pearson:2017jck}
and~\cite{Castro:2017yxe}.

It has been shown that in the Standard Model as is (i.e.~assuming no new physics
effects up to the Planck scale), the vacuum is stable if the top mass, $\mt$,
is below $171$~GeV (i.e.~very close to its present value), metastable up to
$176$~GeV, and unstable above this value~\cite{Degrassi:2012ry,
  Buttazzo:2013uya, Andreassen:2017rzq, Chigusa:2017dux}.  The current value
is safely below the instability region. However, it should not be forgotten
that the absence of new physics up to the Planck scale is a very strong
assumption. The only conclusion we can draw from these results is that there
is no indication of new physics below the Planck scale coming from the
requirement of vacuum stability.  On the other hand, the fact that the Higgs
boson quartic coupling almost vanishes at the Planck scale may have some deep
meaning that we are as yet unable to unveil.

Besides the issues related to electroweak tests and the stability of the
vacuum, the question on how precisely we can measure the top mass at hadron
colliders also has its own significance, related to our understanding of QCD
and collider physics. In view of the large abundance of top-pair production
at the LHC, it is likely that precise measurements will be performed with
very different methods, and that comparing them will give us confidence in
our ability to handle hadron-collider physics problems.

Top-mass measurements are generally performed by fitting $\mt$-dependent
kinematic distributions to Monte Carlo predictions. The most precise ones,
generally called \emph{direct measurements}, rely upon the full or partial
reconstruction of the system of the top-decay products.
The ATLAS and CMS measurements of Refs.~\cite{Aaboud:2016igd}
and~\cite{Khachatryan:2015hba}, yielding the value $172.84 \pm 0.34~{\rm
(stat)} \pm 0.61$~(syst)~GeV and $\mt=172.44 \pm 0.13~{\rm (stat)} \pm
0.47$~(syst)~GeV respectively, fall into this broad category.

The top mass cannot be defined in terms of the mass distribution of the
system of its decay products: since the top quark is a coloured object, no
final-state particle system can be unambiguously associated with it.  On the
other hand, the top mass is certainly related to the mass distribution of the
system of objects arising from top decay, i.e.~hard leptons,
neutrinos and hard, $b$-flavoured hadronic jets. The mass distribution of
this system can be computed and measured, and the top mass enters this
computation as a parameter.  By extracting its value from a fit to the
measured distributions, we are unavoidably affected by theoretical errors
that must be carefully assessed. In particular, these errors will depend upon
the accuracy of the modelling of these distributions.

The absence of a ``particle truth level'' for the top-decay products has led
to speculations that the top mass cannot be extracted reliably in the direct
measurements.  The extracted mass is unavoidably a parameter in the
theoretical calculation or in the Monte Carlo generator that is used to
compute the relevant distributions.  It has thus been argued that, because of
this, and since shower Monte Carlo~(SMC) models are accurate at leading
order~(LO) only, the extracted mass cannot be identified with a theoretically
well-defined mass, such as the pole mass or the \MSB{} mass (that differ
among each other only at the NLO level and beyond).

In the present work, we use NLO-accurate generators, so that the previously
mentioned objection does not actually apply. Moreover, it can be argued that,
in the narrow width approximation and at the perturbative level, the mass
implemented in Monte Carlo generators corresponds to the pole
mass~\cite{Nason:2017cxd} even if we do not use NLO-accurate generators.

It was also argued in Ref.~\cite{Hoang:2014oea} that the Monte Carlo mass
parameter differs from the top pole mass by an amount of the order of a
typical hadronic scale, that was there quantified to be near 1~GeV.  It was
further argued that this difference is, in fact, intrinsic in the uncertainty
with which the pole mass can even be defined, because of the presence of a
renormalon in the relation of the pole to the \MSB~mass~\cite{Bigi:1994em,
  Beneke:1994sw}.

Recent studies~\cite{Beneke:2016cbu, Hoang:2017btd} have shown that the
renormalon ambiguity in the top-mass definition is not as large as previously
anticipated, being in fact well below the current experimental
error.\footnote{In fact, values in this range were obtained much earlier in
  Refs.~\cite{Pineda:2001zq, Bali:2013pla}, mostly in a bottom physics
  context, but since the renormalon ambiguity does not depend upon the heavy
  quark mass, they also apply to top.} The fact remains, however, that
non-perturbative corrections to top-mass observables (not necessarily related
to the mass renormalon) are present, can affect a top-mass determination, and
are likely to be parametrically of the order of a typical hadronic scale. We
believe, however, that this does not justify the introduction of a ``Monte
Carlo mass'' concept, since it is unlikely that non-perturbative effects,
affecting top-mass observables, can be parametrized as a universal shift of
the top-mass parameter.  The real question to answer is whether these
non-perturbative effects are of the order of 100 MeV, 1 GeV, or more.  While
a top-mass determination from threshold production at an $e^+e^-$ collider would
be free of such uncertainties~\cite{Beneke:2015kwa, Simon:2016htt},\footnote{
  This is also the case for a top-mass determination based upon the spectrum
  of $\gamma\gamma$ production near the $t\bar{t}$
  threshold~\cite{Kawabata:2016aya}, that however is likely to be
  statistically limited, even at the high luminosity LHC.}  at hadron
colliders, non-perturbative effects of this order are likely to affect, to
some extent, most top-mass observables that have been proposed so
far.\footnote{For a recent discussion of all these issues see
  Ref.~\cite{Nason:2017cxd}.}

The theoretical problems raised upon the top-quark mass measurement issues
have induced several theorists to study and propose alternative methods.  The
total cross section for $t\bar{t}$ production is sensitive to the top mass,
and has been computed up to the NNLO order in QCD~\cite{Czakon:2013goa}, and
can be used to extract a top mass value~\cite{Khachatryan:2016mqs,
  Aad:2014kva, Langenfeld:2009wd}.

In Ref.~\cite{Alioli:2013mxa}, observables related to the $t{\bar t}+{\rm jet}$
kinematics are considered.  The authors of Ref.~\cite{Kawabataa:2014osa}
presented a method based upon the charged-lepton energy spectrum, that is not
sensitive to top production kinematics, but only to top decay,
arguing that, since this has been computed at NNLO accuracy~\cite{Gao:2012ja,
  Brucherseifer:2013iv}, a very accurate measurement may be achieved.  Some
authors have advocated the use of boosted top jets (see
Ref.~\cite{Hoang:2017kmk} and references therein).  In
Ref.~\cite{Agashe:2016bok}, the authors make use of the \bjet{} energy peak position,
that is claimed to have a reduced sensitivity to production dynamics.  In
Ref.~\cite{Frixione:2014ala}, the use of lowest Mellin moments of lepton
kinematic distributions is discussed.  In the leptonic channel, it is also
possible to use distributions based on the ``stransverse'' mass
variable~\cite{Sirunyan:2017idq}, which generalizes the concept of transverse
mass for a system with two identical decay branches~\cite{Lester:1999tx,
  Barr:2009jv}.

Some of these methods have in fact been exploited~\cite{CMS-PAS-TOP-13-006,
  CMS-PAS-TOP-15-002, Aad:2015waa, Sirunyan:2017idq, Aaboud:2017ujq} to yield
alternative determinations of $\mt$.  It turns out, however, that the direct
methods yield smaller errors at the moment, and it is likely that alternative
methods, when reaching the same precision level, will face similar
theoretical problems.

\subsection{Goals of this work}
In this work, we exploit the availability of the new
\POWHEGBOX{}~\cite{Nason:2004rx, Frixione:2007vw, Alioli:2010xd} generators
for top-pair production, i.e.~the \ttbnlodec{}~\cite{Campbell:2014kua} and
\bbfourl{}~\cite{Jezo:2016ujg} ones, in order to perform a theoretical study
of uncertainties in the top-mass determination. In particular, we are in a
position to assess whether NLO corrections in top decay, that are implemented
in both the \ttbnlodec{} and \bbfourl{} generators, and finite width effects,
non-resonant contributions and interference of radiation generated in
production and decay, that are implemented in \bbfourl{}, can lead to sizeable
corrections to the extracted value of the top mass. Since the \hvq{}
generator~\cite{Frixione:2007nw}, that implements NLO corrections only in
production, is widely used by the experimental collaborations in top-mass
analyses, we are particularly interested in comparing it with the new
generators, and in assessing to what extent it is compatible with
them.\footnote{The \hvq{} and \ttbnlodec{} generators can be found under the
  {\tt User-Processes-V2} directory of the \VTWO{} repository in the {\tt
    hvq} and {\tt ttb\_NLO\_dec} directories, respectively.  The \bbfourl{}
  generator can be found under the {\tt User-Processes-RES/b\_bbar\_4l}
  directory of the \RES{} code. Detailed instructions are found at
  \url{powhegbox.mib.infn.it}.}  We will consider variations in the scales,
parton distribution functions~(PDFs) and the jet radius parameter to better
assess the level of compatibility of the different generators.

We are especially interested in effects that can be important in the
top-mass determination performed in direct measurements. Thus, the main focus
of our work is upon the mass of a reconstructed top, that we define as a
system comprising a hard lepton, a hard neutrino and a hard $b$ jet. We will
assume that we have access to the particle truth level, i.e.~that we can also
access the flavour of the $b$ jet, and the neutrino momentum and
flavour. We are first of all interested in understanding to what extent the
mass peak of the reconstructed top depends upon the chosen NLO+PS
generator. This would be evidence that the new features introduced in the
most recent generators are mandatory for an accurate mass extraction.

We will also consider the inclusion of detector effects in the form of a
smearing function applied to our results. Although this procedure is quite
crude, it gives a rough indication of whether the overall description of the
process, also outside of the reconstructed resonance peak, affects the
measurement.

Besides studying different NLO+PS generators, we have also attempted to give
a first assessment of ambiguities associated with shower and non-perturbative
effects, by interfacing our NLO+PS generators to two shower Monte Carlo
programs: \PythiaEightPtwo{}~\cite{Sjostrand:2014zea} and
\HerwigSevenPone{}~\cite{Bahr:2008pv, Bellm:2015jjp}. Our work focuses upon
NLO+PS and shower matching.  We thus did not consider further variations of
parameters and options within the same parton shower, nor variations on the
observables aimed at reducing the dependence upon those.\footnote{An
  interesting example of work along this direction can be found in
  Refs.~\cite{Wicke:2008iz} and~\cite{Sjostrand:2013cya}, where the impact of
  the colour reconnection model on top-mass measurement is analyzed.  In
  Ref.~\cite{Andreassen:2017ugs}, a study is performed to determine whether
  the use of jet-grooming techniques in top-mass measurement can reduce the
  Monte Carlo tune dependence.}

We have also considered two alternative proposals for top-mass measurements:
the position of the peak in the $b$-jet energy~\cite{Agashe:2016bok} and the
leptonic observables of Ref.~\cite{Frixione:2014ala}.  The first proposal is
an example of a hadronic observable that should be relatively insensitive to
the production mechanism, but may be strongly affected by NLO corrections in
decay.  The second proposal is an example of observables that depend only
upon the lepton kinematics, and that also depend upon production dynamics,
thus stronger sensitivity to scale variations and PDFs may be expected. It is
also generally assumed that leptonic observables should be insensitive to the
$b$-jet modeling.  One should remember, however, that jet dynamics affects
lepton momenta via recoil effects, so it is interesting to study whether
there is any ground to this assumption.

The impact of NLO corrections in decays and finite-width effects were also
considered in Ref.~\cite{Heinrich:2017bqp} for a number of top-mass related
observables, and in Ref.~\cite{Bevilacqua:2017ipv} for the method relying
upon the $t\bar{t}j$ final state.  Here we are more interested in observables
related to direct measurements, that are not considered there. Furthermore,
we focus our studies upon the differences with respect to the widely-used
\hvq{} generator.

\subsection{Preamble}
The study presented in this work was triggered by the availability of new
NLO+PS generators describing top decay with increasing accuracy. As such, its
initial aim was to determine whether and to what extent these new generators,
and the associated new effects that they implement, may impact present
top-mass measurements.  As we will see, had we limited ourselves to the study
of the NLO+PS generators interfaced to \PythiaEightPtwo{}, we would have
found a fairly consistent picture and a rather simple answer to this
question.

Since another modern shower generator that can be interfaced to our NLO+PS
calculation is available, namely \HerwigSevenPone{}, we have developed an
appropriate interface to it, and have also carried out our study using it as
our shower model.  Our results with \HerwigSevenPone{} turn out to be quite
different from the \PythiaEightPtwo{} ones, to the point of drastically
altering the conclusions of our study. In fact, variations in the extracted
top mass values due to switching between \PythiaEightPtwo{} and
\HerwigSevenPone{} prevail over all variations that can be obtained within
\PythiaEightPtwo{} by switching among different NLO+PS generators, or by
varying scales and matching parameters within them. Moreover, the comparison
of the various NLO+PS generators, when using \HerwigSevenPone{}, does not
display the same degree of consistency that we find within
\PythiaEightPtwo{}. If, as it seems, the differences found between
\PythiaEightPtwo{} and \HerwigSevenPone{} are due to the different shower
models (the former being a dipole shower, and the latter an angular-ordered
one), the very minimal message that can be drawn from our work is that, in
order to assess a meaningful theoretical error in top-mass measurements, the
use of different shower models, associated with different NLO+PS generators,
is mandatory.



Our results are collected in tables and figures that are presented and
discussed by giving all details that are necessary to reproduce them.  We
present a large number of results that show the effect of changing parameter
settings and matching methods in the NLO+PS calculations, some of which are
very technical.  Since this may obscure the logical development of our work,
we have written our Summary (Sec.~\ref{sec:Summary}) in such a way that the
main logical developments and findings are presented in a concise way.  In
fact, the summary section can be read independently of the rest of the paper,
and may be used to navigate the reader through the rest of the material.

\subsection{Outline}
The paper is organized as follows.  In Sec.~\ref{sec:generators} we briefly
review the features of the \hvq{}, \ttbnlodec{} and \bbfourl{} generators.
We also discuss the interfaces to the parton-shower programs
\PythiaEightPtwo{} and \HerwigSevenPone{}.

In Sec.~\ref{sec:pheno}, we detail the setup employed for the
phenomenological studies presented in the subsequent sections.

In Sec.~\ref{sec:reconstructedpeak}, we perform a generic study of the
differences of our generators focusing upon the mass distribution of the $W\,
b$-jet system. The aim of this section is to show how this distribution is
affected by the different components of the generators by examining results
at the Born level, after the inclusion of NLO corrections, after the parton
shower, and at the hadron level.

In Sec.~\ref{sec:methodology} we describe how we relate the computed
value of our observables to the corresponding value of the top mass that
would be extracted in a measurement.

In Sec.~\ref{sec:mwbj} we consider as our top-mass sensitive observable the
peak position in the mass distribution of the reconstructed top, defined as
the mass of the system comprising the hardest lepton and neutrino, and the
jet with the hardest $b$-flavoured hadron, all of them with the appropriate
flavour to match a $t$ or a $\bar{t}$.  We study its dependence upon the
NLO+PS generator being used, the scale choices, the PDFs, the value of $\as$
and the jet radius parameter.  Furthermore, we present and compare results
obtained with the two shower Monte Carlo generators \PythiaEightPtwo{} and
\HerwigSevenPone{}.

We repeat these studies for the peak of the $b$-jet energy
spectrum~\cite{Agashe:2016bok} in Sec.~\ref{sec:Ebjet}, and for the leptonic
observables~\cite{Frixione:2014ala} in Sec.~\ref{sec:LepObs}.

In Sec.~\ref{sec:Summary} we summarize our results, and in
Sec.~\ref{sec:Conc} we present our conclusions. In the appendices we give
some technical details.

\section{NLO+PS generators}
\label{sec:generators}
In this section we summarize the features of the \POWHEGBOX{} generators used
in the present work, i.e.~the \hvq{}, the \ttNLOdec{} and the \bbfourl{}
generators.

The \hvq{} program~\cite{Frixione:2007nw} was the first top-pair production
generator implemented in \POWHEG{}. It uses on-shell matrix elements for NLO
production of $t\bar{t}$ pairs. Off-shell effects and top decays, including
spin correlations, are introduced in an approximate way, according to the
method presented in Ref.~\cite{Frixione:2007zp}.  Radiation in decays is
fully handled by the parton-shower generators.  The ones that we consider,
\PythiaEightPtwo{} and \HerwigSevenPone{}, implement internally matrix-element
corrections for top decays, with \HerwigSevenPone{} also optionally
including a \POWHEG{}-style hardest-radiation generation. In these cases,
their accuracy in the description of top decays is, for our purposes,
equivalent to the NLO level.

The \ttNLOdec{} code~\cite{Campbell:2014kua} implements full spin
correlations and NLO corrections in production and decay in the narrow-width
approximation.  Off-shell effects are implemented via a reweighting method,
such that the LO cross section includes them exactly.  As such, it also contains
contributions of associated top-quark and $W$-boson production at LO. It
does not include, however, interference of radiation generated in production
and decay.

In \ttbnlodec{} the \POWHEG{} method has been adapted to deal with radiation
in resonance decays.  Radiation is generated according to the \POWHEG{}
Sudakov form factor both for the production and for all resonance decays that
involve coloured partons.  This feature also offers the opportunity to modify
the standard \POWHEG{} single-radiation approach.  Rather than picking the
hardest radiation from one of all possible origins (i.e.~production and
resonance decays), the \POWHEGBOX{} can generate simultaneously the hardest
radiation in production and in each resonance decay.  The LH events
generated in this way can thus carry more radiated partons, one for
production and one for each resonance.  Multiple-radiation events have to be
completed by a shower Monte Carlo program, that has to generate radiation
from each origin without exceeding the hardness of the corresponding
\POWHEG{} one, thus requiring an interface that goes beyond the simple Les
Houches standard~\cite{Boos:2001cv}.

A general procedure for dealing with decaying resonances that can radiate by
strong interactions has been introduced and implemented in a fully general
and automatic way in a new version of the \POWHEGBOX{} code, the
\RES~\cite{Jezo:2015aia}. This framework allows for the treatment of
off-shell effects, non-resonant subprocesses including full interference, and
for the treatment of interference of radiation generated in production and
resonance decay.\footnote{A related approach within the \MCatNLO{} framework
  has been presented in Ref.~\cite{Frederix:2016rdc}.}  In
Ref.~\cite{Jezo:2016ujg} an automated interface of the \RES{} code to the
\Openloops{}~\cite{Cascioli:2011va} matrix-element generator has been
developed and used to build the \bbfourl{} generator, that implements the
process $pp\to b\bar{b}\,\fourl$, including all QCD NLO corrections in the
4-flavour scheme, i.e.~accounting for finite $b$-mass effects.  So, double-top,
single-top and non-resonant\footnote{By non-resonant we mean processes that
  do not contain an intermediate top quark, e.g.~$pp \to b\,\bar{b} \,Z\to
  b\,\bar{b} \,W^+\, W^- \to b\, \bar{b}\,\fourl $.}  diagrams are all
included with full spin-correlation effects, radiation in production and
decays, and their interference.

As for the \ttNLOdec{} generator, \bbfourl{} can generate LH events including
simultaneous radiation from the production process and from the top and
anti-top decaying resonances.  It thus requires a non-standard interface to
parton-shower Monte Carlo programs, as for the case of the \ttNLOdec{}
generator.

\subsection{Interface to shower generators} 
According to the \POWHEG{} method, the PS program must complete the event
only with radiation softer than the \POWHEG{} generated one. In the standard Les
Houches Interface for User Processes~(LHIUP)~\cite{Boos:2001cv}, each
generated event has a hardness parameter associated with it, called {\tt
  scalup}. This parameter is set in \POWHEG{} to the relative transverse
momentum of the generated radiation and each emission attached by the parton
shower must have a $\pt$ smaller than its value.  The LHIUP treats all
emissions on an equal footing, and has no provision for handling radiation
from decaying resonances. This drives a standard PS to allow showering to
start from scales of the order of the resonance mass.

\subsubsection{Generic method}
\label{sec:genericmethod}
References~\cite{Campbell:2014kua} and~\cite{Jezo:2016ujg} introduce a
generic method for interfacing \POWHEG{} processes that include radiation in
decaying resonances with PS generators. According to this method, shower
radiation from the resonance is left unrestricted, and a veto is applied
\emph{a posteriori}: if any radiation in the decaying resonance shower is
harder than the \POWHEG{} generated one, the event is discarded, and the same
LH event is showered again. We also stress that the standard PS
implementations conventionally preserve the mass of the resonance, as long as
the resonance decay products, including eventually the radiation in decay,
have the resonance as mother particle in the LH event record.

The hardness of the radiation associated with the decaying top ($t\to
W\,b\,g$) in \POWHEG{} is given by
\begin{equation}
  \label{eq:bg_splitting}
  t=2\,\frac{E_g}{E_b} \,p_g\cdot p_b = 2\,E_g^2 \(1-\beta_b\cos\theta_{bg}\),
\end{equation}
where $p_{g/b}$ and $E_{g/b}$ are the four momentum and energy of the gluon
and of the bottom quark, $\beta_b$ is the velocity of the bottom quark and
$\theta_{bg}$ is the angle between the bottom and gluon momenta, all
evaluated in the top rest frame.  This hardness definition is internally used
to define the corresponding Sudakov form factor.  The same should be also
used to limit the transverse momentum generated by the PS in the resonance
decay.

The practical implementation of the veto procedure depends on whether we are
using a dipole, as in \PythiaEightPtwo{}, or an angular-ordered shower, as in
\HerwigSevenPone{}.  If we are using a dipole ($\pt$-ordered) shower, it is
sufficient to check the first shower-generated emission from the bottom quark
and (if present at the LH level) from the gluon arising in top decay.  The
hardness $t_b$ of the shower-generated emission from the bottom is evaluated
using eq.~(\ref{eq:bg_splitting}), while the one from the gluon is taken to
be
\begin{equation}
  \label{eq:gg_splitting}
  t_g=2\,E_1^2\,E_2^2\,\frac{(1-\cos\theta_{12})}{(E_1+E_2)^2}\,,
\end{equation}
where $E_{1,2}$ are the energies of the two gluons arising from the
splitting, and $\theta_{12}$ is the angle between them. Both $t_g$ and $t_b$
are computed in the top frame. If they are smaller than $t$, the event is
accepted, otherwise it is showered again.

In the case of angular-ordered showers, as in \HerwigSevenPone{}, it is not
enough to examine the first emission, because the hardest radiation may take
place later.  As shown in Ref.~\cite{Nason:2004rx}, in the leading
logarithmic approximation, the hardest emission in an angular-ordered shower
can be always found by following either the quark line in a $q \to q g$
splitting, or the most energetic line in a $g \to gg$ splitting.  Thus, when
inspecting the sequence of splittings, in order to find the hardest
radiation, if the parton that generates the shower is a fermion (in our case,
the $b/\bar{b}$ quark), we simply follow the fermionic line; in case of a
gluon splitting, we follow the most energetic gluon. We go on until either
the shower ends, or we reach a $g\to q \bar{q}$ splitting. Since this last
process is not soft-singular, configurations with the hardest emission
arising after it are suppressed.

\subsubsection{Standalone implementations in \PythiaEightPtwo{}}
\label{sec:PY8_different_showers}
The \PythiaEightPtwo{} generator provides facilities for implementing the
above-described method to internally veto radiation in resonance decays. We
prepared two implementations, each based on a different facility, and now we
describe them in turn.

\begin{enumerate}

\item
At every radiation generated by \PythiaEightPtwo{}, a function is called internally
using the \verb!UserHooks! facility. The function inspects the radiation
kinematics. If the radiation comes from top decays, it computes its
transverse momentum, according to eqs.~(\ref{eq:bg_splitting})
and~(\ref{eq:gg_splitting}). If the transverse momentum is larger than the
one of the radiation generated by \POWHEG{} in the resonance decay, the
emission is vetoed, and \PythiaEightPtwo{} tries to generate another splitting.
The process is repeated until an acceptable splitting is generated. This
behaviour is achieved by implementing the method
\begin{verbatim}
UserHooks::doVetoFSREmission,
\end{verbatim}  
whose description can be found in the \PythiaEightPtwo{} manual~\cite{pythiamanual}.
It is activated by setting the \PythiaEightPtwo{} flag
\begin{verbatim}
POWHEG:bb4l:FSREmission:veto = on.
\end{verbatim}
\item
The {\tt UserHooks} facility also allows us to set the initial scale of
final-state shower evolution (for the shower arising from the decaying
resonances) equal to the transverse momentum of the top radiation in decay.
This is achieved using the method
\begin{verbatim}
UserHooks::scaleResonance,
\end{verbatim}
and is activated by setting the \PythiaEightPtwo{} flag
\begin{verbatim}
POWHEG:bb4l:ScaleResonance:veto = on.
\end{verbatim}
This method has the disadvantage of relying upon the assumption that the
hardness definition used by \PythiaEightPtwo{} is compatible with the \POWHEG{} one.
\end{enumerate}
Both methods are implemented in the file
\begin{verbatim}
PowhegHooksBB4L.h
\end{verbatim}
in the \bbfourl{} subprocess directory.

We have chosen implementation 1 as our default, and compared it with the
other implementations in order to validate it and estimate matching
uncertainties.

\subsubsection{Standalone implementations in \HerwigSevenPone{}}
\label{sec:HW7_different_showers}
Also in the case of \HerwigSevenPone{} we have prepared two implementations that
use the MC internal facilities to perform the veto:
\begin{enumerate}

\item
After the whole time-like shower has been developed, but before hadronization
has been carried out, the showers from the $b$ and from the \POWHEG{}
radiated gluon in top decay are examined.  In the case of the $b$, the quark
line is followed, and the transverse momentum of the radiation is computed
(in the top frame) according to eq.~(\ref{eq:bg_splitting}). In the case of
the gluon, the hardest line is followed, and the transverse momentum of the
radiation is computed according to eq.~(\ref{eq:gg_splitting}). If a
radiation is found with transverse momentum harder than the \POWHEG{}
generated one, the full event is reshowered, starting from the same LH event.
The corresponding method is called
\begin{verbatim}
FullShowerVeto::vetoShower,
\end{verbatim}
and we have implemented it in the files
\begin{verbatim}
bb4lFullShowerVeto.h, bb4lFullShowerVeto.cc.
\end{verbatim}

\item
We veto each radiation in resonance decay if its transverse momentum is
harder than the \POWHEG{} generated one. In this case, \HerwigSevenPone{} tries again
to generate radiation starting from the (angular ordering) hardness parameter
of the vetoed one. As in \PythiaEightPtwo{} second method, we have to rely in this
case upon the \HerwigSevenPone{} definition of the radiation transverse momentum.  The
corresponding method is called
\begin{verbatim}
ShowerVeto::vetoTimeLike
\end{verbatim}
and we implemented it in the files
\begin{verbatim}
bb4lShowerVeto.h, bb4lShowerVeto.cc.
\end{verbatim}

\end{enumerate}
We will adopt implementation~2 as our \HerwigSevenPone{} default, and compare
with the other one in order to validate it, and also in order to get an
indication of the size of matching uncertainties.

\section{Phenomenological analysis setup}
\label{sec:pheno}
We simulate the process $p\,p \to b\, \bar{b}\,\fourl$, which is available in
all three generators. It is dominated by top-pair production, with a smaller
contribution of $Wt$ topologies. For the observables we consider, the decay
of one of the two top quarks is mostly irrelevant, so that our result will
also hold for semileptonic decays.

In the \hvq{} and \ttbnlodec{} generators we renormalize the top mass in the
pole-mass scheme, while in the \bbfourl{} one we adopt the complex mass
scheme~\cite{Jezo:2016ujg}, with the complex mass defined as
$\sqrt{\mt^2-i\,\mt \,\Gamma_t}$.

We perform our simulations for a center-of-mass energy of $\sqrt{s}=8$~TeV.
We have used the {\tt MSTW2008nlo68cl} PDF set~\cite{Martin:2009iq} and we
have chosen as central renormalization and factorization scale ($\muR$ and
$\muF$) the quantity $\mu$, defined, following Ref.~\cite{Jezo:2016ujg}, as
the geometric average of the transverse masses of the top and anti-top
\begin{equation}
\label{eq:centralscale}  
\mu= \sqrt[4]{\left(E^2_t -p_{z,t}^2\right)\left(E^2_{\bar{t}}
  -p_{z,\bar{t}}^2\right)}\,,
\end{equation}
where the top and anti-top energies $E_{t/\bar{t}}$ and longitudinal momenta
$p_{z,t/\bar{t}}$ are evaluated at the underlying-Born level.

In the \bbfourl{} case, there is a tiny component of the cross section given
by the topology
\begin{equation}
pp\to Z g \to (W^+ \to e^+ \nu_e) (W^- \to \mu^-
\bar{\nu}_\mu) (g \to b \bar{b}).
\end{equation}
In this case we define $\mu$ as
\begin{equation}
  \label{eq:centralscaleZ}
  \mu= \frac{\sqrt{p_{\sss Z}^2}}{2}\,,
\end{equation}
where $p_{\sss Z}=p_{\mu^-}+p_{\bar{\nu}_\mu} + p_{e^+} +p_{\nu_{\sss e}}$.
This case is however very rare and unlikely to have any significance. 

The parameter {\tt hdamp} controls the separation of remnants
(see \writeApp\ref{app:remnants}) in the production of $t\bar{t}$ pairs with large
transverse momentum.  We set it to the value of the top mass.

\subsection{Physics objects}
\label{sec:physicsObjects}
In our simulations we make the $B$ hadrons stable, in order to simplify the
definitions of $b$ jets.  Jets are reconstructed using the
Fastjet~\cite{Cacciari:2011ma} implementation of the anti-$k_{\rm\sss T}$
algorithm~\cite{Cacciari:2008gp} with $R=0.5$.  We denote as $B$~(${\bar B}$)
the hardest (i.e.~largest \pT{}) $b$~($\bar{b}$) flavoured hadron. The
$b$~($\bar{b}$) jet is the jet that contains the hardest
$B$~($\bar{B}$).\footnote{Note that this notation is the opposite of what is
  commonly adopted for $B$ mesons, where $B$ refers to the meson containing
  the $\bar{b}$ quark.}  It will be indicated as $j_B$~($j_{\bar{B}}$).  We
discard events where the $b$ jet and $\bar{b}$ jet coincide.  The hardest
$e^+$~($\mu^-$) and the hardest $\nu_e$~($\bar{\nu}_{\mu}$) are paired to
reconstruct the $W^+$~($W^-$).  The reconstructed top~(anti-top) quark is
identified with the corresponding $W^+j_B$ ($W^-j_{\bar{B}}$) pair.  In the
following we will refer to the mass of this system as \mwbj{}.

We require the two $b$ jets to have
\begin{equation}
 \pT>30\mbox{~GeV}\,, \qquad |\eta|<2.5\,.
\end{equation}
These cuts suppress the single-top topologies.
The hardest $e^+$ and the hardest $\mu^-$ must satisfy
\begin{equation}
 \pT>20\mbox{~GeV}\,, \qquad |\eta|<2.4\,.
\end{equation}

\subsection{Generated sample}

For each generator under study, we have produced three samples of events,
each sample computed with a top mass of 169.5, 172.5 and 175.5~GeV,
respectively, with the corresponding decay width computed at NLO.  Using the
reweighting feature of the \POWHEGBOX{}, we have computed the event weights
obtained by varying the parton distribution functions and the renormalization
and factorization scales, for a total of 12 weights (see
Secs.~\ref{sec:fac_ren_scales} and~\ref{sec:PDF_dependence} for more
details).

In the reweighting procedure, only the inclusive \POWHEG{} cross section is
recomputed.  The Sudakov form factor is not recomputed, so that the radiated
partons retain the same kinematics.  For this reason, the change of the
renormalization and factorization scales do not affect the emission of
radiation.
Thus, in order to investigate the sensitivity of the result on the intensity
of radiation, where we are particularly concerned with emissions from the
final-state $b$ quarks, we have also generated samples with the
{\tt NPDF30\_nlo\_as115} and {\tt NNPDF30\_nlo\_as121}, with $\as(\mZ)=0.115$
and $\as(\mZ)=0.121$ respectively, for each generator,
for the central value of the top
mass, i.e.~172.5~GeV.  The number of events for each generated sample,
together with an indicative computational time, are reported in
Tab.~\ref{tab:samples}.
\begin{table*}[tb]
  \centering
  \resizebox{\textwidth}{!}
{  \begin{tabular}{l|c|c||c|c||c|c||c|c||c|c|}
    \cline{2-11}
    &\multicolumn{10}{|c|}{Generated samples} \\
    \cline{2-11}
     &\multicolumn{6}{|c||}{$m_t$~[GeV]} & \multicolumn{4}{|c|}{$\as(\mZ)$} \\
    \cline{2-11}
      & \multicolumn{2}{|c||}{$172.5$} &  \multicolumn{2}{|c||}{$169.5$} &  \multicolumn{2}{|c||}{$175.5$} &
    \multicolumn{2}{|c||}{$0.115$} & \multicolumn{2}{|c|}{$0.121$} \\
    \cline{2-11}
    &  \phantom{\Big|}\# events & time    &\# events  & time    & \# events & time    & \# events & time    & \# events & time \\
    \cline{1-11}
    \multicolumn{1}{|c|}{\hvq{}} &  \phantom{\Big|}12M & 10~h     & 3M  & 2.5~h     & 3M & 2.5~h  & 12M & 9~h & 12M & 9~h \\
    \cline{1-11}
    \multicolumn{1}{|c|}{\ttbnlodec{}} &  \phantom{\Big|}12M & 46~d     & 3M  & 11.5~d     & 3M & 11.5~d  & 12M & 25~d & 12M & 25~d \\
    \cline{1-11}
    \multicolumn{1}{|c|}{\bbfourl{}} &  \phantom{\Big|}20M & 4600~d     & 1.7M  & 390~d     & 1.7M & 390~d  & 3M & 64~d & 3M & 64~d \\
    \cline{1-11}
  \end{tabular}
  }
  \caption{Number of events and total CPU time of the generated samples.  The
    samples used for the $\as$ variations were obtained in a relatively
    smaller time, since in this case only the central weight was
    computed. This leads to a difference that can be sizeable, depending upon
    the complexity of the virtual corrections.}
\label{tab:samples}
\end{table*}

\section{Anatomy of the reconstructed top mass distribution at NLO+PS}
\label{sec:reconstructedpeak}

In this section we investigate the impact of individual ingredients in a
typical NLO+PS calculation on the kinematic distribution of the reconstructed
top mass \mwbj{}.  On the perturbative side, we examine the impact of the
different level of accuracy in the treatment of top production and decay
provided by the three generators we are considering, and the impact of
parton-shower effects.  On the non-perturbative side, we illustrate the
effect of including hadronization and underlying event in the simulation.

\subsection{Les Houches event level comparison of the generators}
We begin by comparing the three generators at the Les Houches event~(LHE) level.
\begin{figure}[tb]
\centering
     \includegraphics[width=\wfigsing]{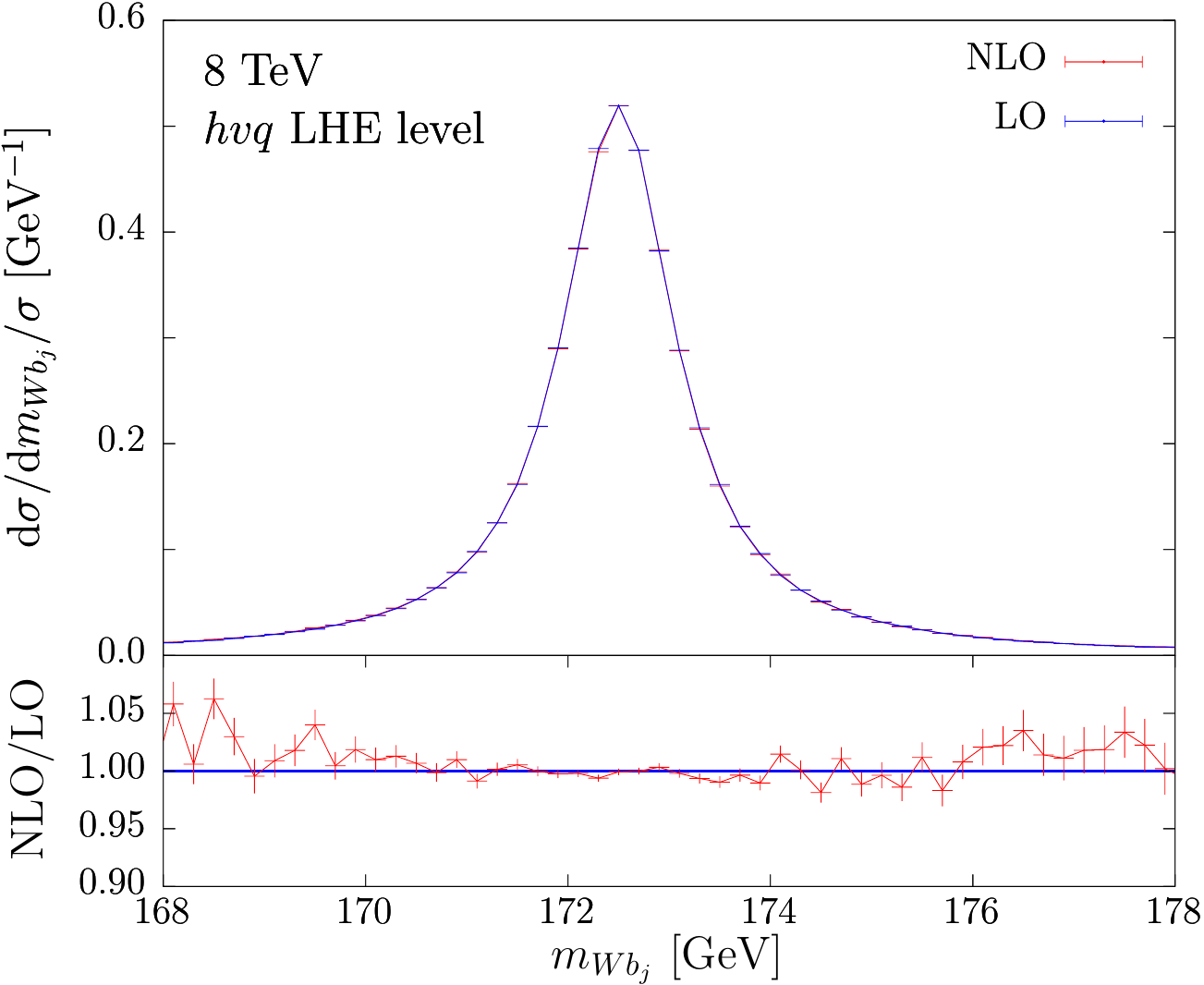}
\caption{${d\sigma}/{d \mwbj}$ distribution at LO~(blue) and at NLO~(red)
  obtained with the \hvq{} generator, normalized to 1 in the displayed range. In the
  bottom panel the ratio with the LO prediction is shown. }
\label{fig:hvqLHE}
\end{figure}
\begin{figure}[tb]
\centering
     \includegraphics[width=\wfigsing]{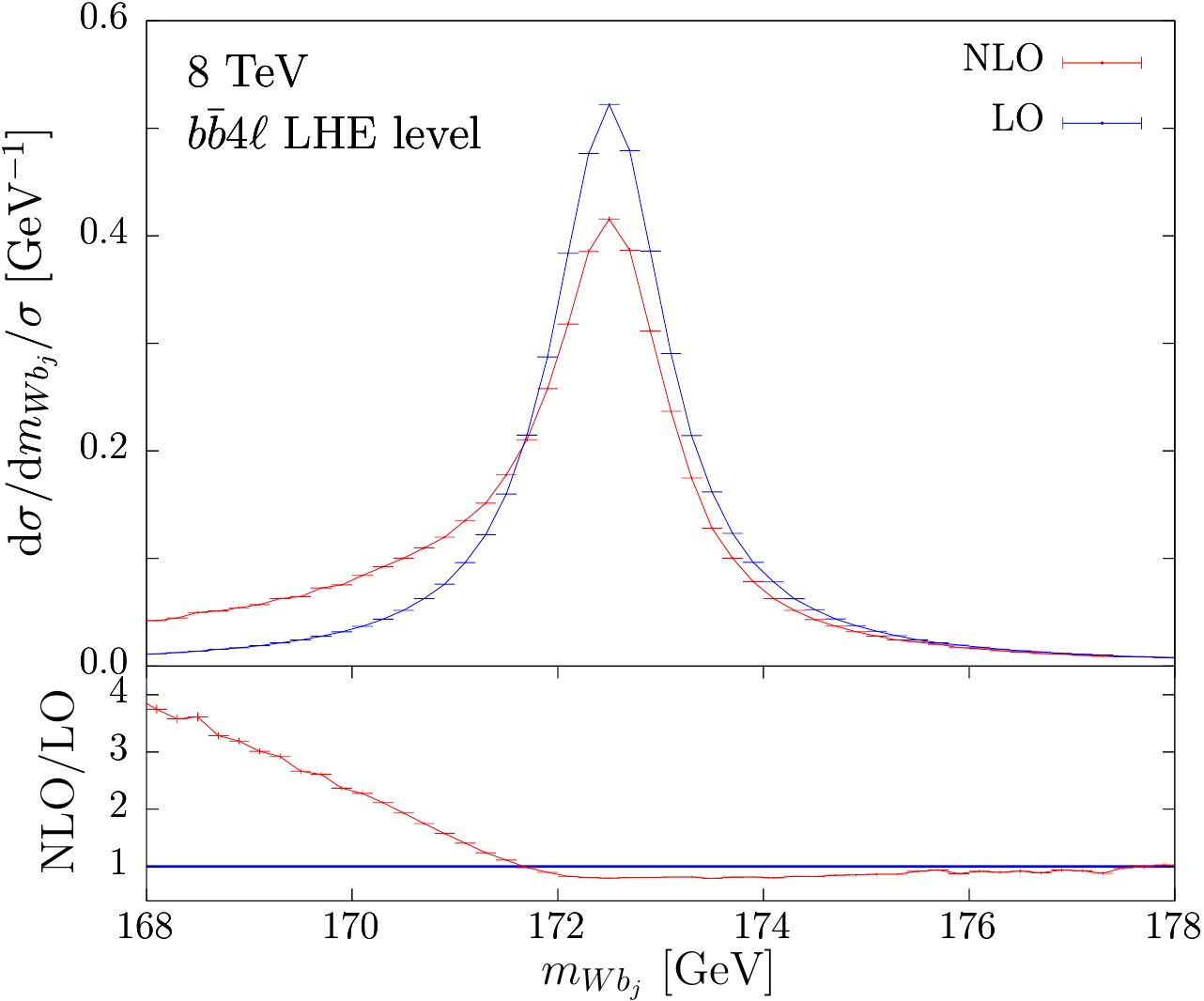}
\caption{${d\sigma}/{d \mwbj}$ distribution at LO~(blue) and at NLO~(red)
  obtained with the \bbfourl{} generator, normalized to 1 in the displayed range. In
  the bottom panel the ratio with the LO prediction is
  shown. }
\label{fig:bb4lLHE}
\end{figure}
In Figs.~\ref{fig:hvqLHE} and~\ref{fig:bb4lLHE} we compare \mwbj{},
normalized to 1 in the displayed range, at LO and NLO accuracy using the
\hvq{} and the \bbfourl{} generators respectively.  The \hvq{} generator
includes NLO corrections only in the production process. Thus the \mwbj{}
distributions at LO and NLO are very similar.  On the other hand, in the case
of the \bbfourl{} generator~(Fig.~\ref{fig:bb4lLHE}), we observe large
differences below the peak region. These differences are easily interpreted
as due to radiation outside the \bjet{} cone in the top-decay process.

\begin{figure}[tb]
\centering \includegraphics[width=\wfigsing]{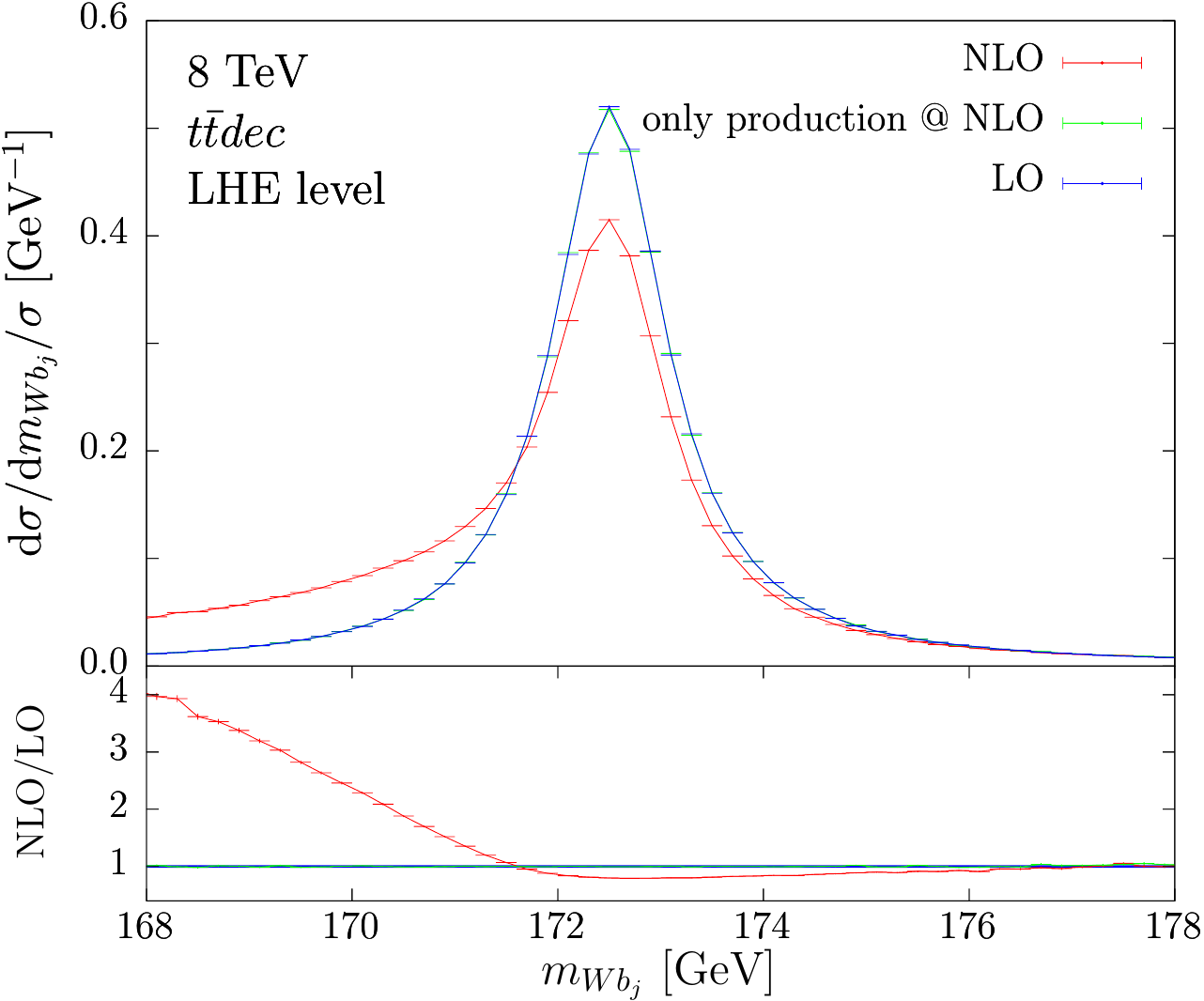}
   \caption{${d\sigma}/{d \mwbj}$ distribution with NLO accuracy in
     production and decay~(red), only in production~(green) and with LO
     accuracy~(blue) obtained with the \ttNLOdec{} generator, normalized to 1 in the
     displayed range. In the bottom panel the ratio with the LO prediction is
     shown.}
   \label{fig:ttdecLHE}
\end{figure}

The \ttNLOdec{} generator allows us to specify whether NLO accuracy is
required both in production and decay (default behaviour), or just in
production (by using the \pwgopt{nlowhich 1} option). In
Fig.~\ref{fig:ttdecLHE}
we compare the two options.
We see that our previous observation is confirmed: the impact of NLO
corrections in production leads to a roughly constant $K$-factor, while the
radiation from top decay affects the shape of the distribution below the peak
region.

A remaining important difference between the \hvq{} and the other two
generators has to do with the way the distribution of the top virtuality is
modeled. The \bbfourl{} and \ttNLOdec{} generators are guaranteed to yield
the correct virtuality distribution at the NLO and LO level, respectively.
This is not the case for the \hvq{} generator, where the resonance structure is
recovered by a reweighting procedure that does not guarantee LO accuracy.
\begin{figure}
\centering \includegraphics[width=\wfigsing]{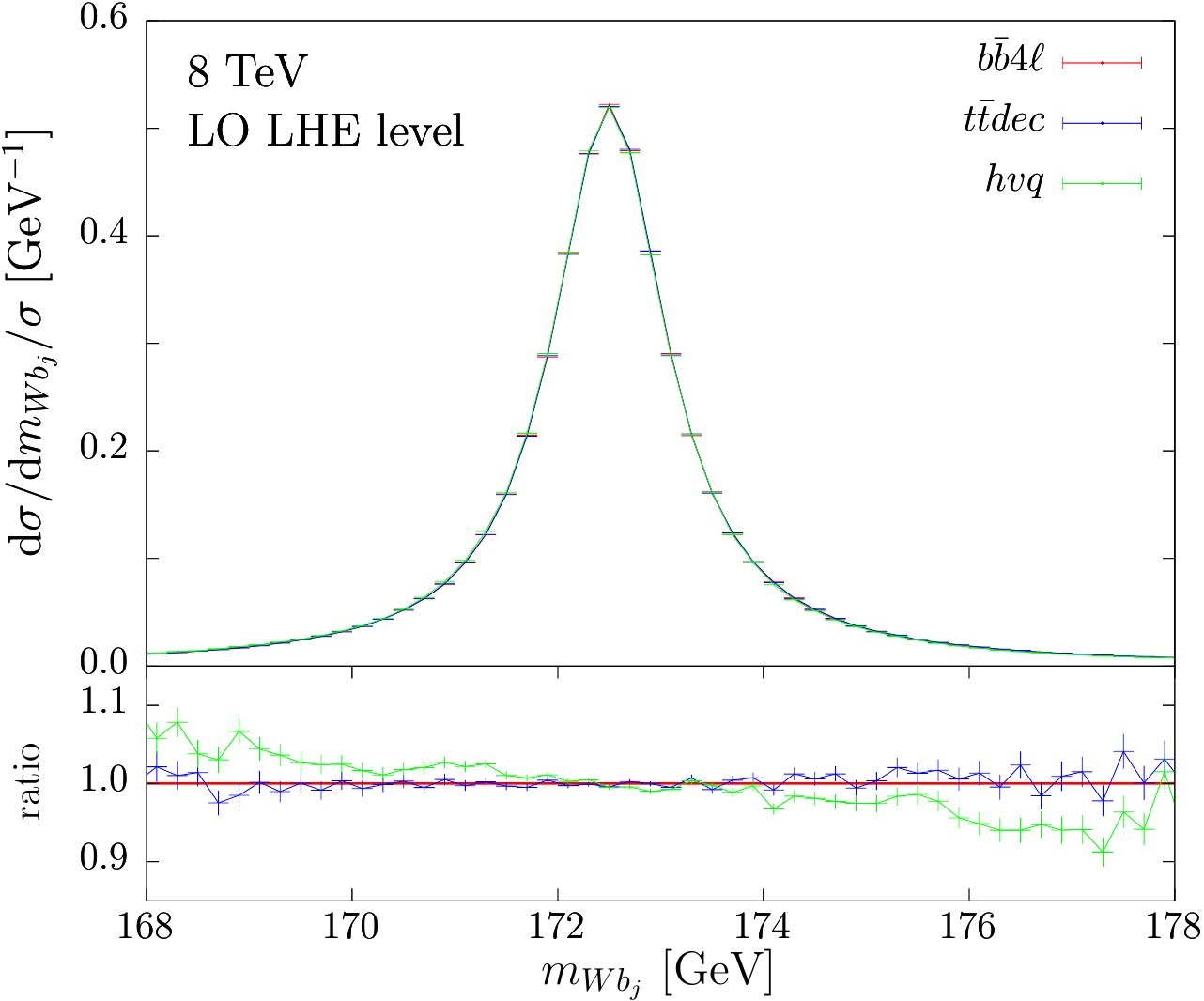}
   \caption{${d\sigma}/{d \mwbj}$ distribution at LO obtained with
     \bbfourl{}~(red), \ttNLOdec{}~(blue) and \hvq{}~(green), normalized to 1
     in the displayed range. In the bottom panel the ratio with the
     \bbfourl{} prediction is shown.}
   \label{fig:allLO}
\end{figure}
This is illustrated in Fig.~\ref{fig:allLO}, where we see that a
non-negligible (although not dramatic) difference in shape is present also at
the LO level between the \hvq{} and the other two generators.

\subsection{Shower effects}
\begin{figure}[tb]
\centering
\includegraphics[width=\wfigsing]{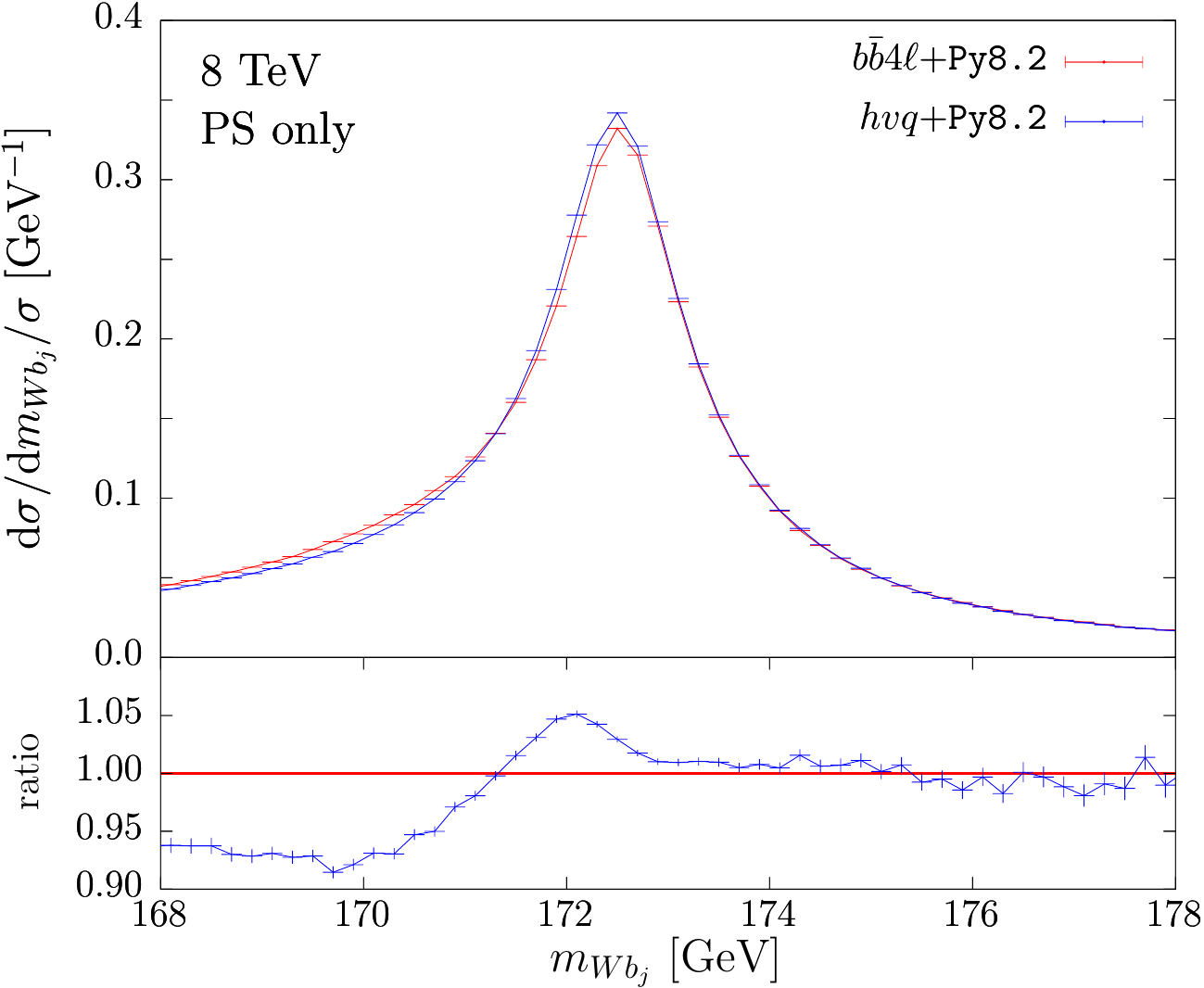}%
\caption{${d\sigma}/{d \mwbj}$ distribution, normalized to 1 in the displayed range,
  obtained with \bbfourl{}~(red) and \hvq{}~(blue) at the NLO+PS level using
  \PythiaEightPtwo{}.}
\label{fig:bb4l+hvq-PS}
\end{figure}

We now examine how the shower, i.e.~the radiation beyond the hardest one,
affects our distributions. First of all, we anticipate an important effect in
\hvq{}, since in this case radiation in decay is fully generated by the
shower. We thus expect a raise of the low mass tail in the \mwbj{}
distribution, comparable in size to the one observed in the \bbfourl{} and
\ttNLOdec{} generators at the LHE level.  Conversely, in the \bbfourl{} and
\ttNLOdec{} cases, we expect smaller shower corrections, since the hardest
radiation in decay is already included at the LHE level.
\begin{figure}[tb]
\centering
\includegraphics[width=\wfigdoub]{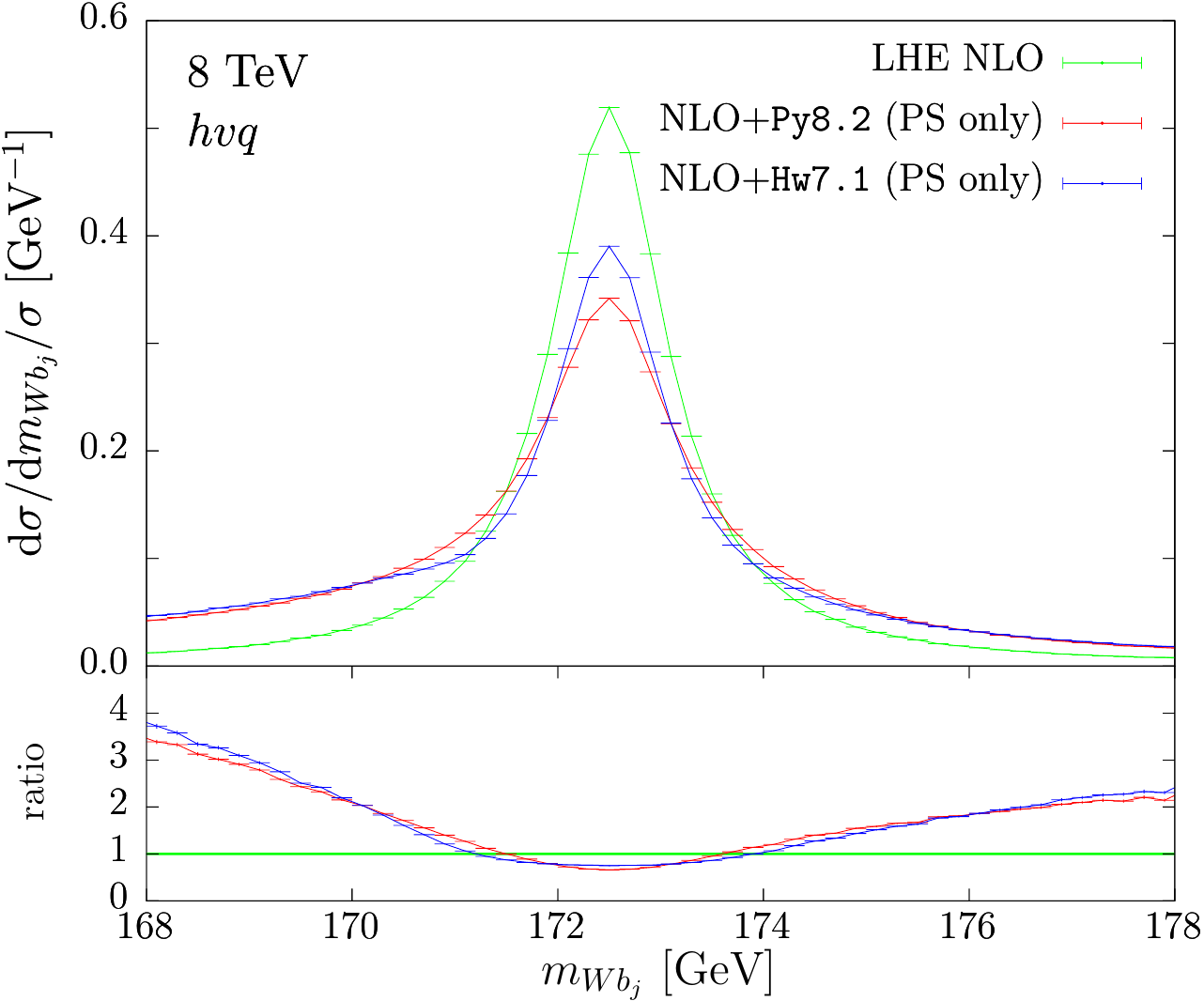} \fignewline
\includegraphics[width=\wfigdoub]{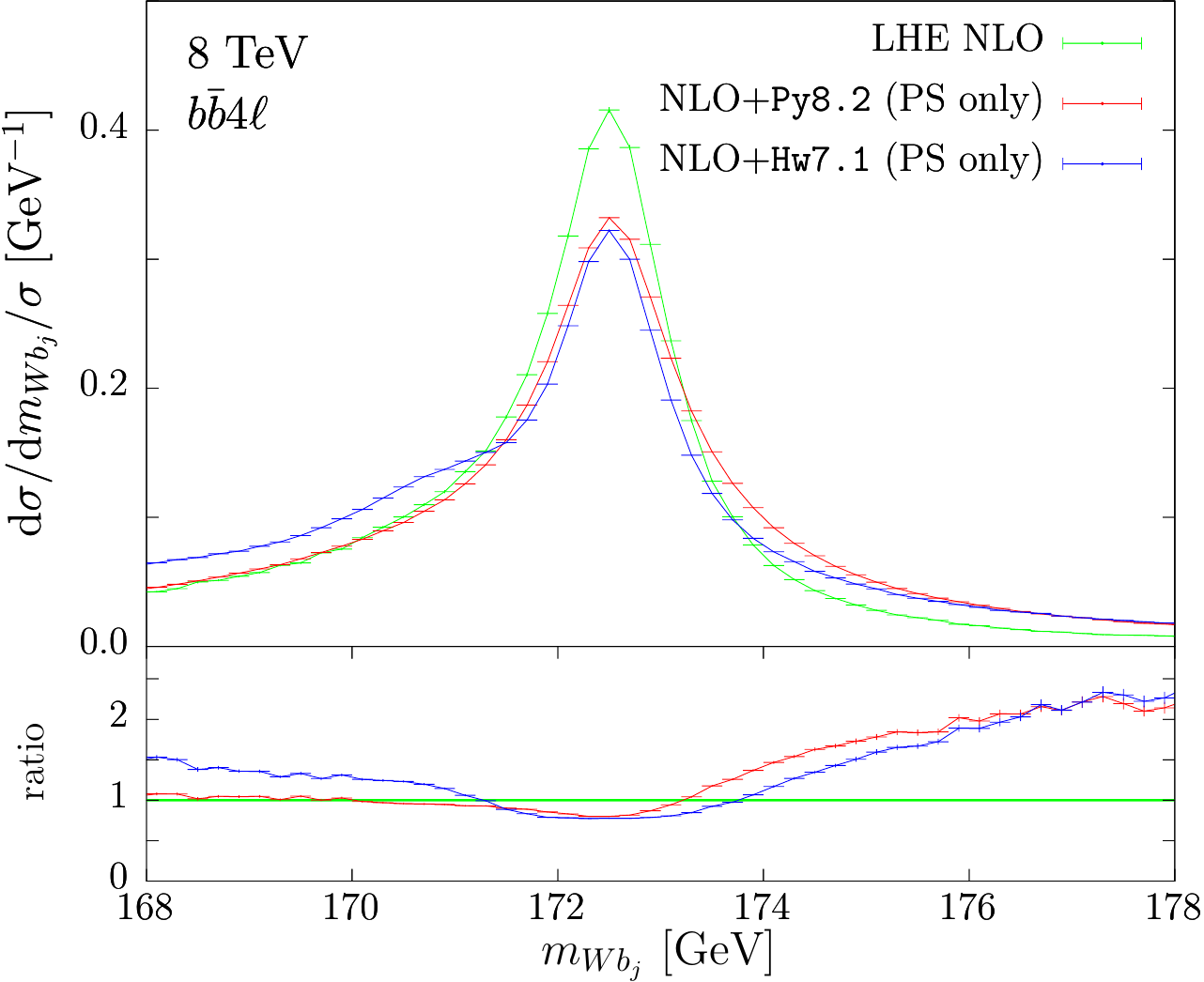}
\caption{${d\sigma}/{d \mwbj}$ distribution obtained with \hvq{}~(left pane) and
  \bbfourl{}~(right pane) at the NLO LHE level~(green), and at NLO+shower (in
  red \PythiaEightPtwo{} and in blue \HerwigSevenPone{}), normalized to 1 in
  the displayed range. In the bottom panel the ratio with the NLO LHE is
  shown.}
\label{fig:NLO-PS}
\end{figure}
This is illustrated in Fig.~\ref{fig:NLO-PS}, where we clearly see that in the
\hvq{} case there is an important increase of the cross section below the
peak.  On the other hand, in the \bbfourl{} case this increase is minor or
even absent, depending upon which shower program is used. In both cases, we
see an enhancement in the region above the peak. This is attributed to shower
radiation that is captured by the \bjet{} cone.  We observe that, after
shower, the \hvq{} result becomes qualitatively very similar to the
\bbfourl{} one, as shown in Fig.~\ref{fig:bb4l+hvq-PS}.

The inclusion of the shower in \ttNLOdec{} leads to effects
similar to those observed in \bbfourl{}.

\subsection{Hadronization and underlying events}

\begin{figure}[tb]
\centering
\includegraphics[width=\wfigdoub]{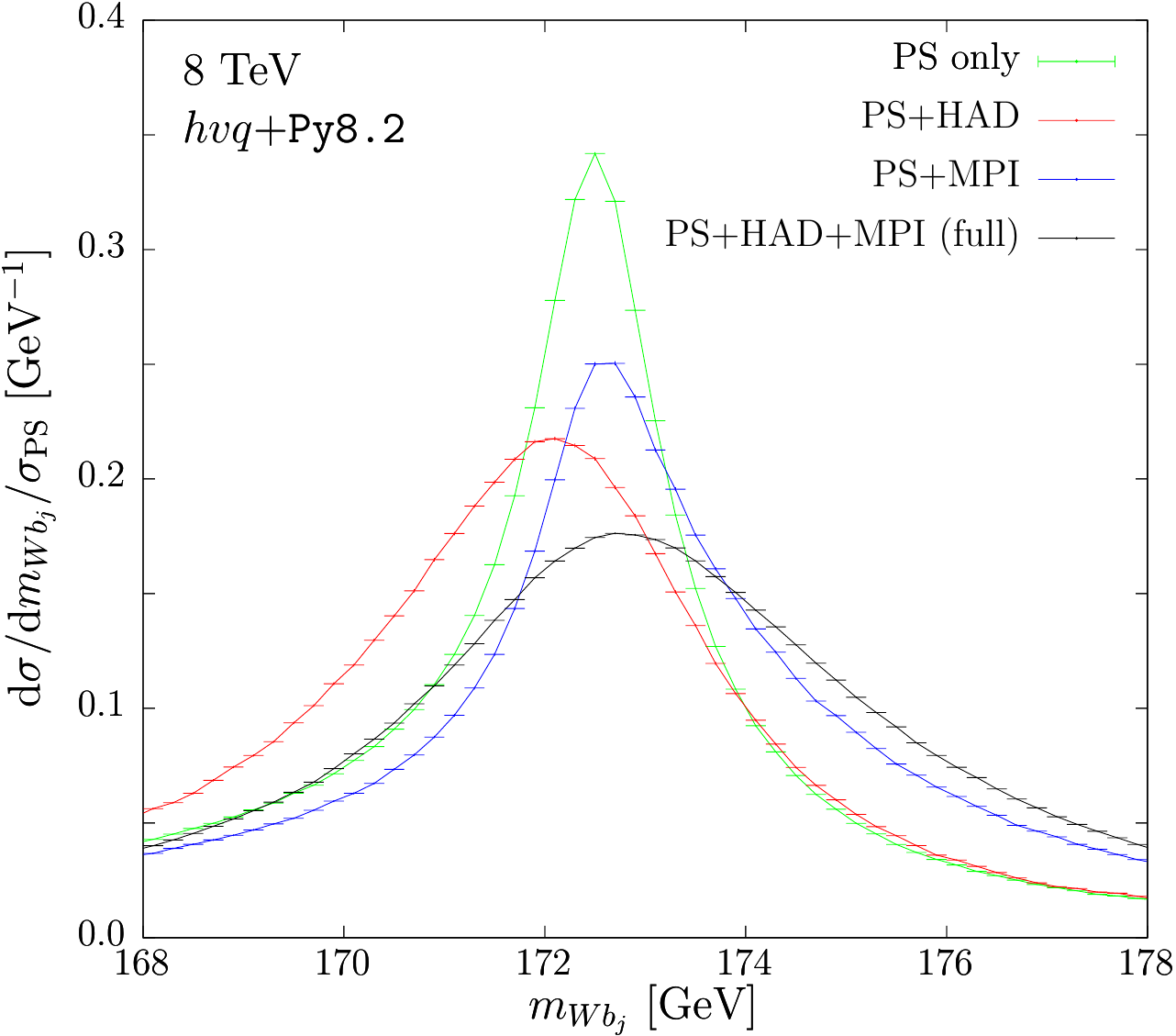}\fignewline
\includegraphics[width=\wfigdoub]{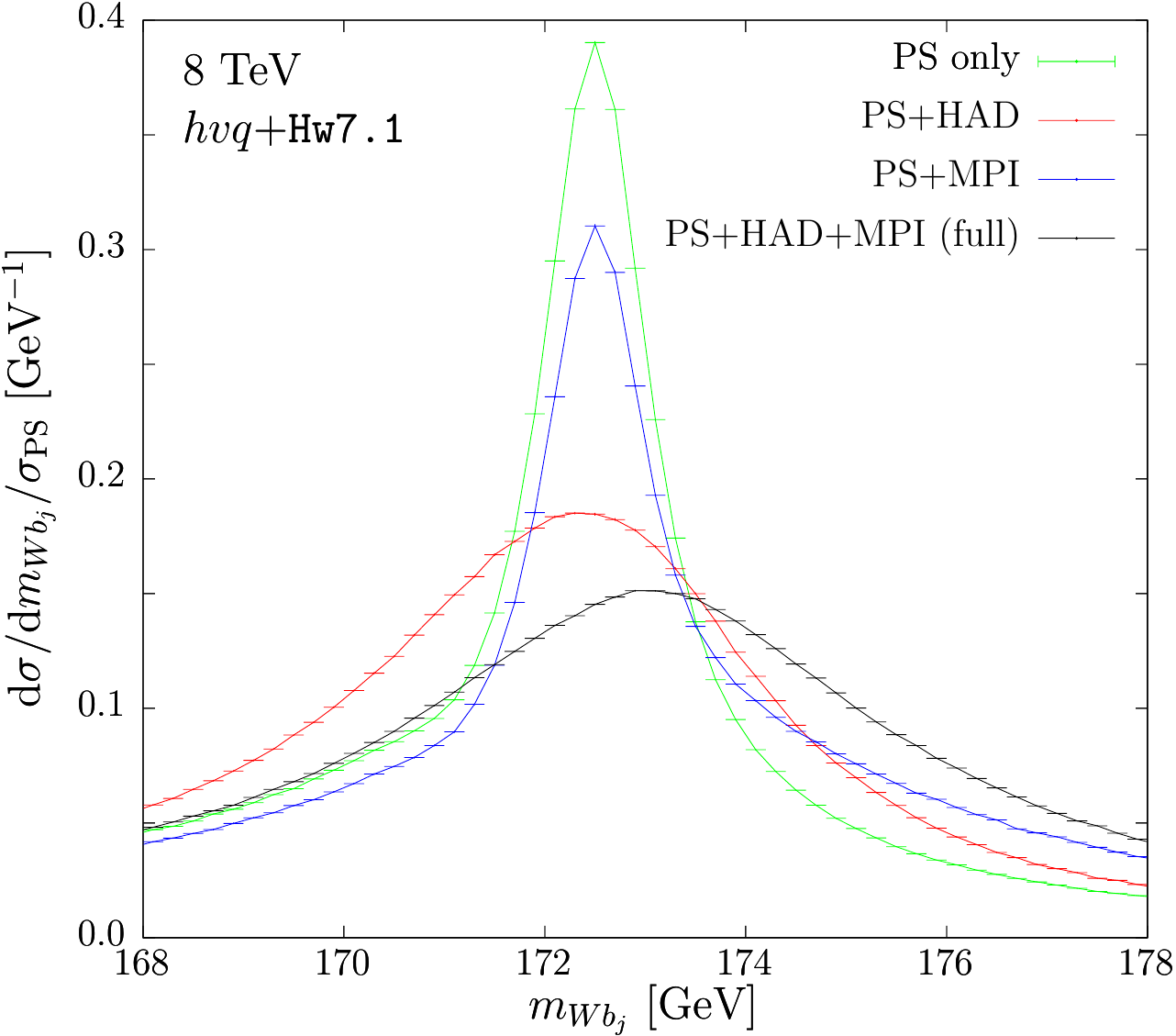}
\caption{${d\sigma}/{d \mwbj}$ distribution obtained with \hvq{} interfaced with
  \PythiaEightPtwo{}~(left panel) and \HerwigSevenPone{}~(right panel). In
  green, the NLO+PS results; in red, hadronization effects are included; in
  blue, NLO+PS with multi-parton interactions~(MPI); and in black, with
  hadronization and MPI effects. The curves are normalized using the NLO+PS
  cross section in the displayed range.}
\label{fig:hvq-PS-NP}
\end{figure}
In Fig.~\ref{fig:hvq-PS-NP} we show the effect of hadronization and
multi-parton interactions~(MPI), as modeled by \PythiaEightPtwo{} and
\HerwigSevenPone{}, when interfaced to the \hvq{} generator.  We can see the
large effect of the hadronization on the final distribution.  This effect is
also considerably different between \PythiaEightPtwo{} and \HerwigSevenPone{}. There
are two main features that emerge in these plots. First of all, as expected,
the MPI raise the tail of the distributions above the peak. In fact,
MPI-generated particles are deposited in the \bjet{} cone, thus increasing
the \bjet{} energy.  Hadronization widens the peak for both
generators. However, in the \PythiaEightPtwo{} case, we also observe a clear
enhancement of the low mass region, that is not as evident in the
\HerwigSevenPone{} case.  In the combined effect of hadronization and MPI,
\HerwigSevenPone{} has a wider peak. On the other hand, the high tail enhancement
seems similar in the two generators.

We remark that the different mechanisms that lead to an increased cross
section above and below the top peak depend on the jet radius parameter
$R$. By increasing (or decreasing) $R$, the peak position is shifted to the left
(or right).  Furthermore, differences in the implementation of radiation from
the resonances, the hadronization model and the underlying events can also
shift the peak, leading eventually to a displacement of the extracted top
mass, that should be carefully assessed.

\section{Methodology}
\label{sec:methodology}
In the following sections we will examine various sources of theoretical
errors in the top-mass extraction, focusing upon three classes of
observables: the reconstructed mass peak, the peak of the \bjet{} energy
spectrum~\cite{Agashe:2016bok}, and the leptonic observables of
Ref.~\cite{Frixione:2014ala}.

The reconstructed mass observable bears a nearly direct relation with the top
mass. If two generators with the same $\mt$ input parameter yield a
reconstructed mass peak position that differ by a certain amount, we can be
sure that if they are used to extract the top mass they will yield results
that differ by roughly the same amount in the opposite direction. Of course, this is
not the case for other observables.  In general, for an observable $O$
sensitive to the top mass, we will have
\begin{equation}
  O = O_c + B \(\mt-\mtc\) + {\cal O}\left(\(\mt-\mtc\)^2\right),
  \label{eq:linearfitfunc}
\end{equation}
where $\mt$ is the input mass parameter in the generator, and
$\mtc=172.5$~GeV is our reference central value for the top mass.  $O_c$ and
$B$ differ for different generators or generator setups.  Given an
experimental result for $O$, $O_{\rm exp}$, the extracted mass value is
\begin{equation}
  \mt = \mtc+\frac{O_{\rm exp}-O_{c}}{B}\,.
\end{equation}
By changing the generator setup, $O_{\rm c}$ and $B$ will assume the
values $O_{\rm c}'$ and $B'$, and will yield a different extracted mass $m'_t$.
We will thus have
\begin{equation}
  {m'_t-\mt} = \frac{O_{\rm c}-O_{\rm c}'}{B} + \(O_{\rm exp}-O_{\rm c}'\)\frac{B-B'}{BB'}\,.
\end{equation}
The second term is parametrically smaller, of one order higher in the
deviation between the two generators, if we assume that at least one of them
yields a $\mt$ value sufficiently close to $\mtc$. We thus have
\begin{equation}
  \label{eq:delta_mt}
  m'_t-\mt \approx \frac{O_{\rm c}-O_{\rm c}'}{B}\,.
\end{equation}
In practice, in the following, we will compute the $B$ parameter
using the \hvq{} generator, that is the fastest one. We also
checked that using the other generators for this purpose yields
results that differ by at most 10\%{}.

\section{Reconstructed top mass distribution $\boldsymbol{\mwbj}$}
\label{sec:mwbj}
The peak of the reconstructed mass \mwbj{}, defined in
Sec.~\ref{sec:physicsObjects}, is a representative of all the direct
measurement methods.  Our simplifying assumptions, that the $b$ jets are
unambiguously identified and the neutrinos are fully reconstructed, including
their sign, lead to an ideal resolution on the top peak that is not
realistic. We thus compute these distributions also introducing a smearing
that mimics the experimental systematics. This very crude approach allows us
to concentrate more on theoretical issues rather then experimental ones. For
example, if by using two different generators (or the same generator with
different settings) we find differences in the extracted mass using our ideal
\mwbj{} observable, we would be forced to conclude that there is an
irreducible theoretical error (i.e.~an error that cannot be reduced by
increasing the experimental accuracy) on the mass measurement.  The same
problem in case of the smeared distribution should instead be considered less
severe, since the corresponding error may be reduced if the experimental
resolution is improved.

We remark that also ``irreducible'' errors (according to the definition given
above) may in fact be reduced in practice. This is the case if one of the
generators at hand does not fit satisfactorily measurable distributions
related to top production. As an example, a generator may not fit reasonably
the profile of the $b$ jet, and we may be forced to change the allowed range
for the parameters that control it, possibly reducing the error.

In the following, we will compare our three generators interfaced to
\PythiaEightPtwo{}, and consider scale variation effects and PDF dependence.  In
order to investigate the sensitivity to the intensity of radiation from the
$b$ quark, we also consider different values of $\as$ as input.  We will then
investigate the \HerwigSevenPone{} and \PythiaEightPtwo{}
differences.\footnote{Unless specified otherwise, \PythiaEightPtwo{} and
  \HerwigSevenPone{} are setup to run in full hadron mode including shower,
  hadronization and multi-parton interactions.}

It is quite obvious that the coefficient $B$ of eq.~(\ref{eq:linearfitfunc})
should be very near 1 for the \mwbj{} observable.  The values for the $B$
coefficients that we have obtained with the three generators showered with
\PythiaEightPtwo{}, by a linear fit of the $\mt$ dependence of the \mwbj{}
distribution, are collected in Tab.~\ref{tab:mwbj_B_values}, and confirm our
expectation.
\begin{table}[tb]
  \centering
 { \begin{tabular}{l|c|c|}
    \cline{2-3}
   $\phantom{\Big|}$ & $B$, no smearing & $B$, smearing \\
    \cline{1-3}
    \multicolumn{1}{ |c|}{\hvq{}} &  $\phantom{\Big|}\Bfrommwbjhvq \pm \Berrfrommwbjhvq$ & $\Bfrommwbjsmearhvq \pm \Berrfrommwbjsmearhvq$ \\
    \cline{1-3}
    \multicolumn{1}{ |c|}{\ttbnlodec{}} & $\phantom{\Big|} \Bfrommwbjttdec \pm \Berrfrommwbjttdec$ & $\Bfrommwbjsmearttdec \pm \Berrfrommwbjsmearttdec$ \\
    \cline{1-3}
    \multicolumn{1}{ |c|}{\bbfourl{}} & $\phantom{\Big|} \Bfrommwbjbbfourl \pm \Berrfrommwbjbbfourl$ &  $\Bfrommwbjsmearbbfourl \pm \Berrfrommwbjsmearbbfourl$ \\
    \cline{1-3}
  \end{tabular}}
  \caption{Values for the $B$ coefficients of eq.~(\ref{eq:linearfitfunc})
    for the \mwbj{} peak position, for the non-smeared and smeared
    distributions~(see Sec.~\ref{sec:mwbj-cmp} for details), obtained with
    the \hvq{}, \ttbnlodec{} and \bbfourl{} generators showered with
    \PythiaEightPtwo{}.}
  \label{tab:mwbj_B_values}
\end{table}

\subsection{Comparison among the different NLO+PS generators}
\label{sec:mwbj-cmp}
We begin by showing comparisons of our three generators, interfaced with
\PythiaEightPtwo{}, for our reference top-mass value of 172.5~GeV.
\begin{figure}[tb]
\centering
\includegraphics[width=\wfigsing]{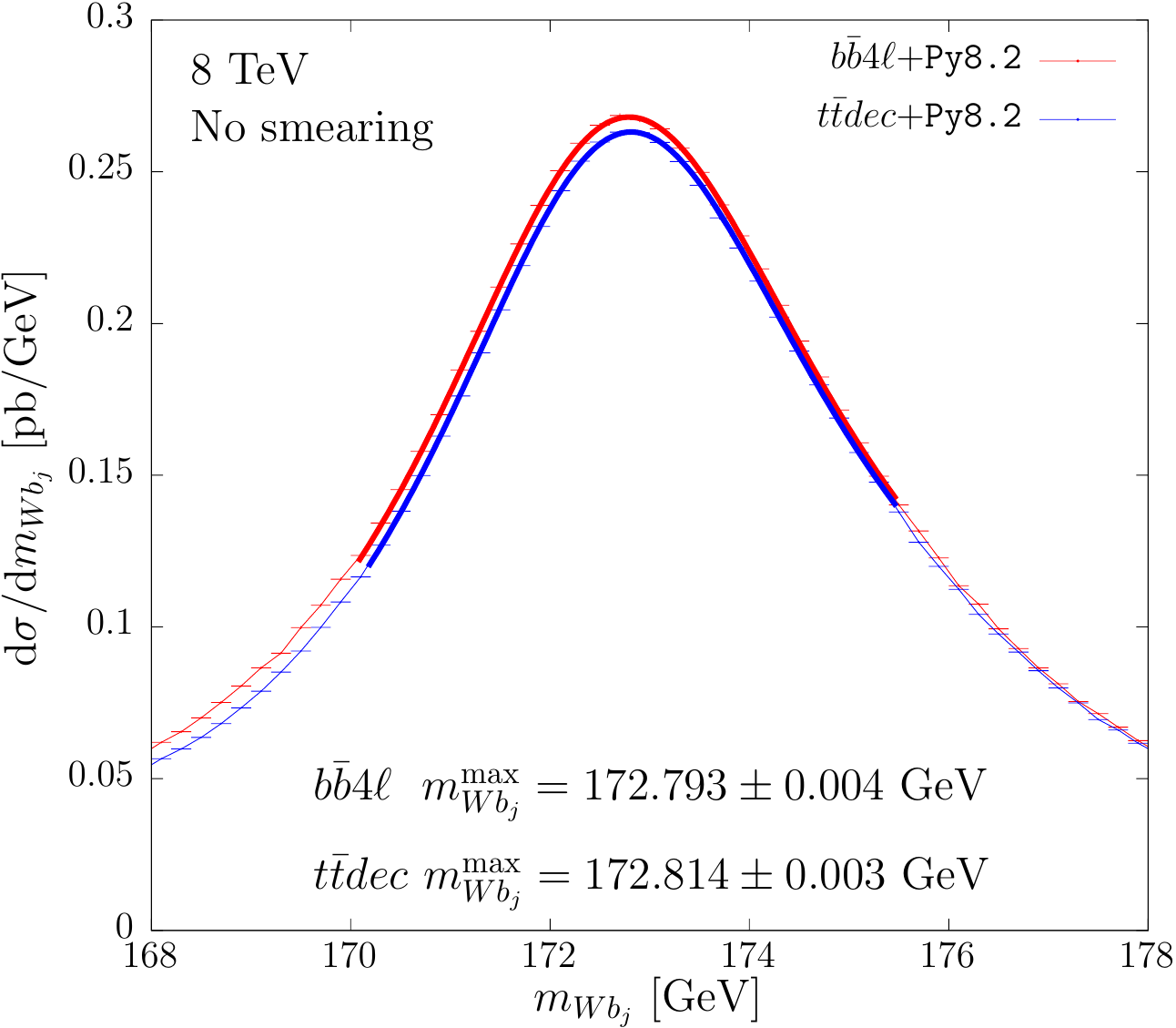}
\caption{${d\sigma}/{d \mwbj}$ distribution obtained with the \bbfourl{} and \ttNLOdec{}
  generators interfaced with \PythiaEightPtwo{}, for $\mt=172.5$~GeV.}
\label{fig:MassPeaks-py8-bb4l-ttb}
\end{figure}
We show in Fig.~\ref{fig:MassPeaks-py8-bb4l-ttb} the \mwbj{} distribution for
the \bbfourl{} and \ttNLOdec{} generators.  We see that the two generators
yield a very similar shape. We have extracted the position of the maximum by
fitting the distribution with a skewed Lorentzian function of the form
\begin{equation}
  y(\mwbj)=\frac{b[1+d(\mwbj-a)]}{(\mwbj-a)^2+c^2}+e\,.
\end{equation}
The peak $\mwbjmax$ is defined by
\begin{equation}
  \frac{ d \, y(\mwbj)}{ d \mwbj} \Big|_{\mwbj=\, \mwbjmax}
  = 0\,.
\end{equation}
The fitting procedure is described in \writeApp\ref{app:fit}.

As we can see from Fig.~\ref{fig:MassPeaks-py8-bb4l-ttb}, the \bbfourl{} and
\ttNLOdec{} results are very close to each other.  We take this as an
indication that interference effects in radiation and other off-shell
effects, that are included in \bbfourl{} but not in \ttNLOdec{}, have a very
minor impact on the peak position, at least if we consider a measurement with
an ideal resolution.

In order to mimic experimental resolution effects, we smear our distribution
with a Gaussian of width $\sigma=15$~GeV (that is the typical experimental
resolution on the reconstructed top mass)
\begin{equation}
f_{\rm smeared}(x) = \mathcal{N} \int dy \, f(y)\,
   \exp\left(-\frac{(y-x)^2}{2\sigma^2}\right)\,,
\end{equation}
where $\mathcal{N}$ is a normalization constant.
\begin{figure}[tb]
\centering
\includegraphics[width=\wfigsing]{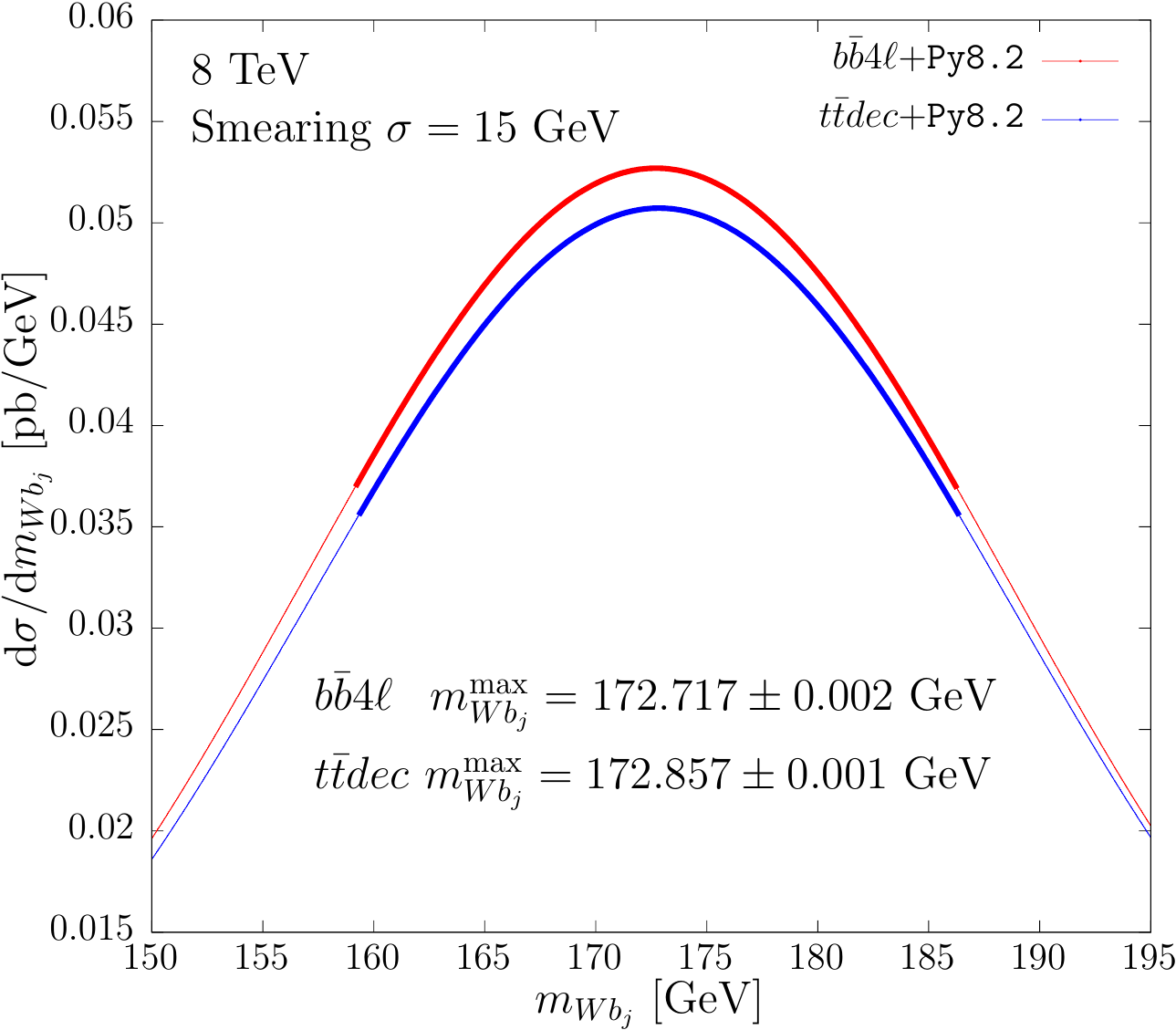}
\caption{Smeared ${d\sigma}/{d \mwbj}$ distribution obtained with the \bbfourl{}
  and \ttNLOdec{} generators interfaced with \PythiaEightPtwo{}, for
  $\mt=172.5$~GeV.}
\label{fig:MassPeaks-py8-bb4l-ttb-smeared}
\end{figure}
The results, obtained with the same fitting procedure, are shown in
Fig.~\ref{fig:MassPeaks-py8-bb4l-ttb-smeared}.  Smearing effects are such
that more importance is given to the region away from the peak, where there
are larger differences between the two generators, leading to a difference in
the peak position of \diffttdecbbfourl~MeV.

\begin{figure}[tb]
\centering
\includegraphics[width=\wfigsing]{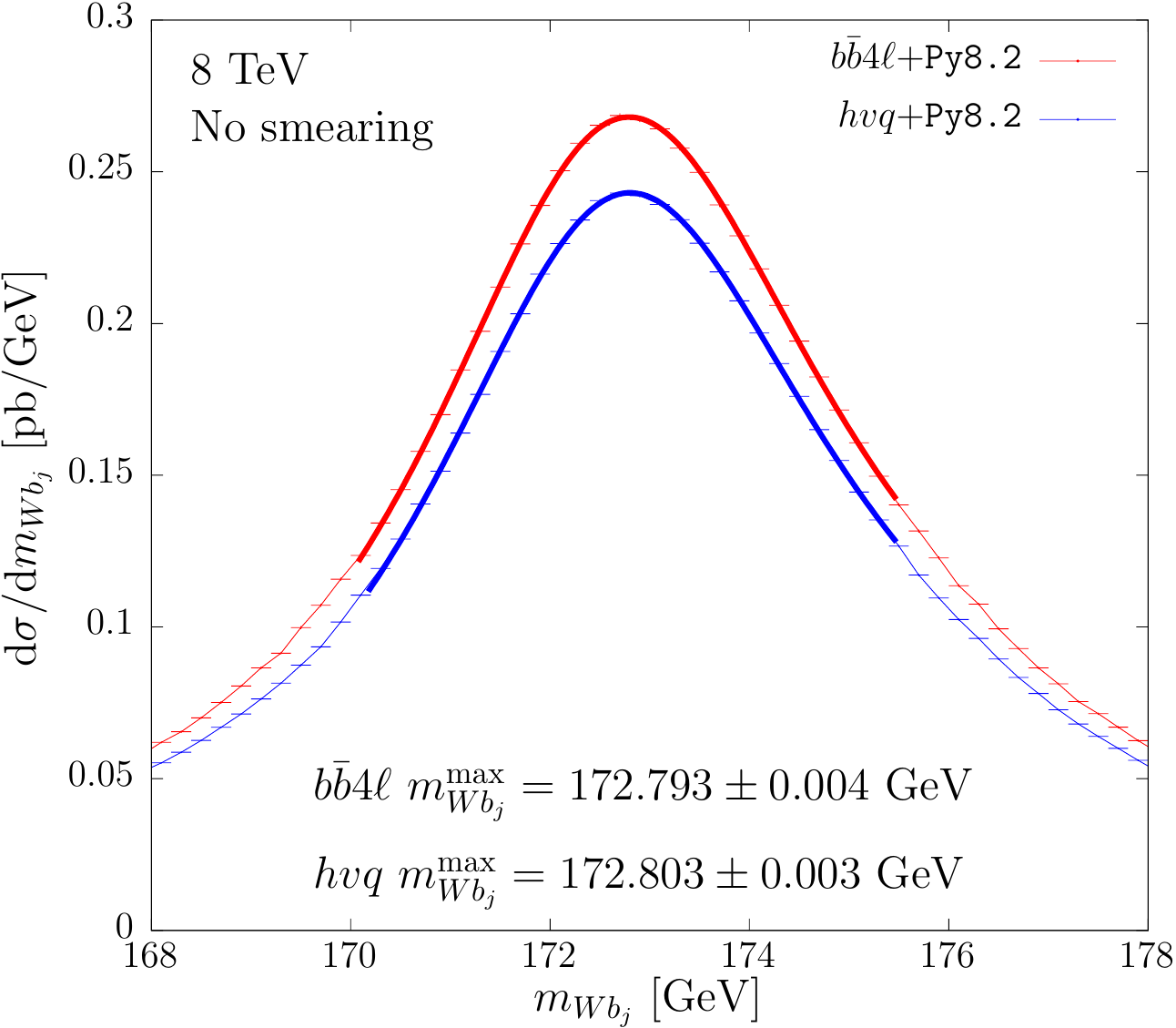}
\caption{${d\sigma}/{d \mwbj}$ distribution obtained with the
  \bbfourl{} and \hvq{} generators
  interfaced with \PythiaEightPtwo{}, for
  $\mt=172.5$~GeV.}
\label{fig:MassPeaks-py8-bb4l-hvq}
\end{figure}
\begin{figure}[tb]
\centering
\includegraphics[width=\wfigsing]{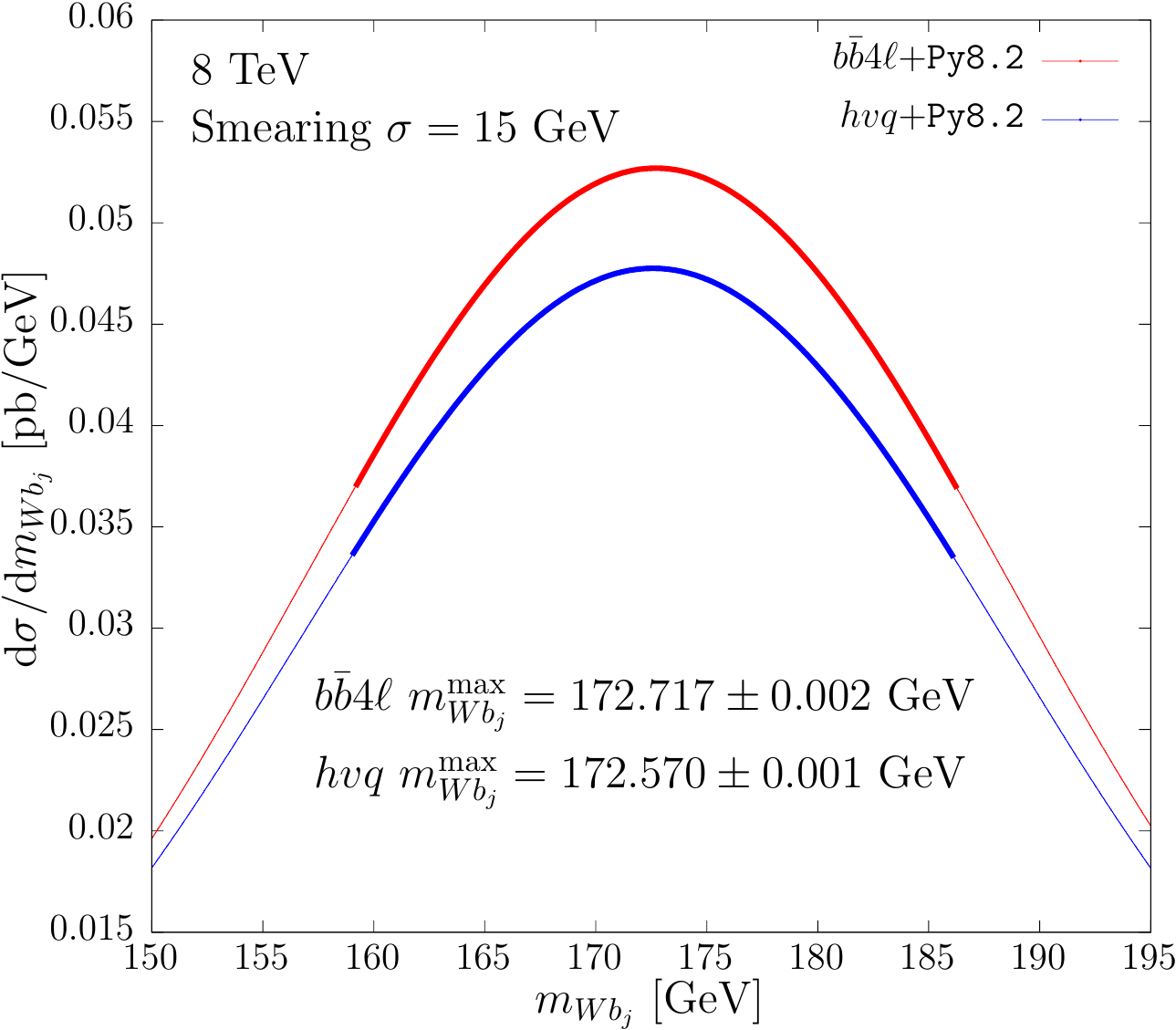}
\caption{Smeared ${d\sigma}/{d \mwbj}$ distribution obtained with the
  \bbfourl{} and \hvq{} generators interfaced with \PythiaEightPtwo{}, for
  $\mt=172.5$~GeV.}
\label{fig:MassPeaks-py8-bb4l-hvq-smeared}
\end{figure}

In Figs.~\ref{fig:MassPeaks-py8-bb4l-hvq} and
\ref{fig:MassPeaks-py8-bb4l-hvq-smeared},
we compare the \bbfourl{} and the \hvq{} generators in the non-smeared and
smeared case respectively.  We see a negligible difference in the peak
position in the non-smeared case, while, in the smeared case, the \hvq{}
generator differs from \bbfourl{} by \diffhvqbbfourl~MeV, similar in
magnitude to the case of \ttbnlodec{}, but with opposite sign.
These findings are summarized in
Tab.~\ref{tab:mwbj_showerOnly},
\begin{table*}[tb]
\centering
\resizebox{\textwidth}{!}
{ \begin{tabular}{l|c|c|c|c|}
 \cline{2-5}
 &  \multicolumn{2}{ |c|}{PS only}
 &  \multicolumn{2}{ |c|}{ \phantom{\Big|} full}\\
 \cline{2-5}
 & \phantom{\Big|} No smearing & 15~GeV smearing
 & \phantom{\Big|} No smearing & 15~GeV smearing \\
 \cline{1-5}
 \multicolumn{1}{ |c|  }{ \phantom{\Big|}  \bbfourl{}}
 & $172.522\pm  0.002$~GeV
 & $171.403\pm  0.002$~GeV
 & $172.793\pm  0.004$~GeV
 & $172.717\pm  0.002$~GeV
 \\ \cline{1-5}
 \multicolumn{1}{ |c|  }{ \phantom{\Big|}\ttbnlodec{} ${}-$ \bbfourl{}}
  &  $         -18 \pm            2 $~MeV
  &  $+         191 \pm            2 $~MeV
  &  $+          21 \pm            6 $~MeV
  &  $+         140 \pm            2 $~MeV
 \\ \cline{1-5}
 \multicolumn{1}{ |c|  }{ \phantom{\Big|}\hvq{} ${}-$ \bbfourl{}}
  &  $         -24 \pm            2 $~MeV
  &  $         -89 \pm            2 $~MeV
  &  $+          10 \pm            6 $~MeV
  &  $        -147 \pm            2 $~MeV
 \\ \cline{1-5}
\end{tabular}
}
\caption{Differences in the $\mwbj{}$ peak position for $\mt$=172.5~GeV for
  \ttbnlodec{} and \hvq{} with respect to \bbfourl{}, showered with
  \PythiaEightPtwo{}, at the NLO+PS level and at the full hadron level.}
\label{tab:mwbj_showerOnly}
\end{table*}
where we also include results obtained at the shower level.

We notice that \hvq{}, in spite of the fact that it does not implement NLO
corrections in top decay, yields results and distributions that are quite
close to those of the most accurate \bbfourl{} generator.  This is due to the
fact that \PythiaEightPtwo{} includes matrix-element corrections~(MEC) in top decay by
default, and MEC are equivalent, up to an irrelevant normalization factor, to
next-to-leading order corrections in decay.  This observation is confirmed by
examining, in Tab.~\ref{tab:mwbj_MEC},
\begin{table*}[tb]
\centering
\resizebox{\textwidth}{!}
{ \begin{tabular}{l|c|c|c|c|}
 \cline{2-5}
 &  \multicolumn{2}{ |c|}{ \phantom{\Big|} No smearing}
 &  \multicolumn{2}{ |c|}{15~GeV smearing} \\
 \cline{2-5}
 & \phantom{\Big|} MEC  & MEC ${}-$ no  MEC
 &  MEC   & MEC ${}-$ no MEC\\
 \cline{1-5}
 \multicolumn{1}{ |c|  }{ \phantom{\Big|}\bbfourl{}}
 & $172.793\pm   0.004$~GeV
  &  $         -12 \pm            6 $~MeV
 & $172.717\pm   0.002$~GeV
  &  $+          55 \pm            2 $~MeV
 \\ \cline{1-5}
 \multicolumn{1}{ |c|  }{ \phantom{\Big|}\ttbnlodec{}}
 & $172.814\pm   0.003$~GeV
  &  $          -4 \pm            5 $~MeV
 & $172.857\pm   0.001$~GeV
  &  $         -26 \pm            2 $~MeV
 \\ \cline{1-5}
 \multicolumn{1}{ |c|  }{ \phantom{\Big|}\hvq{}}
 & $172.803\pm   0.003$~GeV
  &  $+          61 \pm            5 $~MeV
 & $172.570\pm   0.001$~GeV
  &  $+         916 \pm            2 $~MeV
 \\ \cline{1-5}
\end{tabular}
}
\caption{$\mwbj{}$ peak position for $\mt$=172.5~GeV obtained with the three
  different generators, showered with \PythiaEightPtwo{}+MEC~(default). We
  also show the differences between \PythiaEightPtwo{}+MEC and
  \PythiaEightPtwo{} without MEC.  }
\label{tab:mwbj_MEC}
\end{table*}
the impact of the MEC setting on our predictions. When MEC are switched off,
we see a considerable shift, near 1~GeV, in the \hvq{} result for the peak
position in the smeared distribution, and a very minor one in the \bbfourl{}
and \ttbnlodec{} generators, that include the hardest emission off $b$
quarks.  Thus, we conclude that the MEC in \PythiaEightPtwo{} do a decent job in
simulating top decay as far as the \mwbj{} distribution is concerned. The
remaining uncertainty of roughly \diffaverage~MeV in the case of both \hvq{}
and \ttbnlodec{} generators, pulling in opposite directions, is likely due to
the approximate treatment of off-shell effects.

\subsubsection{Renormalization- and factorization-scale dependence}
\label{sec:fac_ren_scales}
In this section, we study the dependence of our results on the
renormalization and factorization scales ($\muR$ and $\muF$), that gives an
indication of the size of higher-orders corrections.  We varied $\muR$ and
$\muF$ around the central scale $\mu$ defined in eqs.~\eqref{eq:centralscale}
and~(\ref{eq:centralscaleZ}) according to
\begin{equation}
  \muR = \KR\, \mu \, , \quad  \muF = \KF \, \mu \, ,
\end{equation}
where $(\KR,\KF)$ are varied over the following combinations
\beq \label{eq:scalechoices}
\bigg\{  (1,1), \, (2,2),\, \left( \frac{1}{2}, \frac{1}{2} \right),
  \, (1,2), \, \left( 1, \frac{1}{2} \right), \, (2,1),
  \, \left( \frac{1}{2},1 \right) \bigg\}.  
\eeq
We take $\KR=\KF=1$ as our central prediction.  We find that for \bbfourl{}
there is a non-negligible scale dependence, that in the smeared case yields a
theoretical uncertainty of~${}^{+\varscalemax}_{-\varscalemin}$~MeV.  For
\ttbnlodec{} and \hvq{} this uncertainty is smaller than
\varscaleothers~MeV. This is due to the fact that, in the last two
generators, the NLO corrections are performed for on-shell tops, and the top
width is subsequently generated with a smearing procedure. Thus, NLO
corrections remain constant around the top peak, leading to a constant scale
dependence. This leads to an underestimate of scale uncertainties in
\ttbnlodec{} and \hvq{}.

\subsubsection{PDF set dependence}
\label{sec:PDF_dependence}
We evaluated the dependence from the PDFs
by considering the central member of the following PDF sets:
\begin{itemize}
\item {\tt MSTW2008nlo68cl ($\as(\mZ)=0.120179$)} (default)~\cite{Martin:2009iq},
\item {\tt PDF4LHC15\_nlo\_30\_pdfas ($\as(\mZ)=0.118$)}~\cite{Butterworth:2015oua}\,,
\item {\tt CT14nlo ($\as(\mZ)=0.118$)}~\cite{Dulat:2015mca}\,,
\item {\tt MMHT2014nlo68cl ($\as(\mZ)=0.120$)}~\cite{Harland-Lang:2014zoa}\,,
\item {\tt NNPDF30\_nlo\_as\_0118 ($\as(\mZ)=0.118$)}~\cite{Ball:2014uwa}\,.
\end{itemize}
We generated the events by using the {\tt MSTW2008nlo68cl} set, and obtained
all other predictions using the internal reweighting facility of the
\POWHEGBOX{}. We find that the corresponding differences in the \mwbj{} peak
position are typically below \varPDF~MeV and the variations are very similar
for all the NLO+PS generators.

We also generated a sample using the central parton-distribution function of
the {\tt PDF4LHC15\_nlo\_30\_pdfas} set, and, by reweighting, all its
members, within the \hvq{} generator.  In this case, our error is given by the
sum in quadrature of all deviations.  We get a variation of
\pdferrorhvqnosmear~MeV in the non-smeared case, and \pdferrorhvqsmear~MeV for
the smeared distribution.  We find that the variation band obtained in this
way contains the central value results for the different PDF sets that we
have considered. It thus makes sense to use this procedure for the estimate
of PDF uncertainties.  On the other hand, reweighting for the 30 members of
the set in the \bbfourl{} case is quite time consuming, since the virtual
corrections are recomputed for each weight. We thus assume that the PDF
uncertainties computed in the \hvq{} case are also valid for the \bbfourl{}
and \ttbnlodec{} cases, since the dependence on the PDF is mostly due to the
implementation of the production processes, and all our generators describe
it at NLO accuracy, and since we have previously observed that by reweighting
to several PDF sets we get very similar variations for all generators.

In general, PDF uncertainties are rather small. This is probably due to the
fact that, in order to shift the position of the peak, some differences must
be present in the modeling of final-state radiation~(FSR). These differences
may arise from differences in $\as$. However, reweighting in \POWHEG{} only
affects the inclusive cross section, and not the radiation, and thus
final-state radiation is not modified by these changes.

\subsubsection{Strong-coupling dependence}
\label{sec:as_dependence}
In \POWHEGBOX{} the scale used to generate the emissions is the transverse
momentum of the radiation (with respect to the emitter). At the moment,
facilities to study uncertainties due to variations of this scheme are not
available. On the other hand, these uncertainties would lead to a different
radiation pattern around the $b$ jet, that can in turn have a non-negligible
effect on the reconstructed mass.

The simplest way at our disposal for studying the sensitivity of the
reconstructed mass to the intensity of radiation from the $b$ quark is by
varying the value of $\as$. To this end we use the {\tt NNPDF30\_nlo\_as115}
and {\tt NNPDF30\_nlo\_as121} sets, where $\as(\mZ)$=0.115 and
$\as(\mZ)$=0.121, respectively. As stated earlier, we cannot use the
\POWHEG{} reweighting facility in order to study this effect, and thus we
generated two dedicated samples (see Tab.~\ref{tab:samples}).

We found that the extracted peak positions in the smeared \mwbj{}
distributions for the two extreme values of $\as$ differ by
\deltaalphasbbfourl~MeV for the \bbfourl{} generator, by
\deltaalphasttdec~MeV for the \ttbnlodec{} generator and by
\deltaalphashvq~MeV for \hvq{}. The small $\as$-sensitivity in the \hvq{}
case is expected, since, in this case, radiation in decays is handled by the
shower, and thus should be studied by varying shower parameters.  In the
\bbfourl{} and \ttbnlodec{} case, the variation is very similar, since they
both include NLO radiation in decay, and the direction of the variation is as
expected, i.e.~the peak position is larger for the smaller $\as$ value, due
to the reduced loss of energy outside the jet cone. Differences in the case
of non-smeared distributions are in all cases not larger than
\deltaalphasunsmear~MeV.

We can estimate the typical scale of radiation in top decay as being of the
order of 30~GeV, i.e.~one-half of the typical $b$ energy in the top rest
frame.  The ratio of the upper to lower $\as(\mZ)$ values that we have
considered is 1.052, and it becomes 1.06 at a scale of $30$~GeV. On the other
hand, a scale variation of a factor of two above and below $30$~GeV yields a
variation in $\as$ of about 26\%.  This can be taken as a rough indication
that a standard scale variation would yield to a variation in the peak
position that is more than a factor four larger than the one obtained by
varying $\as$.

\subsubsection{Matching uncertainties}
The {\tt FSREmission} veto procedure (i.e.~implementation 1 of
Sec.~\ref{sec:PY8_different_showers}) represents the most accurate way to
perform the vetoed shower on the \POWHEGBOX{} generated events, because it
uses the \POWHEG{} definition of transverse momentum rather than the
\PythiaEightPtwo{} one. The {\tt ScaleResonance} procedure (i.e.~method 2) introduces
a mismatch (see Sec.~\ref{sec:PY8_different_showers}) that we take as an
indication of the size of the matching uncertainties.  The extracted peak
position for the \bbfourl{} and \ttbnlodec{} with the two matching procedures
are summarized in Tab.~\ref{tab:mass_extraction-matching}.
\begin{table*}[tb]
\centering
{ \begin{tabular}{l|c|c|c|c|}
 \cline{2-5}
 &  \multicolumn{2}{ |c|}{ \phantom{\Big|} No smearing}
 &  \multicolumn{2}{ |c|}{15~GeV smearing} \\
 \cline{2-5}
 & \phantom{\Big|} {\tt SR} & {\tt SR} ${}{-}$ {\tt FSR}
 & \phantom{\Big|} {\tt SR} & {\tt SR} ${}{-}$ {\tt FSR}\\
 \cline{1-5}
 \multicolumn{1}{ |c|  }{ \phantom{\Big|}\bbfourl{}}
 & $172.816\pm   0.004$~GeV 
  &  $+          23 \pm            6 $~MeV
 & $172.737\pm   0.002$~GeV
  &  $          20 \pm            2 $~MeV
 \\ \cline{1-5}
 \multicolumn{1}{ |c|  }{ \phantom{\Big|}\ttbnlodec{}}
 & $172.812\pm   0.004$~GeV 
  &  $          -1 \pm            5 $~MeV
 & $172.878\pm   0.001$~GeV
  &  $          21 \pm            2 $~MeV
 \\ \cline{1-5}
\end{tabular}
}
\caption{$\mwbj{}$ peak position for $\mt$=172.5~GeV obtained with the
  \bbfourl{} and \ttbnlodec{} generators, showered with \PythiaEightPtwo{},
  for the {\tt ScaleResonance}~({\tt SR}) veto procedure. The differences
  with {\tt FSREmission}~({\tt FSR}), that is our default, are also shown.}
\label{tab:mass_extraction-matching}
\end{table*}
We can see that these differences are roughly 20~MeV in \bbfourl{} for both
the no-smearing and smearing case, and in \ttbnlodec{} they are a few MeV for
the no-smearing case, and 20~MeV with smearing.  When using the generic veto
method of Sec.~\ref{sec:genericmethod} we find differences of comparable
size.

\subsubsection[Summary of scale, PDF and $\as$ variations]{Summary of scale, PDF and $\boldsymbol{\as}$ variations}%
\label{sec:mwbjSummary}

\begin{table*}[tb]
\centering
\resizebox{\textwidth}{!}
{ \begin{tabular}{l|c|c|c|c|c|c|c|c|}
 \cline{2-9}
 &  \multicolumn{4}{ |c|}{\phantom{\Big|} No smearing} &   \multicolumn{4}{ |c|}{15 GeV smearing} \\
 \cline{2-9}
 & \phantom{\Big|} $\%$ ${}-{}$ \bbfourl{} & $(\muR, \muF)$ & PDF & $\as$    
 & \% ${}-{}$ \bbfourl{} & $(\muR, \muF)$ & PDF & $\as$  \\
 \cline{1-9}
 \multicolumn{1}{ |c|  }{ \phantom{\Big|}\bbfourl{}}
 & $+           0 $~MeV
 & $ {}_{-          17 }^{+          26 }$~MeV & -                                                                                                 & $\pm            8 $~MeV
 & $+           0 $~MeV
 & ${}_{-          53 }^{+          86 }$~MeV & -                                                                                                 & $\pm           64 $~MeV\\ \cline{1-9}
 \multicolumn{1}{ |c|  }{ \phantom{\Big|}\ttbnlodec{}}
 & $+          21 $~MeV
 & $ {}_{-          10 }^{+           2 }$~MeV & -                                                                                                 & $\pm            8 $~MeV
 & $+         140 $~MeV
 & ${}_{-           6 }^{+           6 }$~MeV & -                                                                                                 & $\pm           54 $~MeV\\ \cline{1-9}
 \multicolumn{1}{ |c|  }{ \phantom{\Big|}\hvq{}}
 & $+          10 $~MeV
 & $ {}_{-           6 }^{+           2 }$~MeV & $\pm      3$~MeV                                                                                  & $\pm            2 $~MeV
 & $        -147 $~MeV
 & ${}_{-           7 }^{+           7 }$~MeV & $\pm      5$~MeV                                                                                  & $\pm            9 $~MeV\\ \cline{1-9}
\end{tabular}
}
\caption{Theoretical uncertainties associated with the $\mwbj{}$ peak
  position extraction for $\mt$=172.5~GeV for the three different generators,
  showered with \PythiaEightPtwo{}. The PDF uncertainty on the \bbfourl{} and
  \ttbnlodec{} generators is assumed to be equal to the \hvq{} one, as
  explained in Sec.~\ref{sec:PDF_dependence}.}
\label{tab:mass_extraction-errors}
\end{table*}
In Tab.~\ref{tab:mass_extraction-errors} we summarize the uncertainties due
to scale, PDF and strong-coupling variations, connected with the extraction
of the \mwbj{} peak position, for the input mass $\mt=172.5$~GeV, for all the
generators showered with \PythiaEightPtwo{}.

The upper (lower) error due to scale variation reported in the table is
obtained by taking the maximum (minimum) position of the \mwbj{} peak for
each of the seven scales choices of eq.~(\ref{eq:scalechoices}),
minus the one obtained for the central scale.

In the PDF case, as discussed in Sec.~\ref{sec:PDF_dependence}, we compute the
PDF uncertainties only for the \hvq{} generator, and assume that they are the
same for \bbfourl{} and \ttbnlodec{}.

We consider a symmetrized strong-coupling dependence uncertainty, whose
expression is given by
\begin{equation}
\delta { \mwbj\left(\as(\mZ)\right)} = \pm \frac{\left| \mwbj(0.115) -\mwbj(0.121)\right|}{2}\,.
\end{equation}
We stress that these variations have only an indicative meaning. In a
realistic analysis, experimental constraints may reduce these uncertainties.
We also stress that these are not the only theoretical uncertainties. Others
may be obtained by varying Monte Carlo parameters. Here we focus specifically
on those uncertainties that are associated with the NLO+PS generators.

As we have already discussed, the use of the \hvq{} and the \ttbnlodec{}
generators would lead to a negligible bias in the \mwbj{} distribution if we
were able to measure it without any resolution effects. However, if we
introduce a smearing to mimic them, the description of the region away from
the peak plays an important role, and the \hvq{} and \ttbnlodec{} generators
yield predictions for the mass peak position that are shifted by roughly
\diffaverage~MeV in the downward and upward direction respectively with
respect to \bbfourl{}.

We also notice that the \bbfourl{} generator is the most affected by
theoretical uncertainties. In particular, the \ttbnlodec{} and \hvq{}
generators have an unrealistically small scale dependence of the peak shape,
due to the way in which off-shell effects are approximately described.  The
\ttbnlodec{} generator displays a non-negligible sensitivity only to the
strong-coupling constant.  The theoretical errors that we have studied here
lead to very small effects for the \hvq{} generator, since it does not
include radiative corrections in the top decay. On the other hand, the \hvq{}
generator is bound to be more sensitive to variation of parameters in
\PythiaEightPtwo{}, that in this case fully controls the radiation from the
$b$ quark.

\subsubsection{Radius dependence}
In this section we investigate the stability of the previous results with
respect to the choice of the jet radius. The results are summarized in
Tab.~\ref{tab:mass_extraction-radius}.
\begin{table*}[tb]
\centering
\resizebox{\textwidth}{!}
{ \begin{tabular}{l|c|c|c|c|c|c|}
 \cline{2-7}
 &  \multicolumn{2}{ |c|}{ \phantom{\Big|} $R=0.4$}
 &  \multicolumn{2}{ |c|}{ \phantom{\Big|} $R=0.5$}
 &  \multicolumn{2}{ |c|}{ \phantom{\Big|} $R=0.6$} \\
 \cline{2-7}
 & \phantom{\Big|} No smearing & 15~GeV smearing
 & \phantom{\Big|} No smearing & 15~GeV smearing
 & \phantom{\Big|} No smearing & 15~GeV smearing \\
 \cline{1-7}
 \multicolumn{1}{ |c|  }{ \phantom{\Big|} \bbfourl{} [GeV]}

 & $ 172.156\pm  0.004$ & $ 171.018\pm  0.002$

 & $ 172.793\pm  0.004$ & $ 172.717\pm  0.002$

 & $ 173.436\pm  0.005$ & $ 174.378\pm  0.002$
\\ \cline{1-7}
\multicolumn{1}{ |c|  }{ \phantom{\Big|}\ttbnlodec{} ${}-$  \bbfourl{}}
 & $ +    35
\pm      5$~MeV
 & $ +   195
\pm      2$~MeV
 & $ +    21
\pm      6$~MeV
 & $ +   140
\pm      2$~MeV
 & $ +     1
\pm      7$~MeV
 & $ +    97
\pm      2$~MeV
\\ \cline{1-7}
\multicolumn{1}{ |c|  }{ \phantom{\Big|}\hvq{} ${}-$  \bbfourl{}}
 & $ +    47
\pm      5$~MeV
 & $    -113
\pm      2$~MeV
 & $ +    10
\pm      6$~MeV
 & $    -147
\pm      2$~MeV
 & $      -7
\pm      6$~MeV
 & $    -174
\pm      2$~MeV
\\ \cline{1-7}
\end{tabular}

}
\caption{ \mwbj{} peak position obtained with the \bbfourl{} generator for
  three choices of the jet radius. The differences with the \ttbnlodec{} and
  the \hvq{} generators are also shown.}
\label{tab:mass_extraction-radius}
\end{table*}
For the distributions without smearing, the differences between the three
generators are small and decrease as $R$ increases.  For the smeared
distributions, the differences between \ttbnlodec{} and \bbfourl{} decrease
as the radius increases, while the difference between the \hvq{} and the
\bbfourl{} generator increases.

The small differences in $R$ dependence among the three generators in the
non-smeared cases can be understood if we consider that differences in the $b$
radiation do not affect much the peak position in the non-smeared
distribution, but rather they affect the strength of the tail on the left
side of the peak. On the other hand, the peak position is affected by
radiation in production and by the underlying-event structure, that is very
similar in the three generators.

It should be noticed that the difference between the displacements of the
\ttbnlodec{} and \hvq{} with respect to \bbfourl{} is less than
\diffdiffRttdec~MeV and \diffdiffRhvq~MeV, respectively, below the current
statistical precision of top-mass measurements. Thus, the good agreement
found among the three generators persists also for different $R$ values.

\subsection{Comparison with \HerwigSevenPone{}}
In order to assess uncertainties due to the showering program, in this
section we compare the results obtained using \HerwigSevenPone{} and
\PythiaEightPtwo{}.
\begin{table*}[tb]
\centering
\resizebox{\textwidth}{!}
{ \begin{tabular}{l|c|c|c|c|}
 \cline{2-5}
 &  \multicolumn{2}{ |c|}{ \phantom{\Big|} No smearing}
 &  \multicolumn{2}{ |c|}{15~GeV smearing} \\
 \cline{2-5}
 & \phantom{\Big|} {\HerwigSevenPlot} & {\PythiaEightPlot} ${}-$ {\HerwigSevenPlot} 
 & \phantom{\Big|} {\HerwigSevenPlot} & {\PythiaEightPlot} ${}-$ {\HerwigSevenPlot}\\
 \cline{1-5}
 \multicolumn{1}{ |c|  }{ \phantom{\Big|}\bbfourl{}}
 & $172.727\pm   0.005$~GeV 
  &  $+          66 \pm            7 $~MeV
 & $171.626\pm   0.002$~GeV 
  &  $+        1091 \pm            2 $~MeV
 \\ \cline{1-5}
 \multicolumn{1}{ |c|  }{ \phantom{\Big|}\ttbnlodec{}}
 & $172.775\pm   0.004$~GeV 
  &  $+          39 \pm            5 $~MeV
 & $171.678\pm   0.001$~GeV 
  &  $+        1179 \pm            2 $~MeV
 \\ \cline{1-5}
 \multicolumn{1}{ |c|  }{ \phantom{\Big|}\hvq{}}
 & $173.038\pm   0.004$~GeV 
  &  $        -235 \pm            5 $~MeV
 & $172.319\pm   0.001$~GeV 
  &  $+         251 \pm            2 $~MeV
 \\ \cline{1-5}
\end{tabular}
}
\caption{$\mwbj{}$ peak position for $\mt$=172.5~GeV obtained with the three
  different generators, showered with \HerwigSevenPone{}~({\HerwigSevenPlot}). The
  differences with \PythiaEightPtwo{}~({\PythiaEightPlot}) are also shown.}
\label{tab:mass_extraction-shower}
\end{table*}
In Tab.~\ref{tab:mass_extraction-shower} we compare the \mwbj{} peak position
extracted for the input mass $\mt = 172.5$~GeV using the three generators
showered with \PythiaEightPtwo{} and \HerwigSevenPone{}.  For the \hvq{}
generator, the differences are of the order of \pyminushwhvq~MeV for both the
smeared and non-smeared case, but with opposite signs.  In the smeared case,
both the \ttbnlodec{} and \bbfourl{} generators yield much larger
differences, of more than 1~GeV.

In Tab.~\ref{tab:mass_extraction-shower-showerOnly}
\begin{table*}[tb]
\centering
{ \begin{tabular}{l|c|c|c|c|}
 \cline{2-5}
 &  \multicolumn{4}{ |c|}{ $\phantom{\Big|}$ \PythiaEightPtwo{} ${}-$ \HerwigSevenPone{}} \\
 \cline{2-5}
 &  \multicolumn{2}{ |c|}{PS only}
 &  \multicolumn{2}{ |c|}{ \phantom{\Big|} full}\\
 \cline{2-5}
 & \phantom{\Big|} No smearing & 15~GeV smearing
 & \phantom{\Big|} No smearing & 15~GeV smearing \\
 \cline{1-5}
 \multicolumn{1}{ |c|  }{ \phantom{\Big|}\bbfourl{}}
  &  $+          10 \pm            2 $~MeV
  &  $+         984 \pm            2 $~MeV
  &  $+          66 \pm            7 $~MeV
  &  $+        1091 \pm            2 $~MeV
 \\ \cline{1-5}
 \multicolumn{1}{ |c|  }{ \phantom{\Big|}\ttbnlodec{}}
  &  $+           5 \pm            2 $~MeV
  &  $+        1083 \pm            2 $~MeV
  &  $+          39 \pm            5 $~MeV
  &  $+        1179 \pm            2 $~MeV
 \\ \cline{1-5}
 \multicolumn{1}{ |c|  }{ \phantom{\Big|}\hvq{}}
  &  $-           0 \pm            2 $~MeV
  &  $+         113 \pm            2 $~MeV
  &  $        -235 \pm            5 $~MeV
  &  $+         251 \pm            2 $~MeV
 \\ \cline{1-5}
\end{tabular}
}
\caption{Differences between  \PythiaEightPtwo{} and \HerwigSevenPone{}  in the
  extracted $\mwbj{}$ peak position for $\mt$=172.5~GeV obtained with the
  three different generators, at the NLO+PS level (PS only) and including
  also the underlying events, the multi-parton interactions and the
  hadronization~(full).}
\label{tab:mass_extraction-shower-showerOnly}
\end{table*}
we report the differences between the \HerwigSevenPone{} and \PythiaEightPtwo{}
predictions for all the generators, at the NLO+PS level and at the full
hadron level.  We notice that at the NLO+PS level and without smearing, the
differences between the two parton-shower programs are negligible. For the
smeared distributions, at both the NLO+PS and full level, the differences are
roughly 1~GeV for the \bbfourl{} and the \ttbnlodec{} generator. For \hvq{}
the differences are considerably smaller, although not quite
negligible. Furthermore, accidental compensation effects seem to emerge in
this case if we compare the peak displacement in the distributions with and
without smearing.

\begin{table*}[tb]
\centering
\resizebox{\textwidth}{!}
{ \begin{tabular}{l|c|c|c|c|c|c|}
 \cline{2-7}
 & \multicolumn{6}{ |c|}{ \phantom{\Big|}  \PythiaEightPtwo{} ${}-$ \HerwigSevenPone{}} \\
 \cline{2-7}
 &  \multicolumn{2}{ |c|}{ \phantom{\Big|} $R=0.4$}
 &  \multicolumn{2}{ |c|}{ \phantom{\Big|} $R=0.5$}
 &  \multicolumn{2}{ |c|}{ \phantom{\Big|} $R=0.6$} \\
 \cline{2-7}
 & \phantom{\Big|} No smearing & 15~GeV smearing
 & \phantom{\Big|} No smearing & 15~GeV smearing
 & \phantom{\Big|} No smearing & 15~GeV smearing \\
 \cline{1-7}
 \multicolumn{1}{ |c|  }{ \phantom{\Big|}\bbfourl{}}
 & $     -98
\pm      7$~MeV
 & $ +   830
\pm      2$~MeV
 & $ +    66
\pm      7$~MeV
 & $ +  1091
\pm      2$~MeV
 & $ +   253
\pm      8$~MeV
 & $ +  1267
\pm      2$~MeV
\\ \cline{1-7}
 \multicolumn{1}{ |c|  }{ \phantom{\Big|}\ttbnlodec{}}
 & $    -100
\pm      5$~MeV
 & $ +   979
\pm      2$~MeV
 & $ +    39
\pm      5$~MeV
 & $ +  1179
\pm      2$~MeV
 & $ +   210
\pm      6$~MeV
 & $ +  1314
\pm      2$~MeV
\\ \cline{1-7}
 \multicolumn{1}{ |c|  }{ \phantom{\Big|}\hvq{}}
 & $    -370
\pm      5$~MeV
 & $ +    73
\pm      2$~MeV
 & $    -235
\pm      5$~MeV
 & $ +   251
\pm      2$~MeV
 & $     -31
\pm      6$~MeV
 & $ +   389
\pm      2$~MeV
\\ \cline{1-7}
\end{tabular}
}
\caption{ Differences in the \mwbj{} peak position obtained matching the
  three generators with \PythiaEightPtwo{} and \HerwigSevenPone{}, for three
  choices of the jet radius.}
\label{tab:mass_extraction-shower-radius}
\end{table*}

\begin{figure*}[tb]
\centering
\includegraphics[width=\wfigdoub]{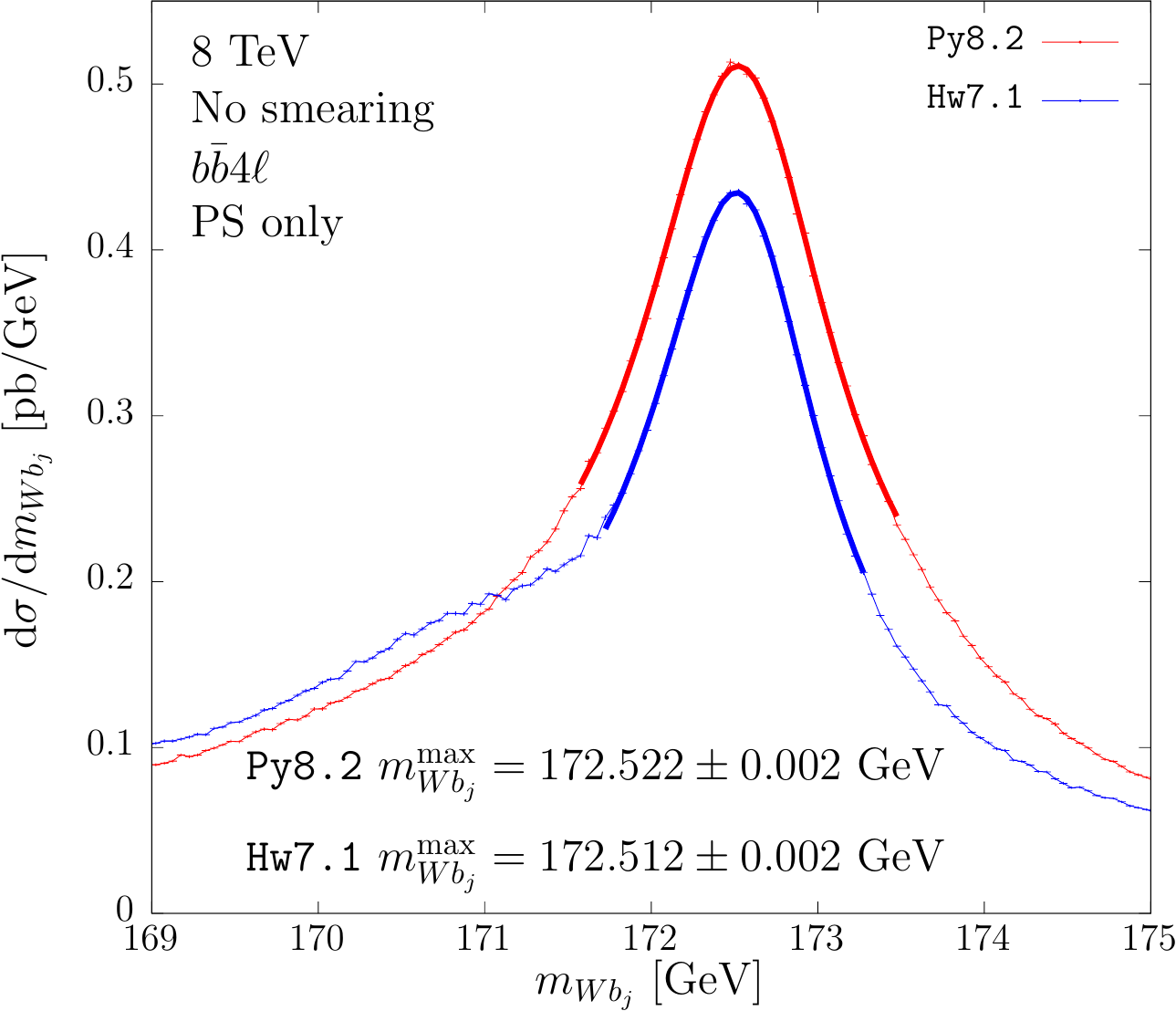}
\includegraphics[width=\wfigdoub]{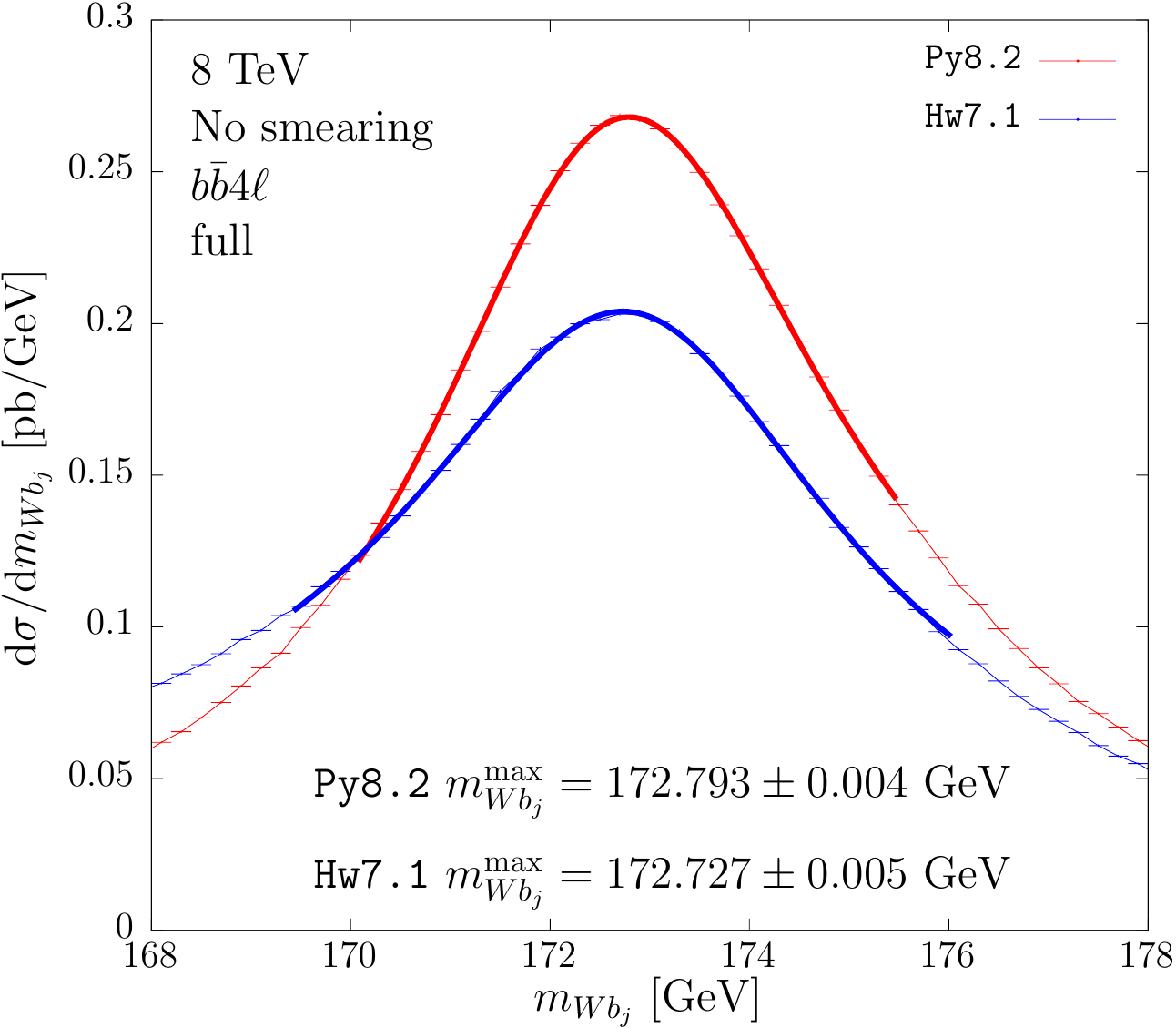}
  \caption{${d\sigma}/{d \mwbj}$ distribution obtained by showering the
    \bbfourl{} results with \PythiaEightPtwo{} and \HerwigSevenPone{},
    at parton-shower level~(left) and with hadronization and underlying events~(right).}
\label{fig:mwbjshapespyh7}
\end{figure*}

\begin{figure}[tb]
\centering
\includegraphics[width=\wfigsing]{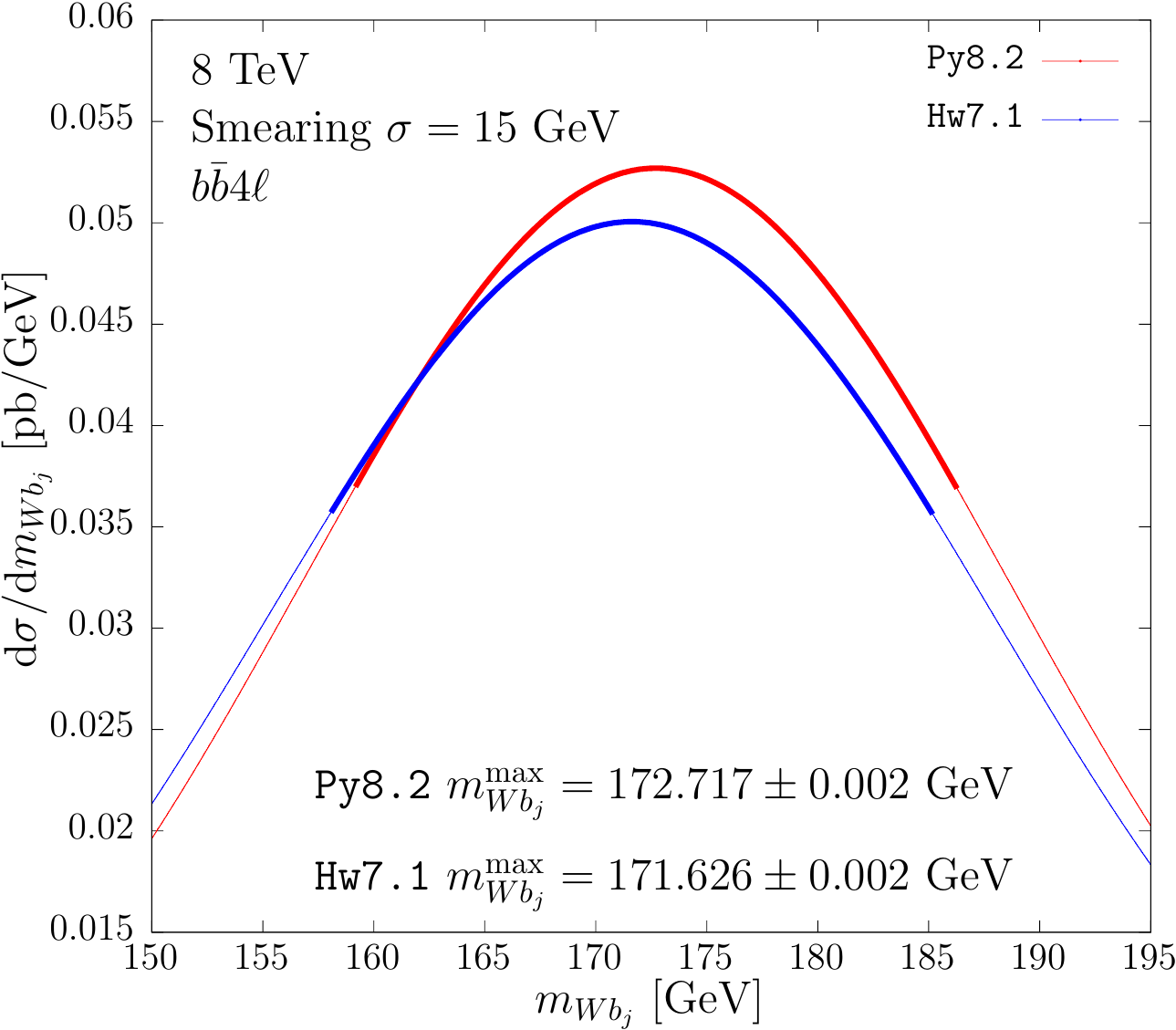}
\caption{Smeared ${d\sigma}/{d \mwbj}$ distribution obtained by matching
  the \bbfourl{} generator with \PythiaEightPtwo{} and \HerwigSevenPone{}.}
\label{fig:mwbjshapespyh7smeared}
\end{figure}

The origin of these large differences are better understood by looking at the
differential cross sections plotted in Figs.~\ref{fig:mwbjshapespyh7}
and~\ref{fig:mwbjshapespyh7smeared}.  In Fig.~\ref{fig:mwbjshapespyh7} we
plot the results for the non-smeared case, at the NLO+PS level~(left) and at
the full hadron level~(right): while the peak position is nearly the same for
both \PythiaEightPtwo{} and \HerwigSevenPone{}, the shape of the curves is very
different around the peak, leading to a different mass peak position when
smearing is applied, as displayed in Fig.~\ref{fig:mwbjshapespyh7smeared}.
We notice that in this last case we see a difference in shape also after
smearing.  This suggests that at least one of the two generators may not
describe the data fairly.

\begin{figure}[tb]
\centering
 \includegraphics[width=\wfigsing]{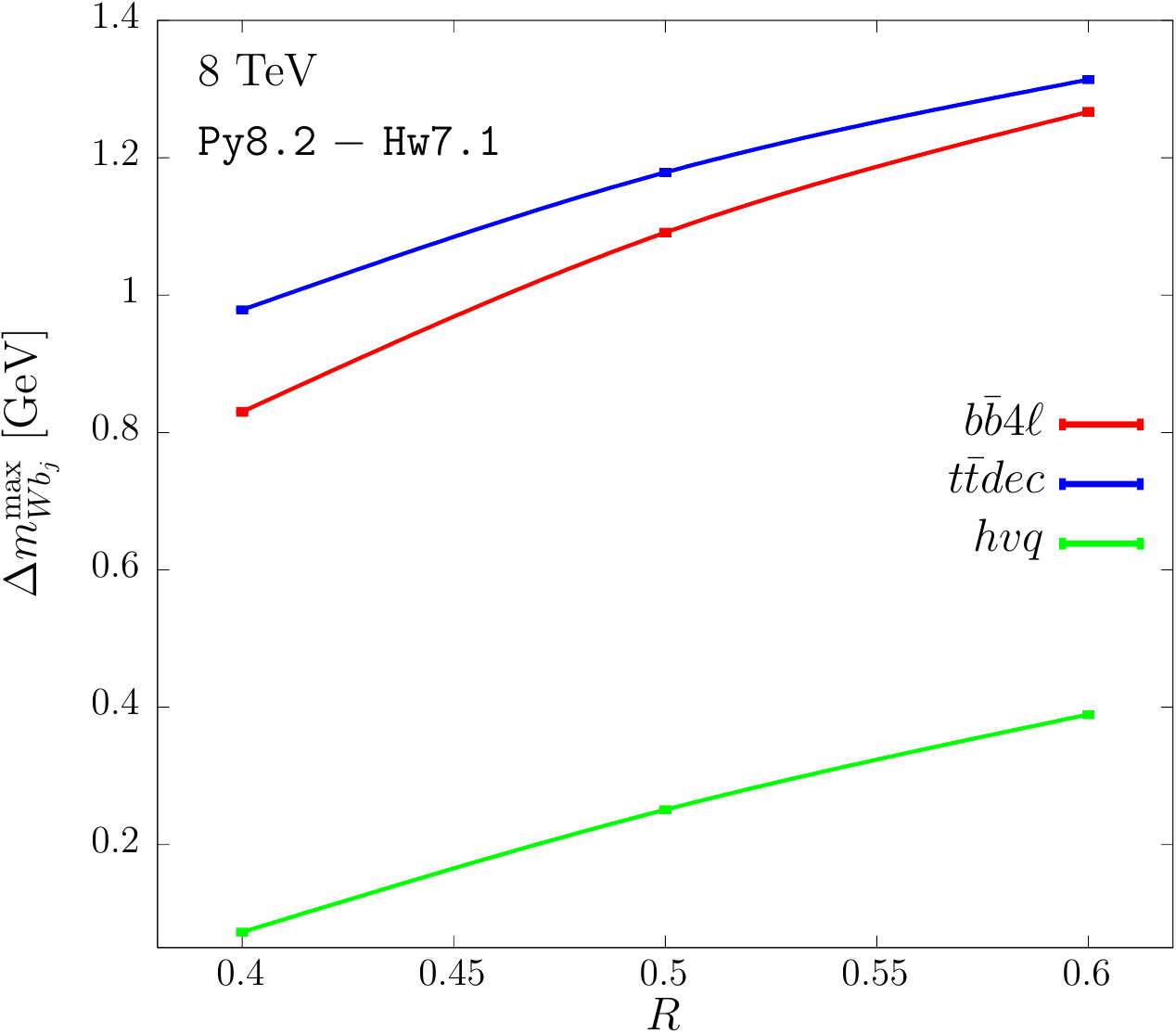}
 \caption{Differences of $\mwbjmax$ between the \PythiaEightPtwo{} and the
     \HerwigSevenPone{} showers, for the three generators, as a function of the
     jet radius.}
 \label{fig:R_mwbj_py8-hw7}
\end{figure}

Since we observe such large differences in the value of $\mwbjmax$ in
\HerwigSevenPone{} and \PythiaEightPtwo{}, we have also studied whether sizeable
differences are also present in the $\mwbjmax$ dependence upon the jet radius
$R$.  The results are shown in Tab.~\ref{tab:mass_extraction-shower-radius},
and displayed in Fig.~\ref{fig:R_mwbj_py8-hw7}.
In the case of the \bbfourl{} generator, the difference between
\PythiaEightPtwo{} and \HerwigSevenPone{} goes from \pyminushwRfour{} to
\pyminushwRsix~MeV. Thus, assuming for instance that \PythiaEightPtwo{} fits
the data perfectly, i.e.~that it extracts the same value of the mass by
fitting the $\mwbjmax$ values obtained with the three different values of
$R$, \HerwigSevenPone{} would extract at $R=0.6$ a mass value that is larger
by \pyminushwdeltaRfoursix{}~MeV from the one extracted at $R=0.4$.  We
stress that the differences in the $R$ behaviour of $\mwbjmax$ may have the same
origin as the difference in the reconstructed mass value, since both effects
may be related to the amount of energy that enters the jet cone, and it is
not unlikely that, by tuning one of the two generators in such a way that they
both have the same $R$ dependence, their difference in $\mwbjmax$ would
also be reduced.\footnote{Similarly, one could fit appropriate calibration
  observables associated to the $b$-jet structure, along the lines of
  Ref.~\cite{Corcella:2017rpt}.} It is unlikely, however that this would lead
to a much improved agreement, since the difference in slope is much less
pronounced than the difference in absolute value.

\subsubsection{Alternative matching prescriptions in \HerwigSevenPone{}}
We have examined several variations in the \HerwigSevenPone{} settings, and in
the interface between \POWHEG{} and \HerwigSevenPone{}, in order to understand
whether the \HerwigSevenPone{} results are reasonably stable, or depend upon our
particular settings.

\subsubsection*{MEC and \POWHEG{} options in \HerwigSevenPone{}}
\HerwigSevenPone{} applies matrix-element corrections by default, but it also
offers the possibility to switch them off. In addition, it allows to
optionally replace the MEC with its internal \POWHEG{} method, when
available, to achieve NLO accuracy in top decays.\footnote{These options are
  activated by the instructions\fignewline {\tt set
    ShowerHandler:HardEmission None} or \fignewline {\tt set
    ShowerHandler:HardEmission POWHEG}, respectively.}  We have verified
that, as expected, switching off the matrix-element corrections does not
significantly affect the \bbfourl{} and \ttbnlodec{} results. In the case of
the \hvq{} generator, we can compare the default case, where MEC is on, with
the cases where \POWHEG{} replaces MEC, and with the case where neither MEC
nor \POWHEG{} is implemented.  The results are shown in
Tab.~\ref{tab:mass_extraction-hvq-mec-powheg-herwig}.
\begin{table}[tb]
\centering
{ \begin{tabular}{|c|c|c|}
 \cline{1-3}
 \hvq & \phantom{\Big|} No smearing  & 15~GeV smearing \\
 \cline{1-3}
 \multicolumn{1}{ |c|  }{ \phantom{\Big|} MEC ${}-$ no MEC}
  & $         307 \pm            6 $~MeV  & $        1371 \pm            2 $~MeV 
 \\ \cline{1-3}
 \multicolumn{1}{ |c|  }{ \phantom{\Big|} MEC ${}-$ {\tt POWHEG}}
  & $         244 \pm            6 $~MeV  & $         356 \pm            2 $~MeV 
 \\ \cline{1-3}
\end{tabular}
}
\caption{Differences in the $\mwbj{}$ peak position for the \hvq{} generator
  showered with \HerwigSevenPone{}, with MEC switched off (no MEC) or using the
  \HerwigSevenPone{} \POWHEG{} option, with respect to our default setting, that
  has MEC switched on.}
\label{tab:mass_extraction-hvq-mec-powheg-herwig}
\end{table}
We notice that the inclusion of MEC enhances by more than 1.3~GeV the peak
position of the smeared distribution. A similar result was found in
\PythiaEightPtwo{} (see Tab.~\ref{tab:mwbj_MEC}), where the difference was
slightly less than 1~GeV.  The difference between the \POWHEG{} and MEC
results is much below the 1~GeV level but not negligible. This fact is hard
to understand, since the \POWHEG{} and MEC procedures should only differ by a
normalization factor.

We have seen previously that the three NLO+PS generators interfaced to
\PythiaEightPtwo{} yield fairly consistent results for the reconstructed top
mass peak. The same consistency is not found when they are interfaced to
\HerwigSevenPone{}. However, the best agreement is found when the internal
\POWHEG{} option for top decay is activated in \HerwigSevenPone{}, as can be
seen in Tab.~\ref{tab:cmp_herwig_table}.
\begin{table*}[tb]
\centering
\resizebox{\textwidth}{!}
{ \begin{tabular}{l|c|c|c|c|}
 \cline{2-5}
 &  \multicolumn{2}{ |c|}{PS only}
 &  \multicolumn{2}{ |c|}{ \phantom{\Big|} full}\\
 \cline{2-5}
 & \phantom{\Big|} No smearing & 15~GeV smearing
 & \phantom{\Big|} No smearing & 15~GeV smearing \\
 \cline{1-5}
 \multicolumn{1}{ |c|  }{ \phantom{\Big|}  \bbfourl{}}
 & $172.512\pm  0.002$~GeV
 & $170.419\pm  0.002$~GeV
 & $172.727\pm  0.005$~GeV
 & $171.626\pm  0.002$~GeV
 \\ \cline{1-5}
 \multicolumn{1}{ |c|  }{ \phantom{\Big|}\ttbnlodec{} ${}-$ \bbfourl{}}
  &  $         -13 \pm            2 $~MeV
  &  $+          92 \pm            2 $~MeV
  &  $+          48 \pm            7 $~MeV
  &  $+          52 \pm            2 $~MeV
 \\ \cline{1-5}
 \multicolumn{1}{ |c|  }{ \phantom{\Big|}\hvq{} ${}-$ \bbfourl{}}
  &  $         -14 \pm            2 $~MeV
  &  $+         782 \pm            2 $~MeV
  &  $+         311 \pm            7 $~MeV
  &  $+         693 \pm            2 $~MeV
 \\ \cline{1-5}
 \multicolumn{1}{ |c|  }{ \phantom{\Big|}\hvq{}+PWG ${}-$ \bbfourl{}}
  &  $         -16 \pm            2 $~MeV
  &  $+         479 \pm            2 $~MeV
  &  $+          67 \pm            7 $~MeV
  &  $+         337 \pm            2 $~MeV
 \\ \cline{1-5}
\end{tabular}
}
\caption{Differences of \hvq{} and \ttbnlodec{} with respect to \bbfourl{},
  all showered with \HerwigSevenPone{}. The result obtained using the
  \HerwigSevenPone{} internal \POWHEG{} implementation of top decay, rather
  than MEC, labelled as \hvq+PWG, is also shown.}
\label{tab:cmp_herwig_table}
\end{table*}
The difference between the \POWHEG{} and MEC or \POWHEG{} \HerwigSevenPone{}
results is puzzling, since they have the same formal accuracy. We will
comment about this issue later on.

\subsubsection*{Alternative veto procedures in \HerwigSevenPone{}}
\label{sec:HW7_different_showers_results}

\begin{table*}[tb]
\centering
\resizebox{\textwidth}{!}
{ \begin{tabular}{l|c|c|c|c|}
 \cline{2-5}
 &  \multicolumn{2}{ |c|}{ \phantom{\Big|} No smearing}
 &  \multicolumn{2}{ |c|}{15~GeV smearing} \\
 \cline{2-5}
 & \phantom{\Big|} {\tt FSV} & {\tt FSV} ${}-$ {\tt SV}
 & \phantom{\Big|} {\tt FSV} & {\tt FSV} ${}-$ {\tt SV}\\
 \cline{1-5}
 \multicolumn{1}{ |c|  }{ \phantom{\Big|}\bbfourl{}}
 & $172.776\pm   0.005$~GeV 
  &  $+          49 \pm            7 $~MeV
 & $171.829\pm   0.002$~GeV 
  &  $+         203 \pm            2 $~MeV
 \\ \cline{1-5}
 \multicolumn{1}{ |c|  }{ \phantom{\Big|}\ttbnlodec{}}
 & $172.810\pm   0.004$~GeV 
  &  $+          35 \pm            6 $~MeV
 & $171.906\pm   0.001$~GeV 
  &  $+         228 \pm            2 $~MeV
 \\ \cline{1-5}
\end{tabular}
}
\caption{$\mwbj{}$ peak position for $\mt$=172.5~GeV for \bbfourl{} and
  \ttbnlodec{} showered with \HerwigSevenPone{} using the {\tt
    FullShowerVeto}~({\tt FSV}) procedure. The differences with {\tt
    ShowerVeto}~({\tt SV}), that represents our default, are also shown. }
\label{tab:mass_extraction-matching-hw7}
\end{table*}
As discussed in Sec.~\ref{sec:HW7_different_showers}, \HerwigSevenPone{}
offers two different classes that implement the veto procedure: the {\tt
  ShowerVeto}, our default one, and the {\tt FullShowerVeto} class.  The
corresponding results are summarized in
Tab.~\ref{tab:mass_extraction-matching-hw7}.  For both the \bbfourl{} and the
\ttbnlodec{} the two procedures lead to a 200~MeV difference in the peak
position for the smeared distributions.  The origin of such difference is not
fully clear to us. In part it may be ascribed to the fact that when using the
{\tt ShowerVeto} class we mix two different definitions of transverse
momentum (the \HerwigSevenPone{} and the \POWHEG{} one), and in part may be
due to the fact that in the {\tt FullShowerVeto} class the vetoing is done on
the basis of the shower structure after reshuffling has been applied.  We
have also checked that the generic procedure of Sec.~\ref{sec:genericmethod},
although much slower, leads to results that are statistically compatible with
the {\tt FullShowerVeto} method.

\subsubsection*{Truncated showers}
It was shown in Ref.~\cite{Nason:2004rx} that, when interfacing a \POWHEG{}
generator to an angular-ordered shower, in order to compensate for the
mismatch between the angular-ordered scale and the \POWHEG{} hardness, that is
taken equal to the relative transverse momentum in radiation, one should
supply appropriate truncated showers. None of our vetoing algorithms take
them into account, but it turns out that \HerwigSevenPone{} provides
facilities to change the settings of the initial showering scale according to
the method introduced in Ref.~\cite{Schofield:2011zi}, that, in our case, are
equivalent to the inclusion of truncated showers (see \writeApp\ref{app:TS}).  This is
done by inserting the following instructions in the \HerwigSevenPone{} input file:
\begin{equation}
  \label{eq:TSsettings}
  \begin{split}
&   \mbox{\tt
      set PartnerFinder:PartnerMethod Maximum} \\
&    \mbox{\tt set
    PartnerFinder:ScaleChoice Different}.
  \end{split}
\end{equation}
The effects of these settings for the \bbfourl{} and \ttbnlodec{} generators
are shown in Tab.~\ref{tab:mass-extraction-truncated-shower-hw7}.
\begin{table*}[tb]
\centering
\resizebox{\textwidth}{!}
{ \begin{tabular}{l|c|c|c|c|}
 \cline{2-5}
 &  \multicolumn{2}{ |c|}{ \phantom{\Big|} No smearing}
 &  \multicolumn{2}{ |c|}{15~GeV smearing} \\
 \cline{2-5}
 & \phantom{\Big|} TS & TS ${}-$ default
 & \phantom{\Big|} TS  & TS ${}-$ default\\
 \cline{1-5}
 \multicolumn{1}{ |c|  }{ \phantom{\Big|}\bbfourl{}}
 & $172.730\pm   0.005$~GeV
  &  $+           3 \pm            8 $~MeV
 & $171.496\pm   0.002$~GeV
  &  $        -130 \pm            2 $~MeV
 \\ \cline{1-5}
 \multicolumn{1}{ |c|  }{ \phantom{\Big|}\ttbnlodec{}}
 & $172.786\pm   0.004$~GeV
  &  $+          12 \pm            6 $~MeV
 & $171.546\pm   0.001$~GeV
  &  $        -132 \pm            2 $~MeV
 \\ \cline{1-5}
\end{tabular}
}
\caption{$\mwbj{}$ peak position for $\mt$=172.5~GeV obtained with the
  \bbfourl{} and \ttbnlodec{} generators showered with \HerwigSevenPone{},
  with the settings of eq.~(\ref{eq:TSsettings}) (labelled as TS). The
  differences with the default results are also shown.}
\label{tab:mass-extraction-truncated-shower-hw7}
\end{table*}
The inclusion of the truncated shower does not introduce dramatic changes in
the peak position: in fact the differences are negligible in the
distributions without smearing, and are roughly \hwTSsmearbbfourlttdec~MeV
when smearing is applied.  It should be noticed that these settings slightly
increase the difference with respect to the results obtained with
\PythiaEightPtwo{}.

\section{The energy of the $\boldsymbol{b}$ jet}
\label{sec:Ebjet}
In Ref.~\cite{Agashe:2016bok} it was proposed to extract $\mt$ using the peak
of the energy spectrum of the $b$ jet.
At leading order, the $b$ jet consists of the $b$ quark alone, and its energy
in the top rest frame, neglecting top-width effects, is fixed and given by
\begin{equation}
\Ebjmax=\frac{\mt^2-m_W^2+m_b^2}{2\,\mt}\,,
\label{eq:ebjlo}
\end{equation}
i.e.~the spectrum is a delta function in the energy.  In the laboratory
frame, because of the variable boost that affects the top, the delta function
is smeared into a wider distribution, but it can be shown that its peak
position remains at $\Ebjmax$. On the basis of this observation we are led to
assume that also after the inclusion of off-shell effects, radiative and
non-perturbative corrections, the relation between $\Ebjmax$ and the top
pole-mass $\mt$ should be largely insensitive to production dynamics.

We performed a study of the $\Ebjmax$ observable along the same lines adopted
for \mwbj{} in the previous section.  If the range of variations of the top
mass around a given central value $\mtc$ is small enough, a linear relation
between $\Ebjmax$ and the top mass must hold, so that we can write
\begin{equation}
  \label{eq:B-for-ebj}
\Ebjmax(\mt)= \Ebjmax(\mtc) +B\,(\mt-\mtc)+\mathcal{O}(\mt-\mtc)^2.
\end{equation}

It was suggested in Ref.~\cite{CMS-PAS-TOP-15-002} that the $\Ebj$
distribution $\mathd \sigma/\mathd \Ebj$ is better fitted in terms of $\log
\Ebj$. Thus, in order to extract the peak position, we fitted the energy
distribution with a fourth order polynomial
\begin{equation}
y=a+b(x-x^{\rm max})^2+c(x-x^{\rm max})^3+d(x-x^{\rm max})^4\,,
\end{equation}
where $x=\log\Ebj$.

The parameter $B$ of eq.~\eqref{eq:B-for-ebj}, extracted from a linear fit of
the three \Ebjmax{} values corresponding to the three different values of
$\mt$ that we have considered (see Tab.~\ref{tab:samples}) using the \hvq{}
generator showered by \PythiaEightPtwo{}, was found to be
\begin{equation}
  \label{eq:B_Ebj}
  B= \BfromEbjhvq \pm \BerrfromEbjhvq \, ,
\end{equation}
compatible with the expected value of 0.5 from
eq.~\eqref{eq:ebjlo}.\footnote{When using the \bbfourl{} generator we obtain
  $B= \BfromEbjbbfourl \pm \BerrfromEbjbbfourl$, while with the \ttbnlodec{}
  one, we get $B= \BfromEbjttdec \pm \BerrfromEbjttdec$.  When using
  \HerwigSevenPone{} instead of \PythiaEightPtwo{}, we find values compatible with the given
  ones within 10\%{}.}

\subsection{Comparison among different NLO+PS generators}
\begin{figure}[tb]
  \centering
  \includegraphics[width=\wfigsing]{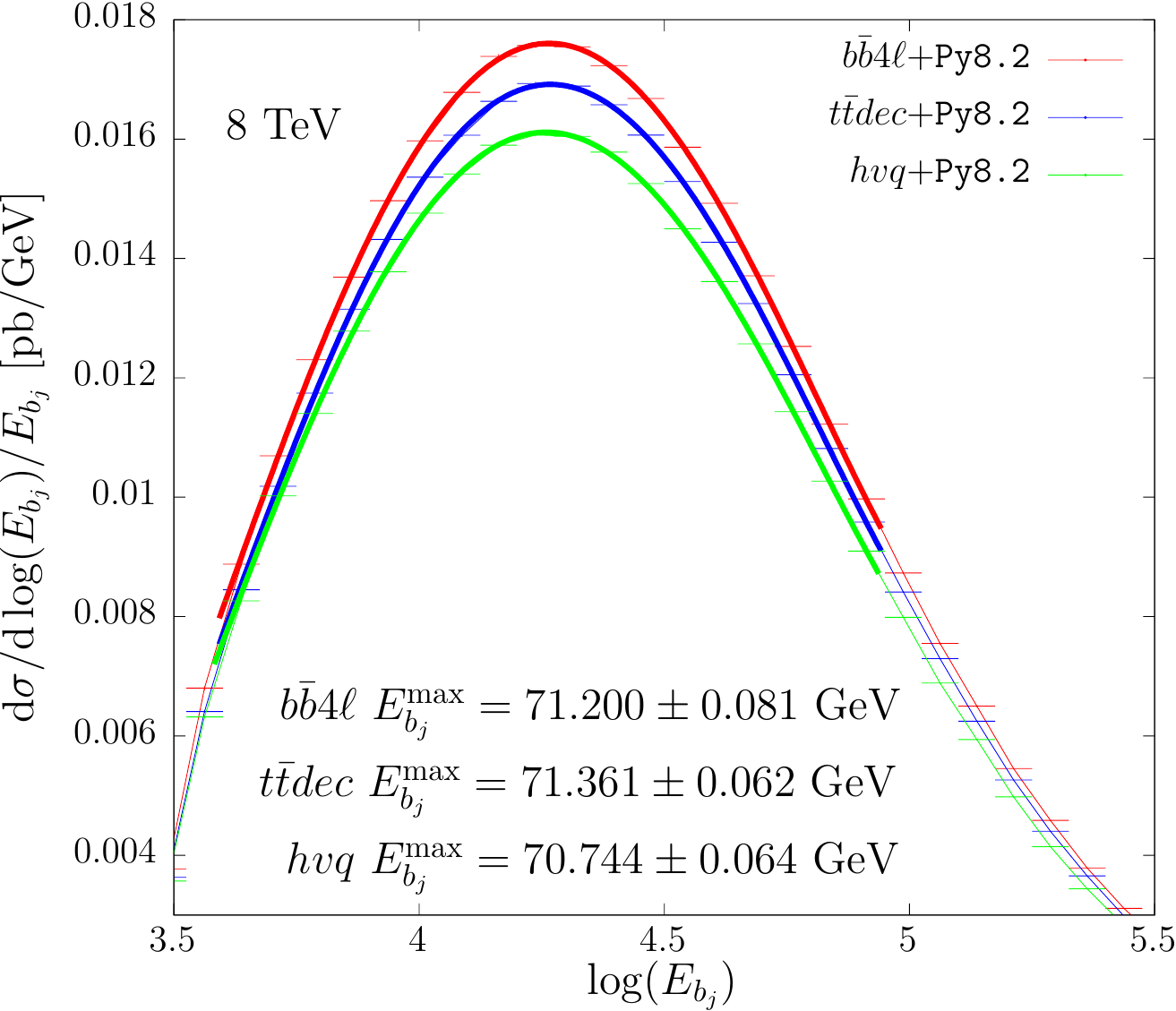}
  \caption{Logarithmic energy distribution obtained with the three generators interfaced
    to \PythiaEightPtwo, together with their polynomial fit, in the range
    displayed in the figure. The value of $\Ebjmax$ for each generator is
    also reported.}
\label{fig:Ebj_bb4l_ttdec_hvq}
\end{figure}

In Fig.~\ref{fig:Ebj_bb4l_ttdec_hvq} we plot the logarithmic energy
distribution for the three generators interfaced to \PythiaEight, together
with their polynomial fit.  The extracted \Ebj{} peaks from the \bbfourl{}
and the \ttbnlodec{} generators are compatible within the statistical
errors. On the other hand, the \hvq{} generator yields a prediction which is
roughly \Ebjbbfourlmhvq{} $\pm$ \Ebjbbfourlmhvqerr~MeV smaller than the
\bbfourl{} one.  We thus observe that the jet modeling implemented by
\PythiaEightPtwo{} with MEC seems to yield slightly less energetic jets. An
effect going in the same direction was also observed for the \mwbj{}
observable (see Tab.~\ref{tab:mass_extraction-errors}, the first column of
the results with smearing), although to a smaller extent.

\begin{table}[tb]
\centering
{ \begin{tabular}{l|c|c|}
 \cline{2-3}
 & $\phantom{\Big|}$ MEC  & MEC ${}-$ no MEC  \\ 
 \cline{1-3}
\multicolumn{1}{ |c|}{ $\phantom{\Big|}$ \bbfourl{}}
 & $ 71.200\pm  0.081 $~GeV
 & $+   170\pm    115 $~MeV
\\ \cline{1-3}
\multicolumn{1}{ |c|}{ $\phantom{\Big|}$ \ttbnlodec{}}
 & $ 71.361\pm  0.062 $~GeV
 & $    -69\pm     87 $~MeV
\\ \cline{1-3}
\multicolumn{1}{ |c|}{ $\phantom{\Big|}$ \hvq{}}
 & $ 70.744\pm  0.064 $~GeV
 & $+  1937\pm     92 $~MeV
\\ \cline{1-3}
\end{tabular}
}
\caption{\Ebj{} peak position obtained with the three generators showered
  with \PythiaEightPtwo{}. The differences between the peak positions extracted
  by switching on and off the matrix-element corrections are also shown.}
\label{tab:Ebj_MEC}
\end{table}
In Tab.~\ref{tab:Ebj_MEC} we have collected the values of $\Ebjmax$ computed
with MEC, and the differences between the results with and without MEC.  We
notice that the MEC setting has little impact in the \bbfourl{} and
\ttbnlodec{} cases. On the other hand, in the \hvq{} case the absence of MEC
would have lead to an $\Ebjmax$ value about 2~GeV smaller than with MEC. We
take this as another indication that the implementation of radiation in top
decay using MEC leads to results that are much closer to the NLO+PS ones.

In Tab.~\ref{tab:Ebj_extraction-errors}
\begin{table*}[tb]
\centering
{ \begin{tabular}{l|c|c|c|c|c|}
 \cline{2-6}
 &$\phantom{\Big|}$  $\%$ ${}-$ \bbfourl{} & $(\muR, \muF)$ & PDF & $\as$ & stat \\
\cline{1-6}
\multicolumn{1}{ |c|}{ $\phantom{\Big|}$ \bbfourl{} } & $+     0 $~MeV & ${}_{-     15}^{+     22}$~MeV & -                                                                                                 & $\pm     35$~MeV  & $\pm     81$~MeV \\
\cline{1-6}
\multicolumn{1}{ |c|}{ $\phantom{\Big|}$ \ttbnlodec{} } & $+   161 $~MeV & ${}_{-     24}^{+     22}$~MeV & -                                                                                                 & $\pm     17$~MeV  & $\pm     62$~MeV \\
\cline{1-6}
\multicolumn{1}{ |c|}{ $\phantom{\Big|}$ \hvq{} } & $   -456 $~MeV & ${}_{-     47}^{+     32}$~MeV & $\pm     30$~MeV                                                                                  & $\pm     25$~MeV  & $\pm     64$~MeV \\
\cline{1-6}
\end{tabular}
}
\caption{Theoretical uncertainties for the $\Ebj${} peak position
  obtained with the three generators showered with \PythiaEightPtwo{}.
  The last column reports the statistical uncertainty of our results.}
\label{tab:Ebj_extraction-errors}
\end{table*}
we summarize our results together with the scale, PDF and $\as$
uncertainties, that are extracted with a procedure analogous to one described
for the \mwbj{} observable.  We also report the corresponding
statistical errors of our results.  We see that scale and PDF variations have
negligible impact on our observable, the only important change being
associated with the choice of the NLO+PS generator.

We notice that our errors on scale and PDF variations are much smaller than
our statistical errors. On the other hand, these variations are performed by
reweighting techniques, that, because of correlations, lead to errors in the
differences that are much smaller than the error on the individual term. In
view of the small size of these variations, we do not attempt to perform a
better estimate of their error. On the other hand, the variation of $\as$ do
not benefit from this cancellation, and are all below the statistical
uncertainties.

As previously done for \mwbj{}, we have also investigated the dependence of
the \bjet{} peak positions on the jet radius. The results are summarized in
Tab.~\ref{tab:Ebj_cmp_bb4l_allradii}.
\begin{table*}[h!tb]
\centering
{ \begin{tabular}{l|c|c|c|}
 \cline{2-4}
 & $\phantom{\Big|}$ $R=0.4$ & $R=0.5$ & $R=0.6$ \\ 
 \cline{1-4}
\multicolumn{1}{ |c|}{ $\phantom{\Big|}$ \bbfourl{}}
 & $ 67.792\pm  0.089 $~GeV
 & $ 71.200\pm  0.081 $~GeV
 & $ 74.454\pm  0.076 $~GeV
\\ \cline{1-4}
\multicolumn{1}{ |c|}{ $\phantom{\Big|}$ \ttbnlodec{} ${}-$ \bbfourl{}}
 & $+   365\pm    110 $~MeV
 & $+   161\pm    102 $~MeV
 & $+    75\pm     97 $~MeV
\\ \cline{1-4}
\multicolumn{1}{ |c|}{ $\phantom{\Big|}$ \hvq{} ${}-$ \bbfourl{}}
 & $   -563\pm    110 $~MeV
 & $   -456\pm    103 $~MeV
 & $   -323\pm     97 $~MeV
\\ \cline{1-4}
\end{tabular}
}
\caption{\Ebj{} peak position obtained with the \bbfourl{} generator showered
  with \PythiaEightPtwo{}, for three choices of the jet radius. The differences
  with the \ttbnlodec{} and the \hvq{} generators are also shown.}
\label{tab:Ebj_cmp_bb4l_allradii}
\end{table*}
While we observe a marked change in \Ebjmax{}, that grows by $\diffEbjcml$
and $\diffEbjumc$~GeV when going from $R=0.4$ to $0.5$ and from $0.5$ to
$0.6$ respectively, \ttbnlodec{} and \hvq{} differ by \bbfourl{} by much
smaller amounts.  It is not clear whether such small differences could be
discriminated experimentally.

According to eqs.~(\ref{eq:delta_mt}) and~(\ref{eq:B_Ebj}), the uncertainties
that affect the value of the extracted top mass are nearly twice the
uncertainties on the \bjet{} energy. Considering the difference for $R=0.5$
between \hvq{} and \bbfourl{} in Tab.~\ref{tab:Ebj_cmp_bb4l_allradii}, we see
that, by using \hvq{} instead of \bbfourl{}, the extracted top mass would be
roughly 900~MeV larger. This should be compared with the corresponding
difference of about~150~MeV, that is shown in
Tab.~\ref{tab:mass_extraction-radius}, for the smeared \mwbj{} case.

As before, we have checked the sensitivity of our result to variations in the
matching procedure in \PythiaEightPtwo{}, by studying the difference between {\tt
  ScaleResonance} and {\tt FSREmission} options. The differences turn out to
be of the order of the statistical error.

\subsection{Comparison with \HerwigSevenPone{}}
\begin{figure}[tb]
  \centering
 \includegraphics[width=\wfigsing]{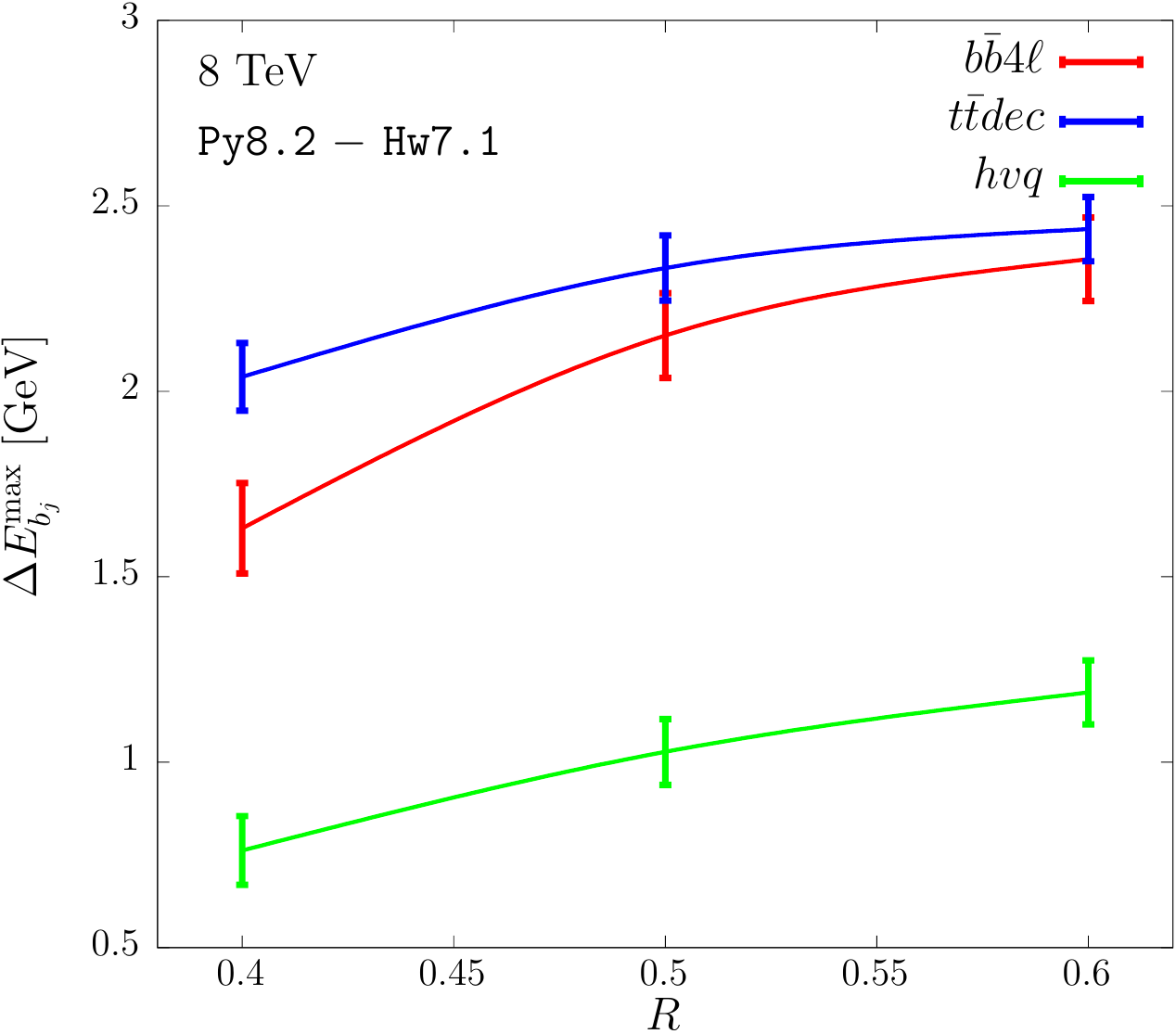}
 \caption{Differences of $\Ebjmax$ between the \PythiaEightPtwo{} and the
     \HerwigSevenPone{} showers, for the three generators, as a function of the
     jet radius.}
 \label{fig:R_Ebj_py8-hw7}
\end{figure}

In this section, we study the dependence of our results on the shower MC
program, comparing \HerwigSevenPone{} and \PythiaEightPtwo{} predictions.  We extract
the differences in the \Ebjmax{} position for three values of the jet radius:
$R=0.4$, 0.5 and 0.6. The results are summarized in
Tab.~\ref{tab:Ebj_py8-hw7}, where we also show the results at the
PS-only level,
\begin{table*}[tb]
  \centering
  \resizebox{\textwidth}{!}
  { \begin{tabular}{l|c|c|c|c|c|c|}
 \cline{2-7}
 & \multicolumn{6}{|c|}{$\phantom{\Big|}$ \PythiaEightPtwo{} ${}-$ \HerwigSevenPone{} [MeV]} \\
 \cline{2-7}
 & \multicolumn{2}{|c|}{$\phantom{\Big|}$ $R=0.4$}  & \multicolumn{2}{|c|}{$R=0.5$}  & \multicolumn{2}{|c|}{$R=0.6$} \\
 \cline{2-7}
 & $\phantom{\Big|}$ PS only & full & $\phantom{\Big|}$ PS only & full & $\phantom{\Big|}$ PS only & full \\ 
 \cline{1-7}
\multicolumn{1}{ |c|}{ $\phantom{\Big|}$ \bbfourl{}}
 & $+  1297\pm    120 $
 & $+  1631\pm    122 $
 & $+  1666\pm    117 $
 & $+  2150\pm    114 $
 & $+  1802\pm    114 $
 & $+  2356\pm    113 $
\\ \cline{1-7}
\multicolumn{1}{ |c|}{ $\phantom{\Big|}$ \ttbnlodec{}}
 & $+  1786\pm     91 $
 & $+  2039\pm     91 $
 & $+  2179\pm     88 $
 & $+  2332\pm     88 $
 & $+  2121\pm     89 $
 & $+  2437\pm     87 $
\\ \cline{1-7}
\multicolumn{1}{ |c|}{ $\phantom{\Big|}$ \hvq{}}
 & $+   515\pm     94 $
 & $+   762\pm     93 $
 & $+   707\pm     90 $
 & $+  1028\pm     89 $
 & $+   779\pm     87 $
 & $+  1188\pm     86 $
\\ \cline{1-7}
\end{tabular}
}
  \caption{Differences in the \Ebj{} peak position between the \PythiaEightPtwo{}
    and the \HerwigSevenPone{} showers applied to the three generators for three
    choices of the jet radius. The results at the NLO+PS level (PS only) are
    also shown.}
  \label{tab:Ebj_py8-hw7}
\end{table*}
and in Fig.~\ref{fig:R_Ebj_py8-hw7}.
From Tab.~\ref{tab:Ebj_py8-hw7} we clearly see that the \bbfourl{} and the
\ttbnlodec{} generators display larger discrepancies. For example, for the
central value $R=0.5$, we would get $\Delta \Ebjmax{}\approx 2$~GeV, that
roughly corresponds to $\Delta \mt=-4$~GeV. In the case of the \hvq{}
generator the difference is near 1~GeV, implying that the extracted mass
using \hvq{}+\HerwigSevenPone{} would be 2~GeV bigger than the one obtained with
\hvq{}+\PythiaEightPtwo{}.

We find that the differences between \HerwigSevenPone{} and \PythiaEightPtwo{}
increases for larger jet radii.  Furthermore, by looking at
Fig.~\ref{fig:R_Ebj_py8-hw7}, we notice that the \bbfourl{} generator
displays a different $R$ dependence, as we have already observed from
Tab.~\ref{tab:Ebj_cmp_bb4l_allradii}.  Figure~\ref{fig:R_Ebj_py8-hw7}
indicates that \bbfourl{} and \ttbnlodec{} are in better agreement for larger
values of the jet radius. This was also observed for the peak of the \mwbj{}
smeared distribution~(Tab.~\ref{tab:mass_extraction-radius}).

We notice that, as in the case of the reconstructed mass peak, the
predominant contribution to the difference arises at the parton shower level.

As for the previous cases, we have examined the variations due to a different
choice of the matching scheme in \HerwigSevenPone{}, that we found to be below
the 200~MeV level, and thus negligible in the present context.

\section{Leptonic observables}
\label{sec:LepObs}
In this section, we investigate the extraction of the top mass from the
leptonic observables introduced in Ref.~\cite{Frixione:2014ala}.  This method
has been recently studied by the ATLAS collaboration in
Ref.~\cite{ATLAS-CONF-2017-044}.

Following Ref.~\cite{Frixione:2014ala}, we consider the subsequent five observables
\begin{equation}
  \begin{array}{l l l }
O_1 =  \pT(\ell^+), &  O_2=  \pT(\ell^+\ell^-), &  O_3=  m(\ell^+\ell^-), \\[2mm] 
O_4=  E(\ell^+\ell^-),\quad  & O_5=  \pT(\ell^+)+\pT(\ell^-), &
  \end{array}
  \nonumber
\end{equation}
i.e.~the transverse momentum of the positive charged lepton, and the
transverse momentum, the invariant mass, the energy and the scalar sum of the
transverse momenta of the lepton pair.  We compute the average value of the
first three Mellin moments for each of the above mentioned observables,
$\langle (O_i)^j\rangle$, with $i=1,\dots,5$ and $j=1,2,3$.  We assume that,
if we do not vary too much the range of the top mass, we can write the linear
relation
\begin{equation}
\langle (O_i)^j \rangle =O_{\rm c}^{(ij)} + B^{(ij)} \lq \(\mt\)^j- \(\mtc\)^j \rq.
\label{eq:leptonicObs}
\end{equation}
For ease of notation, we will refer to $O_{\rm c}^{(ij)}$ and $B^{(ij)}$ as
$O_{\rm c}$ and $B$ in the following.
Their determination will be discussed later.

We choose as reference sample the one generated with \bbfourl{} matched with
\PythiaEightPtwo{}, using $\mtc=172.5$~GeV as input mass and the central
choices for the PDF and scales. We indicate the values of the observables
computed with this generator as ${O}^{b\bar{b}4\ell}$, and with $O_{\rm c}'$
the values of the observable computed either with an alternative generator or
with different generator settings, but using as input parameter the same
reference mass. The mass value that we would extract from the events of the
reference sample using the new generator is then given by
\begin{equation}
\label{eq:mtprime}  
\mt' = \left[\(\mtc\)^j -\frac{O'_{\rm c}-{O}^{b\bar{b}4\ell }_{\rm c}}{B} \right]^{1/j}\,.
\end{equation}

\subsection{Comparison among NLO+PS generators}

\begin{table*}[tb]
  \centering
  \resizebox{\textwidth}{!}
  { \begin{tabular}{lcc|c|c|c|c|}
\cline{1-7}
\multicolumn{1}{ |c|}{observable $\phantom{\Big|}$}  & \multicolumn{1}{ |c|}{gen} & \multicolumn{1}{ |c|}{ $\langle O_{\rm c}\rangle $}
 & $\%$ ${}-$ \bbfourl & $(\muF, \muR )$ & PDF & $\as$  \\
\cline{1-7}
 \\[-1.25em]
\cline{1-7}
\multicolumn{1}{ |c|}{}  & \multicolumn{1}{ |c|}{\bbfourl $\phantom{\Big|}$}
& $ 56.653\pm  0.050$~GeV
& -                                               
& ${}_{-     86}^{+     79}$~MeV
& -                                               
& $\pm     26\,\,( \pm     92) $~MeV   
\\ \cline{2-7}
\multicolumn{1}{ |c|}{$\langle \pT(\ell^+)\rangle $} & \multicolumn{1}{ |c|}{\ttbnlodec $\phantom{\Big|}$}
& $ 56.804\pm  0.033$~GeV
& $+   151\pm     60$~MeV                         
& ${}_{-     86}^{+     84}$~MeV
& -                                               
& $\pm     41\,\,( \pm     23) $~MeV   
\\ \cline{2-7}
\multicolumn{1}{ |c|}{} & \multicolumn{1}{ |c|}{\hvq $\phantom{\Big|}$ }
& $ 56.738\pm  0.032$~GeV
& $+    85\pm     59$~MeV                         
& ${}_{-     86}^{+     82}$~MeV
& $\pm    130$~MeV                                
& $\pm     49\,\,( \pm     23) $~MeV   
\\ \cline{2-7}
\cline{1-7}
 \\[-1.25em]
\cline{1-7}
\multicolumn{1}{ |c|}{}  & \multicolumn{1}{ |c|}{\bbfourl $\phantom{\Big|}$}
& $ 69.759\pm  0.059$~GeV
& -                                               
& ${}_{-    444}^{+    710}$~MeV
& -                                               
& $\pm     85\,\,( \pm    110) $~MeV   
\\ \cline{2-7}
\multicolumn{1}{ |c|}{$\langle \pT(\ell^+\ell^-)\rangle $} & \multicolumn{1}{ |c|}{\ttbnlodec $\phantom{\Big|}$}
& $ 69.660\pm  0.040$~GeV
& $   -100\pm     71$~MeV                         
& ${}_{-    361}^{+    538}$~MeV
& -                                               
& $\pm     78\,\,( \pm     28) $~MeV   
\\ \cline{2-7}
\multicolumn{1}{ |c|}{} & \multicolumn{1}{ |c|}{\hvq $\phantom{\Big|}$ }
& $ 69.201\pm  0.038$~GeV
& $   -558\pm     71$~MeV                         
& ${}_{-    367}^{+    553}$~MeV
& $\pm     95$~MeV                                
& $\pm     95\,\,( \pm     27) $~MeV   
\\ \cline{2-7}
\cline{1-7}
 \\[-1.25em]
\cline{1-7}
\multicolumn{1}{ |c|}{}  & \multicolumn{1}{ |c|}{\bbfourl $\phantom{\Big|}$}
& $108.685\pm  0.099$~GeV
& -                                               
& ${}_{-    341}^{+    234}$~MeV
& -                                               
& $\pm     57\,\,( \pm    191) $~MeV   
\\ \cline{2-7}
\multicolumn{1}{ |c|}{$\langle m(\ell^+\ell^-)\rangle $} & \multicolumn{1}{ |c|}{\ttbnlodec $\phantom{\Big|}$}
& $108.812\pm  0.065$~GeV
& $+   127\pm    119$~MeV                         
& ${}_{-    259}^{+    244}$~MeV
& -                                               
& $\pm     33\,\,( \pm     46) $~MeV   
\\ \cline{2-7}
\multicolumn{1}{ |c|}{} & \multicolumn{1}{ |c|}{\hvq $\phantom{\Big|}$ }
& $109.200\pm  0.064$~GeV
& $+   515\pm    118$~MeV                         
& ${}_{-    265}^{+    247}$~MeV
& $\pm    395$~MeV                                
& $\pm     68\,\,( \pm     45) $~MeV   
\\ \cline{2-7}
\cline{1-7}
 \\[-1.25em]
\cline{1-7}
\multicolumn{1}{ |c|}{}  & \multicolumn{1}{ |c|}{\bbfourl $\phantom{\Big|}$}
& $186.803\pm  0.163$~GeV
& -                                               
& ${}_{-    385}^{+    342}$~MeV
& -                                               
& $\pm    540\,\,( \pm    305) $~MeV   
\\ \cline{2-7}
\multicolumn{1}{ |c|}{$\langle E(\ell^+\ell^-)\rangle $} & \multicolumn{1}{ |c|}{\ttbnlodec $\phantom{\Big|}$}
& $187.005\pm  0.107$~GeV
& $+   201\pm    195$~MeV                         
& ${}_{-    434}^{+    448}$~MeV
& -                                               
& $\pm    474\,\,( \pm     76) $~MeV   
\\ \cline{2-7}
\multicolumn{1}{ |c|}{} & \multicolumn{1}{ |c|}{\hvq $\phantom{\Big|}$ }
& $186.809\pm  0.105$~GeV
& $+     6\pm    194$~MeV                         
& ${}_{-    427}^{+    441}$~MeV
& $\pm   1068$~MeV                                
& $\pm    559\,\,( \pm     74) $~MeV   
\\ \cline{2-7}
\cline{1-7}
 \\[-1.25em]
\cline{1-7}
\multicolumn{1}{ |c|}{}  & \multicolumn{1}{ |c|}{\bbfourl $\phantom{\Big|}$}
& $113.322\pm  0.095$~GeV
& -                                               
& ${}_{-    184}^{+    165}$~MeV
& -                                               
& $\pm     93\,\,( \pm    178) $~MeV   
\\ \cline{2-7}
\multicolumn{1}{ |c|}{$\langle \pT(\ell^+)+\pT(\ell^-)\rangle $} & \multicolumn{1}{ |c|}{\ttbnlodec $\phantom{\Big|}$}
& $113.598\pm  0.063$~GeV
& $+   276\pm    114$~MeV                         
& ${}_{-    174}^{+    165}$~MeV
& -                                               
& $\pm     72\,\,( \pm     44) $~MeV   
\\ \cline{2-7}
\multicolumn{1}{ |c|}{} & \multicolumn{1}{ |c|}{\hvq $\phantom{\Big|}$ }
& $113.425\pm  0.062$~GeV
& $+   104\pm    113$~MeV                         
& ${}_{-    177}^{+    163}$~MeV
& $\pm    259$~MeV                                
& $\pm    101\,\,( \pm     43) $~MeV   
\\ \cline{2-7}
\cline{1-7}
\end{tabular}
}
  \caption{The average values of each leptonic observable computed with
    \bbfourl{}, \ttbnlodec{} and \hvq{}, showered with \PythiaEightPtwo{},
    for $\mt$=172.5~GeV, and their variations with respect to \bbfourl{} are
    shown in the first two columns.  The differences with respect to their
    corresponding central values due to scale and PDF variations are also
    shown in columns three and four.  Their $\as$ uncertainties, computed as
    described in Sec.~\ref{sec:as_dependence} are displayed in column five.
    The statistical errors are also reported, except for the scale and PDF
    variations, where they have been estimated to be below 13\%{} of the
    quoted values.}
\label{tab:Olep-summary-py}
\end{table*}

\begin{table*}[tb]
  \centering
\resizebox{\textwidth}{!}
      { \begin{tabular}{lcc|c|c|c|c|}
\cline{1-7}
\multicolumn{1}{ |c|}{observable $\phantom{\Big|}$}  & \multicolumn{1}{ |c|}{gen} & \multicolumn{1}{ |c|}{ $\langle O_{\rm c}\rangle $}
 & $\%$ ${}-$ \bbfourl & $(\muF, \muR )$ & PDF & $\as$  \\
\cline{1-7}
 \\[-1.25em]
\cline{1-7}
\multicolumn{1}{ |c|}{}  & \multicolumn{1}{ |c|}{\bbfourl $\phantom{\Big|}$}
& $ 56.104\pm  0.049$~GeV
& -                                               
& ${}_{-    106}^{+     92}$~MeV
& -                                               
& $\pm     20\,\,( \pm     91) $~MeV   
\\ \cline{2-7}
\multicolumn{1}{ |c|}{$\langle \pT(\ell^+)\rangle $} & \multicolumn{1}{ |c|}{\ttbnlodec $\phantom{\Big|}$}
& $ 56.199\pm  0.047$~GeV
& $+    95\pm     68$~MeV                         
& ${}_{-    105}^{+     90}$~MeV
& -                                               
& $\pm     23\,\,( \pm     23) $~MeV   
\\ \cline{2-7}
\multicolumn{1}{ |c|}{} & \multicolumn{1}{ |c|}{\hvq $\phantom{\Big|}$ }
& $ 56.399\pm  0.032$~GeV
& $+   295\pm     59$~MeV                         
& ${}_{-    100}^{+     87}$~MeV
& $\pm    222$~MeV                                
& $\pm     45\,\,( \pm     23) $~MeV   
\\ \cline{2-7}
\cline{1-7}
 \\[-1.25em]
\cline{1-7}
\multicolumn{1}{ |c|}{}  & \multicolumn{1}{ |c|}{\bbfourl $\phantom{\Big|}$}
& $ 68.665\pm  0.059$~GeV
& -                                               
& ${}_{-    372}^{+    587}$~MeV
& -                                               
& $\pm     54\,\,( \pm    108) $~MeV   
\\ \cline{2-7}
\multicolumn{1}{ |c|}{$\langle \pT(\ell^+\ell^-)\rangle $} & \multicolumn{1}{ |c|}{\ttbnlodec $\phantom{\Big|}$}
& $ 68.632\pm  0.051$~GeV
& $    -33\pm     78$~MeV                         
& ${}_{-    307}^{+    452}$~MeV
& -                                               
& $\pm     56\,\,( \pm     28) $~MeV   
\\ \cline{2-7}
\multicolumn{1}{ |c|}{} & \multicolumn{1}{ |c|}{\hvq $\phantom{\Big|}$ }
& $ 68.566\pm  0.038$~GeV
& $    -99\pm     70$~MeV                         
& ${}_{-    312}^{+    466}$~MeV
& $\pm    161$~MeV                                
& $\pm     91\,\,( \pm     27) $~MeV   
\\ \cline{2-7}
\cline{1-7}
 \\[-1.25em]
\cline{1-7}
\multicolumn{1}{ |c|}{}  & \multicolumn{1}{ |c|}{\bbfourl $\phantom{\Big|}$}
& $108.497\pm  0.099$~GeV
& -                                               
& ${}_{-    265}^{+    201}$~MeV
& -                                               
& $\pm     24\,\,( \pm    190) $~MeV   
\\ \cline{2-7}
\multicolumn{1}{ |c|}{$\langle m(\ell^+\ell^-)\rangle $} & \multicolumn{1}{ |c|}{\ttbnlodec $\phantom{\Big|}$}
& $108.076\pm  0.072$~GeV
& $   -422\pm    122$~MeV                         
& ${}_{-    250}^{+    240}$~MeV
& -                                               
& $\pm      2\,\,( \pm     46) $~MeV   
\\ \cline{2-7}
\multicolumn{1}{ |c|}{} & \multicolumn{1}{ |c|}{\hvq $\phantom{\Big|}$ }
& $109.056\pm  0.063$~GeV
& $+   559\pm    117$~MeV                         
& ${}_{-    258}^{+    247}$~MeV
& $\pm    683$~MeV                                
& $\pm     52\,\,( \pm     45) $~MeV   
\\ \cline{2-7}
\cline{1-7}
 \\[-1.25em]
\cline{1-7}
\multicolumn{1}{ |c|}{}  & \multicolumn{1}{ |c|}{\bbfourl $\phantom{\Big|}$}
& $185.540\pm  0.162$~GeV
& -                                               
& ${}_{-    380}^{+    337}$~MeV
& -                                               
& $\pm    504\,\,( \pm    304) $~MeV   
\\ \cline{2-7}
\multicolumn{1}{ |c|}{$\langle E(\ell^+\ell^-)\rangle $} & \multicolumn{1}{ |c|}{\ttbnlodec $\phantom{\Big|}$}
& $185.315\pm  0.118$~GeV
& $   -225\pm    200$~MeV                         
& ${}_{-    416}^{+    428}$~MeV
& -                                               
& $\pm    426\,\,( \pm     76) $~MeV   
\\ \cline{2-7}
\multicolumn{1}{ |c|}{} & \multicolumn{1}{ |c|}{\hvq $\phantom{\Big|}$ }
& $186.125\pm  0.104$~GeV
& $+   585\pm    192$~MeV                         
& ${}_{-    410}^{+    420}$~MeV
& $\pm   1842$~MeV                                
& $\pm    520\,\,( \pm     73) $~MeV   
\\ \cline{2-7}
\cline{1-7}
 \\[-1.25em]
\cline{1-7}
\multicolumn{1}{ |c|}{}  & \multicolumn{1}{ |c|}{\bbfourl $\phantom{\Big|}$}
& $112.280\pm  0.095$~GeV
& -                                               
& ${}_{-    218}^{+    188}$~MeV
& -                                               
& $\pm     52\,\,( \pm    177) $~MeV   
\\ \cline{2-7}
\multicolumn{1}{ |c|}{$\langle \pT(\ell^+)+\pT(\ell^-)\rangle $} & \multicolumn{1}{ |c|}{\ttbnlodec $\phantom{\Big|}$}
& $112.455\pm  0.077$~GeV
& $+   174\pm    122$~MeV                         
& ${}_{-    205}^{+    177}$~MeV
& -                                               
& $\pm     36\,\,( \pm     45) $~MeV   
\\ \cline{2-7}
\multicolumn{1}{ |c|}{} & \multicolumn{1}{ |c|}{\hvq $\phantom{\Big|}$ }
& $112.796\pm  0.061$~GeV
& $+   516\pm    112$~MeV                         
& ${}_{-    204}^{+    176}$~MeV
& $\pm    444$~MeV                                
& $\pm     97\,\,( \pm     43) $~MeV   
\\ \cline{2-7}
\cline{1-7}
\end{tabular}
}
  \caption{As in Tab.~\ref{tab:Olep-summary-py} but for \HerwigSevenPone{}.}
\label{tab:Olep-summary-hw}
\end{table*}

We begin by showing in Tabs.~\ref{tab:Olep-summary-py}
and~\ref{tab:Olep-summary-hw} the average values of the leptonic observables
computed with our three NLO+PS generators interfaced with \PythiaEightPtwo{}
and \HerwigSevenPone{}. We show the central values, the differences with
respect to \bbfourl{}, and the upper and lower results induced by scale, PDF
and $\as$ variations.

The scale and PDF variations are performed by reweighting.  As a consequence
of that, the associated error is much smaller than the statistical error on
the cross section. In order to estimate it, we have divided our sample of
events in ten sub-samples, computed the observables for each sub-sample, and
carried out a straightforward statistical analysis on the ten sets of
results. We found errors that never exceed the quoted value by more than
13\%.

For the PDF variation, we have verified that differences due to variations in
our reference PDF sets (see Sec.~\ref{sec:PDF_dependence}) are very similar
among the different generators. On the other hand, a full error study using
the {\tt PDF4LHC15\_nlo\_30\_pdfas} set was only performed with the \hvq{}
generator, and the associated errors exceed by far the variation band that we
obtain with our reference sets. Thus, also in this case we quote the PDF
variations only for \hvq{}, implying that a very similar variation should
also be present for the others.  It is clear from the tables that the PDF
uncertainties are dominant for several observables, and scale variations are
also sizeable.

The large variations in the $\as$ column are not always conclusive because of
the large statistical errors (in parentheses), due to the fact that we cannot
perform this variation by reweighting. However, unlike for the $\mwbj$ case,
here the PDF dependence is not small, and thus we cannot conclude that the
$\as$ variation probes mainly the sensitivity to the intensity of radiation
in decay, since when we vary $\as$ we change also the PDF set.

It is instead useful to look at the effect of MEC on the leptonic
observables, displayed in Tab.~\ref{tab:leptobs_MEC}.
\begin{table*}[tb]
  \centering
  { \begin{tabular}{c|c|c|c|}
 \cline{2-4}
 &  \multicolumn{3}{ |c|}{$\phantom{\Big|}$  MEC ${}-$ no MEC} \\
 \cline{2-4}
 & $\phantom{\Big|}$  \tt \bbfourl & \ttbnlodec & \hvq{}\\
 \cline{1-4}
\multicolumn{1}{ |c|}{$\phantom{\Big|}$ $\langle \pT(\ell^+)\rangle $}
 & $+   117\pm
     74$~MeV 
 & $+    30\pm
     47$~MeV 
 & $+   342\pm
     46$~MeV 
\\ \cline{1-4}
\multicolumn{1}{ |c|}{$\phantom{\Big|}$ $\langle \pT(\ell^+\ell^-)\rangle $}
 & $+   167\pm
     89$~MeV 
 & $+    41\pm
     57$~MeV 
 & $+   544\pm
     55$~MeV 
\\ \cline{1-4}
\multicolumn{1}{ |c|}{$\phantom{\Big|}$ $\langle m(\ell^+\ell^-)\rangle $}
 & $+   171\pm
    149$~MeV 
 & $+   102\pm
     94$~MeV 
 & $+   631\pm
     91$~MeV 
\\ \cline{1-4}
\multicolumn{1}{ |c|}{$\phantom{\Big|}$ $\langle E(\ell^+\ell^-)\rangle $}
 & $+   372\pm
    243$~MeV 
 & $+   159\pm
    153$~MeV 
 & $+  1245\pm
    150$~MeV 
\\ \cline{1-4}
\multicolumn{1}{ |c|}{$\phantom{\Big|}$ $\langle \pT(\ell^+)+\pT(\ell^-)\rangle $}
 & $+   232\pm
    142$~MeV 
 & $+    85\pm
     89$~MeV 
 & $+   699\pm
     88$~MeV 
\\ \cline{1-4}
\end{tabular}
}
  \caption{Impact of MEC in \PythiaEightPtwo{} on the leptonic observables for the
   different NLO+PS generators.}
 \label{tab:leptobs_MEC}
\end{table*}
We observe that in the \bbfourl{} and \ttbnlodec{} case the effect of MEC is
compatible with the statistical uncertainty.  In the \hvq{} case we find
instead sizeable effects. This is expected, since large-angle radiation from
the $b$ quark, by subtracting energy to the whole $Wb$ system, affects
significantly also leptonic observables.

In Ref.~\cite{Frixione:2014ala} it was observed that the observables
$\pt(\ell^+\ell^-)$ and $m(\ell^+\ell^-)$ had larger errors due to a stronger
sensitivity to radiative corrections, and were more sensitive to
spin-correlation effects.  We see a confirmation of this observations in
their larger errors due to scale variation, and in the fact that for \hvq{}
their central value is shifted with respect to the \bbfourl{} and
\ttbnlodec{} generators, that treat spin correlations in a better way.

\begin{table}[tb]
\centering
{ \begin{tabular}{cc|c|}
\cline{1-3}
\multicolumn{1}{|c|}{observable $\phantom{\Big|}$} & \multicolumn{1}{|c|}{generator} &  $B$  \\
\cline{1-3}
 \\[-1.25em]
\cline{1-3}
\multicolumn{1}{ |c|}{} & \multicolumn{1}{ |c|}{\bbfourl $\phantom{\Big|}$}
& $   0.17\pm   0.04$
\\ \cline{2-3}
\multicolumn{1}{ |c|}{$\langle \pT(\ell^+)\rangle $} & \multicolumn{1}{ |c|}{\ttbnlodec $\phantom{\Big|}$}
& $   0.19\pm   0.02$
\\ \cline{2-3}
\multicolumn{1}{ |c|}{} & \multicolumn{1}{ |c|}{\hvq $\phantom{\Big|}$ }
& $   0.19\pm   0.02$
\\ \cline{2-3}
 \cline{1-3}
 \\[-1.25em]
\cline{1-3}
\multicolumn{1}{ |c|}{} & \multicolumn{1}{ |c|}{\bbfourl $\phantom{\Big|}$}
& $   0.30\pm   0.05$
\\ \cline{2-3}
\multicolumn{1}{ |c|}{$\langle \pT(\ell^+\ell^-)\rangle $} & \multicolumn{1}{ |c|}{\ttbnlodec $\phantom{\Big|}$}
& $   0.30\pm   0.02$
\\ \cline{2-3}
\multicolumn{1}{ |c|}{} & \multicolumn{1}{ |c|}{\hvq $\phantom{\Big|}$ }
& $   0.29\pm   0.02$
\\ \cline{2-3}
 \cline{1-3}
 \\[-1.25em]
\cline{1-3}
\multicolumn{1}{ |c|}{} & \multicolumn{1}{ |c|}{\bbfourl $\phantom{\Big|}$}
& $   0.31\pm   0.08$
\\ \cline{2-3}
\multicolumn{1}{ |c|}{$\langle m(\ell^+\ell^-)\rangle $} & \multicolumn{1}{ |c|}{\ttbnlodec $\phantom{\Big|}$}
& $   0.31\pm   0.03$
\\ \cline{2-3}
\multicolumn{1}{ |c|}{} & \multicolumn{1}{ |c|}{\hvq $\phantom{\Big|}$ }
& $   0.33\pm   0.03$
\\ \cline{2-3}
 \cline{1-3}
 \\[-1.25em]
\cline{1-3}
\multicolumn{1}{ |c|}{} & \multicolumn{1}{ |c|}{\bbfourl $\phantom{\Big|}$}
& $   0.55\pm   0.14$
\\ \cline{2-3}
\multicolumn{1}{ |c|}{$\langle E(\ell^+\ell^-)\rangle $} & \multicolumn{1}{ |c|}{\ttbnlodec $\phantom{\Big|}$}
& $   0.56\pm   0.05$
\\ \cline{2-3}
\multicolumn{1}{ |c|}{} & \multicolumn{1}{ |c|}{\hvq $\phantom{\Big|}$ }
& $   0.56\pm   0.05$
\\ \cline{2-3}
 \cline{1-3}
 \\[-1.25em]
\cline{1-3}
\multicolumn{1}{ |c|}{} & \multicolumn{1}{ |c|}{\bbfourl $\phantom{\Big|}$}
& $   0.38\pm   0.08$
\\ \cline{2-3}
\multicolumn{1}{ |c|}{$\langle \pT(\ell^+)+\pT(\ell^-)\rangle $} & \multicolumn{1}{ |c|}{\ttbnlodec $\phantom{\Big|}$}
& $   0.39\pm   0.03$
\\ \cline{2-3}
\multicolumn{1}{ |c|}{} & \multicolumn{1}{ |c|}{\hvq $\phantom{\Big|}$ }
& $   0.39\pm   0.03$
\\ \cline{2-3}
 \cline{1-3}
\end{tabular}
}
\caption{Extracted $B$ coefficients for the three different generators
  showered with \PythiaEightPtwo{}.}
\label{tab:Bcoeffs-lept}
\end{table}
In Tab.~\ref{tab:Bcoeffs-lept} we show the extracted values of the $B$
coefficients for the first Mellin moment of each observable. The $B$ values
corresponding to the different generators are compatible within the
statistical errors.  We thus choose the values computed with the \hvq{}
generator, that have the smallest error.  According to
eq.~(\ref{eq:mtprime}), we can translate a variation in an observable into a
variation of the extracted mass, that for the first Mellin moment is simply
obtained applying a $-1/B$ factor.  The results are illustrated in
Tab.~\ref{tab:mass_average_lept}.
\begin{table*}[tb]
\begin{center}  
\resizebox{\textwidth}{!}
{ \begin{tabular}{c|c|c|c|c|c|c|}
 \cline{2-7}
\multicolumn{1}{c|}{} &\multicolumn{3}{|c|}{$\phantom{\Big|}$ $\mt$ extracted with \PythiaEightPtwo{}} &\multicolumn{3}{|c|}{ $\mt$ extracted with \HerwigSevenPone{}}
 \\ \cline{1-7}
\multicolumn{1}{|c|}{$\phantom{\Big|}$ observable}
 & \bbfourl{} & \ttbnlodec{} & \hvq{} & \bbfourl{} & \ttbnlodec{} & \hvq{}
 \\ \cline{1-7}
\multicolumn{1}{|c|}{$\phantom{\Big|}$$\langle \pT(\ell^+)\rangle $}
& $ 172.500_{-  0.825}^{+  0.845} $
& $ 171.719_{-  0.816}^{+  0.821} $
& $ 172.060_{-  0.811}^{+  0.822} $
& $ 175.340_{-  1.269}^{+  1.298} $
& $ 174.847_{-  1.263}^{+  1.293} $
& $ 173.817_{-  1.244}^{+  1.270} $
\\ \cline{1-7}
\multicolumn{1}{|c|}{$\phantom{\Big|}$$\langle \pT(\ell^+\ell^-)\rangle $}
& $ 172.500_{-  2.515}^{+  1.601} $
& $ 172.848_{-  1.915}^{+  1.315} $
& $ 174.451_{-  1.967}^{+  1.334} $
& $ 176.328_{-  2.141}^{+  1.433} $
& $ 176.442_{-  1.689}^{+  1.227} $
& $ 176.675_{-  1.728}^{+  1.235} $
\\ \cline{1-7}
\multicolumn{1}{|c|}{$\phantom{\Big|}$$\langle m(\ell^+\ell^-)\rangle $}
& $ 172.500_{-  1.419}^{+  1.605} $
& $ 172.116_{-  1.417}^{+  1.441} $
& $ 170.945_{-  1.420}^{+  1.450} $
& $ 173.068_{-  2.171}^{+  2.233} $
& $ 174.342_{-  2.198}^{+  2.208} $
& $ 171.379_{-  2.203}^{+  2.214} $
\\ \cline{1-7}
\multicolumn{1}{|c|}{$\phantom{\Big|}$$\langle E(\ell^+\ell^-)\rangle $}
& $ 172.500_{-  2.037}^{+  2.061} $
& $ 172.138_{-  2.091}^{+  2.081} $
& $ 172.490_{-  2.086}^{+  2.076} $
& $ 174.771_{-  3.378}^{+  3.393} $
& $ 175.176_{-  3.406}^{+  3.401} $
& $ 173.720_{-  3.401}^{+  3.397} $
\\ \cline{1-7}
\multicolumn{1}{|c|}{$\phantom{\Big|}$$\langle \pT(\ell^+)+\pT(\ell^-)\rangle $}
& $ 172.500_{-  0.827}^{+  0.852} $
& $ 171.791_{-  0.806}^{+  0.818} $
& $ 172.233_{-  0.802}^{+  0.821} $
& $ 175.178_{-  1.265}^{+  1.296} $
& $ 174.730_{-  1.246}^{+  1.275} $
& $ 173.851_{-  1.239}^{+  1.267} $
\\ \cline{1-7}
\multicolumn{1}{|c|}{$\phantom{\Big|}$$\langle \pT^2(\ell^+)\rangle $}
& $ 172.500_{-  0.960}^{+  0.977} $
& $ 171.657_{-  1.011}^{+  0.998} $
& $ 172.286_{-  1.007}^{+  0.991} $
& $ 175.816_{-  1.502}^{+  1.515} $
& $ 175.326_{-  1.524}^{+  1.541} $
& $ 174.424_{-  1.497}^{+  1.508} $
\\ \cline{1-7}
\multicolumn{1}{|c|}{$\phantom{\Big|}$$\langle \pT^2(\ell^+\ell^-)\rangle $}
& $ 172.500_{-  3.375}^{+  2.072} $
& $ 172.945_{-  2.585}^{+  1.716} $
& $ 174.738_{-  2.577}^{+  1.694} $
& $ 176.673_{-  2.725}^{+  1.770} $
& $ 176.864_{-  2.170}^{+  1.533} $
& $ 177.253_{-  2.199}^{+  1.532} $
\\ \cline{1-7}
\multicolumn{1}{|c|}{$\phantom{\Big|}$$\langle m^2(\ell^+\ell^-)\rangle $}
& $ 172.500_{-  1.643}^{+  1.787} $
& $ 172.119_{-  1.680}^{+  1.687} $
& $ 171.286_{-  1.695}^{+  1.702} $
& $ 173.511_{-  2.569}^{+  2.573} $
& $ 174.808_{-  2.595}^{+  2.571} $
& $ 172.082_{-  2.644}^{+  2.619} $
\\ \cline{1-7}
\multicolumn{1}{|c|}{$\phantom{\Big|}$$\langle E^2(\ell^+\ell^-)\rangle $}
& $ 172.500_{-  2.462}^{+  2.457} $
& $ 172.072_{-  2.534}^{+  2.490} $
& $ 172.611_{-  2.518}^{+  2.475} $
& $ 175.005_{-  4.067}^{+  3.992} $
& $ 175.339_{-  4.093}^{+  3.996} $
& $ 174.054_{-  4.117}^{+  4.019} $
\\ \cline{1-7}
\multicolumn{1}{|c|}{$\phantom{\Big|}$$\langle (\pT(\ell^+)+\pT(\ell^-))^2\rangle $}
& $ 172.500_{-  1.035}^{+  1.076} $
& $ 171.642_{-  1.004}^{+  1.036} $
& $ 172.198_{-  1.008}^{+  1.043} $
& $ 175.489_{-  1.552}^{+  1.608} $
& $ 174.982_{-  1.536}^{+  1.563} $
& $ 174.145_{-  1.539}^{+  1.566} $
\\ \cline{1-7}
\multicolumn{1}{|c|}{$\phantom{\Big|}$$\langle \pT^3(\ell^+)\rangle $}
& $ 172.500_{-  1.268}^{+  1.269} $
& $ 171.558_{-  1.302}^{+  1.273} $
& $ 172.626_{-  1.299}^{+  1.262} $
& $ 176.472_{-  1.817}^{+  1.801} $
& $ 175.877_{-  1.872}^{+  1.861} $
& $ 175.212_{-  1.823}^{+  1.798} $
\\ \cline{1-7}
\multicolumn{1}{|c|}{$\phantom{\Big|}$$\langle \pT^3(\ell^+\ell^-)\rangle $}
& $ 172.500_{-  4.970}^{+  2.912} $
& $ 173.092_{-  3.825}^{+  2.435} $
& $ 175.316_{-  3.692}^{+  2.333} $
& $ 177.424_{-  3.756}^{+  2.355} $
& $ 177.691_{-  3.038}^{+  2.075} $
& $ 178.410_{-  3.033}^{+  2.046} $
\\ \cline{1-7}
\multicolumn{1}{|c|}{$\phantom{\Big|}$$\langle m^3(\ell^+\ell^-)\rangle $}
& $ 172.500_{-  2.080}^{+  2.172} $
& $ 172.416_{-  2.099}^{+  2.089} $
& $ 171.834_{-  2.140}^{+  2.124} $
& $ 173.978_{-  3.243}^{+  3.170} $
& $ 175.662_{-  3.219}^{+  3.127} $
& $ 172.980_{-  3.339}^{+  3.237} $
\\ \cline{1-7}
\multicolumn{1}{|c|}{$\phantom{\Big|}$$\langle E^3(\ell^+\ell^-)\rangle $}
& $ 172.500_{-  3.022}^{+  2.958} $
& $ 172.003_{-  3.107}^{+  2.998} $
& $ 172.843_{-  3.070}^{+  2.963} $
& $ 175.349_{-  4.944}^{+  4.701} $
& $ 175.515_{-  4.972}^{+  4.704} $
& $ 174.576_{-  5.017}^{+  4.744} $
\\ \cline{1-7}
\multicolumn{1}{|c|}{$\phantom{\Big|}$$\langle (\pT(\ell^+)\!+\!\pT(\ell^-))^3\rangle $}
& $ 172.500_{-  1.428}^{+  1.511} $
& $ 171.431_{-  1.374}^{+  1.417} $
& $ 172.134_{-  1.373}^{+  1.422} $
& $ 175.963_{-  2.022}^{+  2.137} $
& $ 175.379_{-  1.995}^{+  2.011} $
& $ 174.558_{-  2.012}^{+  2.029} $
\\ \cline{1-7}
\multicolumn{1}{|c|}{$\phantom{\Big|}$ {\bf all observables}}
& $ \mathbf{172.500_{-  0.766}^{+  0.784} }$
& $ \mathbf{171.751_{-  0.751}^{+  0.751} }$
& $ \mathbf{172.238_{-  0.748}^{+  0.754} }$
& $ \mathbf{175.392_{-  1.138}^{+  1.045} }$
& $ \mathbf{175.452_{-  1.104}^{+  0.962} }$
& $ \mathbf{174.607_{-  1.097}^{+  0.961} }$
\\ \cline{1-7}
\multicolumn{1}{|c|}{$\phantom{\Big|}$ {\bf 1st moment}}
& $ \mathbf{172.500_{-  0.772}^{+  0.794} }$
& $ \mathbf{171.755_{-  0.756}^{+  0.764} }$
& $ \mathbf{172.247_{-  0.753}^{+  0.766} }$
& $ \mathbf{175.440_{-  1.184}^{+  1.102} }$
& $ \mathbf{175.445_{-  1.141}^{+  1.011} }$
& $ \mathbf{174.756_{-  1.135}^{+  1.010} }$
\\ \cline{1-7}
\end{tabular}
}
\caption{Extracted mass in GeV for all the generators, showered with
  \PythiaEightPtwo{} and \HerwigSevenPone{}, corresponding to the different leptonic
  observables, using as reference sample the \bbfourl{} one generated with
  $\mt=172.5$~GeV and showered with \PythiaEightPtwo{}.  The quoted errors are
  obtained by summing in quadrature the scale, PDF and the statistical
  errors.  The weighted average is also shown, for all the observables and
  considering only their first Mellin moment.}
\label{tab:mass_average_lept}
\end{center}
\end{table*}
The errors shown have been obtained by summing in quadrature the statistical
error and the scale and PDF uncertainties. We have not included the $\as$
variation in the error in order to avoid overcounting, since, in the present
case, is likely to be largely dominated by the change in the associated PDF.

The overall errors on the last two lines of Tab.~\ref{tab:mass_average_lept}
are obtained with the same procedure adopted in Ref.~\cite{Frixione:2014ala}
to account for correlations among the different observables.  We do not see
excessive differences among our three generators showered with the same
Monte Carlo generator, while the differences between the \PythiaEightPtwo{} and
\HerwigSevenPone{} results are considerably large. This is also the case for the
\hvq{} generator, that has a much simpler interface to both \PythiaEightPtwo{}
and \HerwigSevenPone{}.

As we did for \mwbjmax{} and \Ebjmax{}, also in the present case we have
computed the leptonic observables without including hadronization effects,
i.e.~at parton-shower only level, in order to determine whether the
differences between \PythiaEightPtwo{} and \HerwigSevenPone{}  are due to
the shower or to the hadronization. Our findings are summarized in
Tab.~\ref{tab:lept-PSonly}. 
\begin{table}[ht]
  \centering
  {\begin{tabular}{lc|c|c|}
\\ \cline{3-4}
 & & \multicolumn{2}{ |c|}{$\phantom{\Big|}$ \PythiaEightPlot{} ${}-$ \HerwigSevenPlot{} [MeV]}
\\ \cline{1-4}
 \multicolumn{1}{ |c|}{observable\!\!$\phantom{\Big|}$} & \multicolumn{1}{ |c|}{gen} & \multicolumn{1}{ |c|}{full} & \multicolumn{1}{ |c|}{PS only}
\\ \cline{1-4}
\\[-1.25em]
\cline{1-4}
\multicolumn{1}{ |c|}{} & \multicolumn{1}{ |c|}{\bbfourl $\phantom{\Big|}$}
 & $+   549 \pm
     70 $ 
 & $+   563 \pm
     71 $ 
\\ \cline{2-4}
\multicolumn{1}{ |c|}{$\langle \pT(\ell^+)\rangle $} & \multicolumn{1}{ |c|}{\ttbnlodec $\phantom{\Big|}$}
 & $+   605 \pm
     57 $ 
 & $+   609 \pm
     48 $ 
\\ \cline{2-4}
\multicolumn{1}{ |c|}{} & \multicolumn{1}{ |c|}{\hvq $\phantom{\Big|}$ }
 & $+   340 \pm
     45 $ 
 & $+   376 \pm
     46 $ 
\\ \cline{2-4}
\cline{1-4}
\\[-1.25em]
\cline{1-4}
\multicolumn{1}{ |c|}{} & \multicolumn{1}{ |c|}{\bbfourl $\phantom{\Big|}$}
 & $+  1094 \pm
     83 $ 
 & $+  1092 \pm
     84 $ 
\\ \cline{2-4}
\multicolumn{1}{ |c|}{$\langle \pT(\ell^+\ell^-)\rangle $} & \multicolumn{1}{ |c|}{\ttbnlodec $\phantom{\Big|}$}
 & $+  1027 \pm
     65 $ 
 & $+  1020 \pm
     59 $ 
\\ \cline{2-4}
\multicolumn{1}{ |c|}{} & \multicolumn{1}{ |c|}{\hvq $\phantom{\Big|}$ }
 & $+   636 \pm
     54 $ 
 & $+   662 \pm
     55 $ 
\\ \cline{2-4}
\cline{1-4}
\\[-1.25em]
\cline{1-4}
\multicolumn{1}{ |c|}{} & \multicolumn{1}{ |c|}{\bbfourl $\phantom{\Big|}$}
 & $+   188 \pm
    140 $ 
 & $+   286 \pm
    142 $ 
\\ \cline{2-4}
\multicolumn{1}{ |c|}{$\langle m(\ell^+\ell^-)\rangle $} & \multicolumn{1}{ |c|}{\ttbnlodec $\phantom{\Big|}$}
 & $+   736 \pm
     97 $ 
 & $+   814 \pm
     98 $ 
\\ \cline{2-4}
\multicolumn{1}{ |c|}{} & \multicolumn{1}{ |c|}{\hvq $\phantom{\Big|}$ }
 & $+   144 \pm
     90 $ 
 & $+   182 \pm
     91 $ 
\\ \cline{2-4}
\cline{1-4}
\\[-1.25em]
\cline{1-4}
\multicolumn{1}{ |c|}{} & \multicolumn{1}{ |c|}{\bbfourl $\phantom{\Big|}$}
 & $+  1263 \pm
    229 $ 
 & $+  1342 \pm
    232 $ 
\\ \cline{2-4}
\multicolumn{1}{ |c|}{$\langle E(\ell^+\ell^-)\rangle $} & \multicolumn{1}{ |c|}{\ttbnlodec $\phantom{\Big|}$}
 & $+  1690 \pm
    160 $ 
 & $+  1712 \pm
    159 $ 
\\ \cline{2-4}
\multicolumn{1}{ |c|}{} & \multicolumn{1}{ |c|}{\hvq $\phantom{\Big|}$ }
 & $+   684 \pm
    148 $ 
 & $+   719 \pm
    150 $ 
\\ \cline{2-4}
\cline{1-4}
\\[-1.25em]
\cline{1-4}
\multicolumn{1}{ |c|}{} & \multicolumn{1}{ |c|}{\bbfourl $\phantom{\Big|}$}
 & $+  1041 \pm
    134 $ 
 & $+  1091 \pm
    136 $ 
\\ \cline{2-4}
\multicolumn{1}{ |c|}{$\langle \pT(\ell^+)+\pT(\ell^-)\rangle $} & \multicolumn{1}{ |c|}{\ttbnlodec $\phantom{\Big|}$}
 & $+  1143 \pm
     99 $ 
 & $+  1173 \pm
     92 $ 
\\ \cline{2-4}
\multicolumn{1}{ |c|}{} & \multicolumn{1}{ |c|}{\hvq $\phantom{\Big|}$ }
 & $+   629 \pm
     86 $ 
 & $+   690 \pm
     88 $ 
\\ \cline{2-4}
\cline{1-4}
\\[-1.25em]
\cline{1-4}
\end{tabular}
}
\caption{Differences between the \PythiaEightPtwo{} and \HerwigSevenPone{} results
  for the leptonic observables, at full hadron level and at parton-level
  only.}
\label{tab:lept-PSonly}
\end{table}
Most of the differences already arise at the shower level. We also remark
that, within the same SMC generator, they are not large, yielding differences
in the extracted top mass of the same size as the statistical errors.

We observe in Tab.~\ref{tab:mass_average_lept} that the inclusion of higher
moments of the leptonic observables does not modify appreciably the results
from the first moments. This is a consequence of the large error on the
higher moments, and of the strong correlations among different moments.

\begin{figure}
  \centering
  \includegraphics[width=\wfigsingmulti]{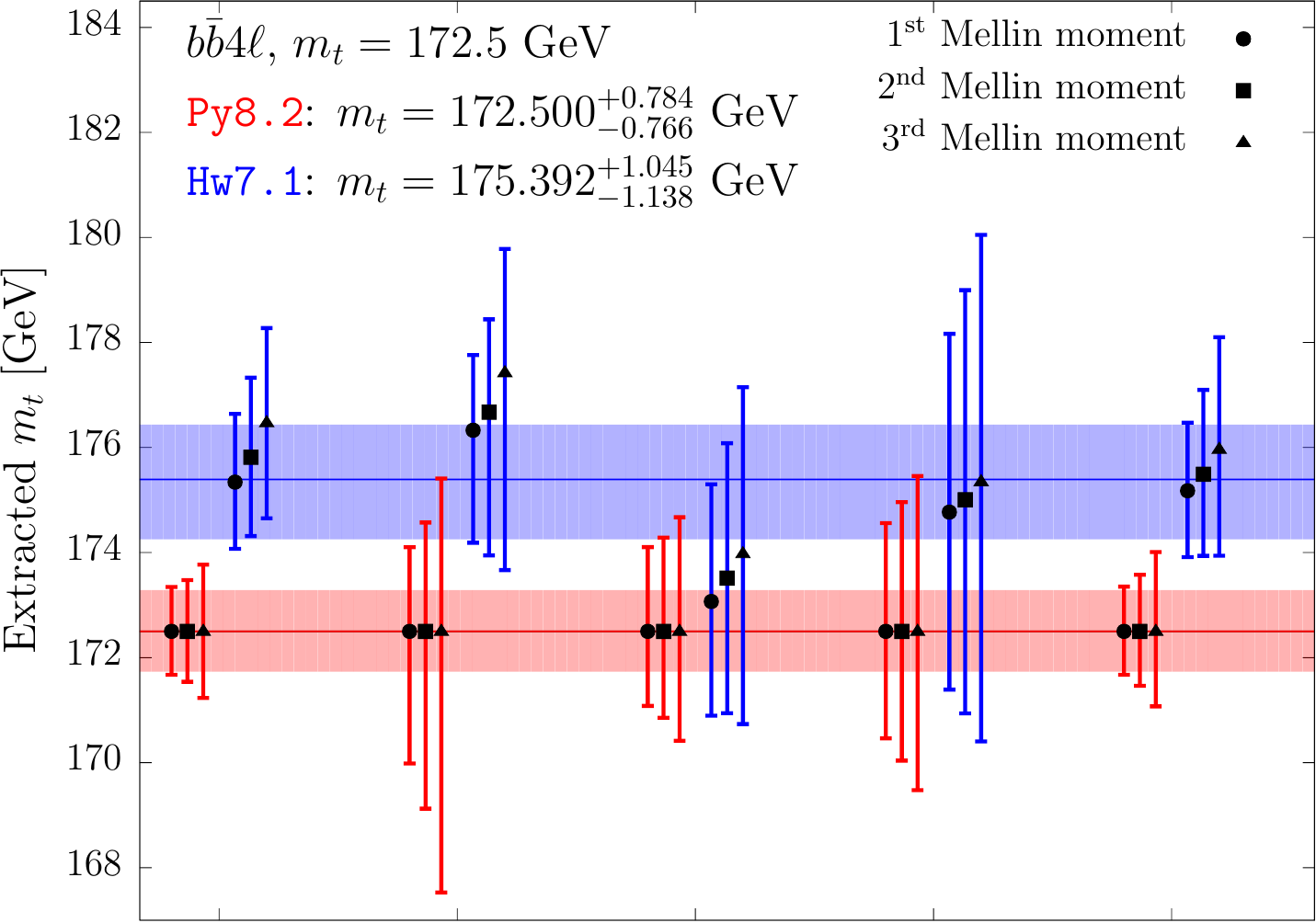}\\
  \includegraphics[width=\wfigsingmulti]{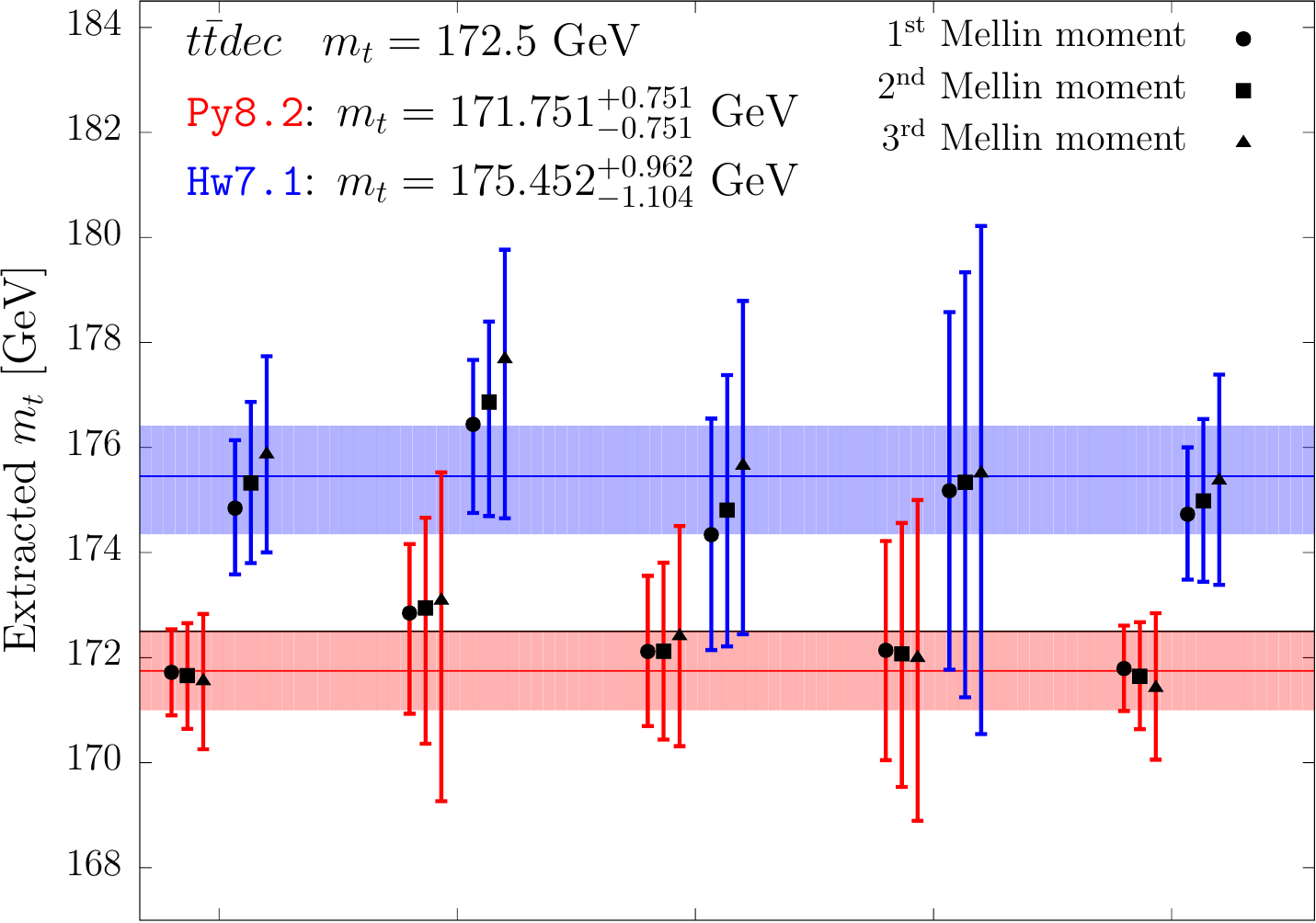}\\
  \includegraphics[width=\wfigsingmulti]{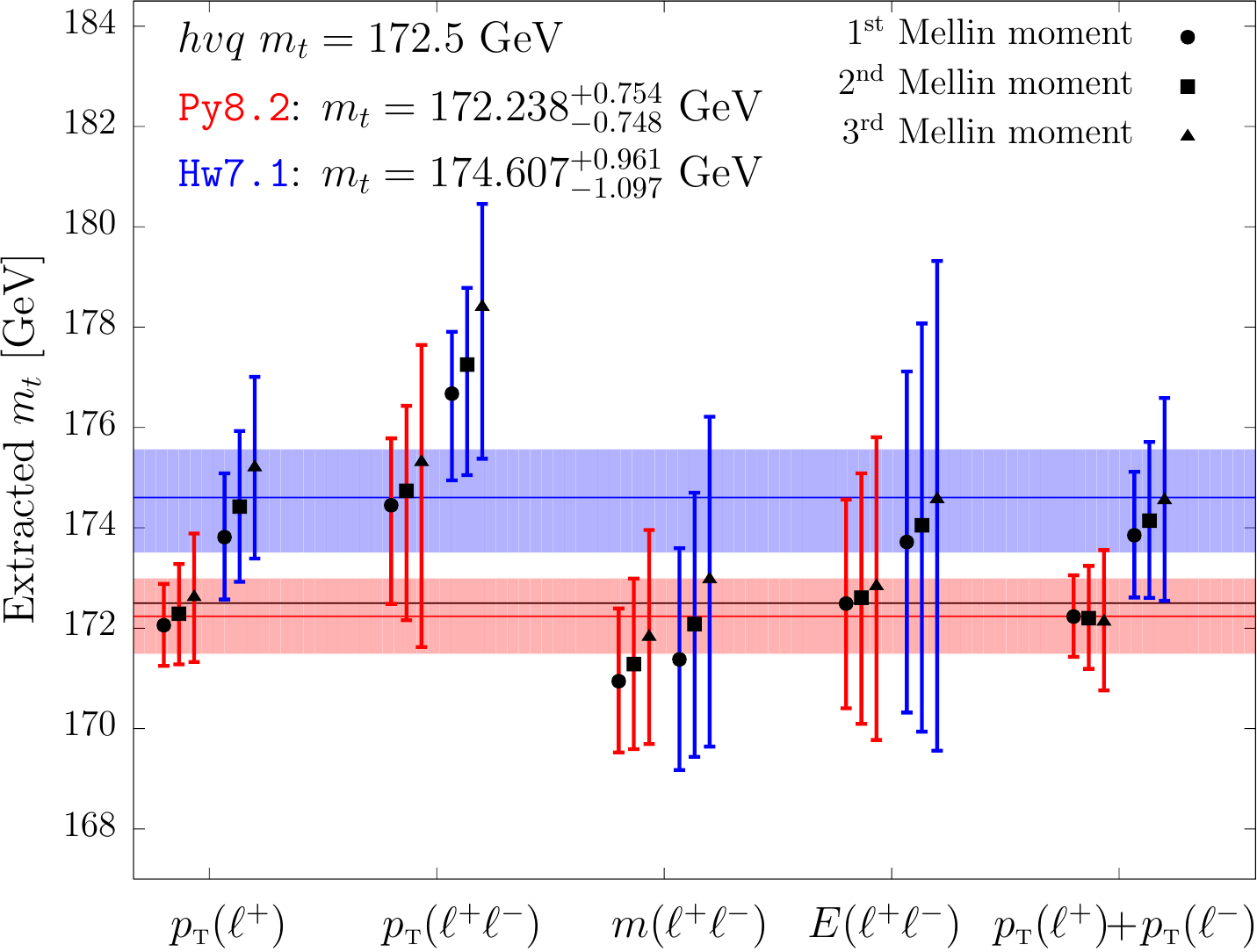}
\caption{ Extracted mass for the three generators matched with
  \PythiaEightPtwo{}~(red) and \HerwigSevenPone{}~(blue) using the first three
  Mellin moments of the five leptonic observables. The horizontal band
  represents the weighted average of the results, and the black
  horizontal line corresponds to $\mt=172.5$~GeV, which is the top
  mass value used in the \bbfourl{}+\PythiaEightPtwo{}
  reference sample.}
\label{fig:leptObs}
\end{figure}

The results in Tab.~\ref{tab:mass_average_lept} are also summarized in
Fig.~\ref{fig:leptObs},
where the discrepancy between \PythiaEightPtwo{} and \HerwigSevenPone{} and the
mutual consistency of the different observables can be immediately
appreciated.

As for the previous observables, we have studied the effect of changing the
matching scheme, by switching between our two alternative matching schemes
with \PythiaEightPtwo{} and \HerwigSevenPone{}, and by considering the settings of
eq.~(\ref{eq:TSsettings}) in \HerwigSevenPone{}. In both cases we find results
that are consistent within statistical errors.

\section{Summary}
\label{sec:Summary}
In this work we have compared generators of increasing accuracy for the
production and decay of $t\bar{t}$ pairs considering observables suitable for
the measurement of the top mass.  The generators that we have considered are:
\begin{itemize}
\item The \hvq{} generator~\cite{Frixione:2007nw}, that implements NLO
  corrections in production for on-shell top quarks, and includes
  finite-width effects and spin correlations only in an approximate way, by
  smearing the on-shell kinematics with Breit-Wigner forms of appropriate
  width, and by generating the angular distribution of the decay products
  according to the associated tree-level matrix
  elements~\cite{Frixione:2007zp}.
\item The \ttbnlodec{} generator~\cite{Campbell:2014kua}, that implements NLO
  corrections in production and decay in the narrow-width approximation. Spin
  correlations are included at NLO accuracy. Finite width effects are
  implemented by reweighting the NLO results using the tree-level matrix
  elements for the associated Born-level process, including however all
  finite width non-resonant and interference effects at the Born level for
  the given final state.
\item The \bbfourl{} generator~\cite{Jezo:2016ujg}, that uses the full matrix
  elements for the production of the given final state, including all
  non-resonant diagrams and interference effects.  This includes interference
  of QCD radiation in production and decay.
\end{itemize}
The main focus of our work has been the study of the mass distribution of a
particle-level reconstructed top, consisting of a lepton-neutrino pair and a
$b$-quark jet with the appropriate flavour. The peak position of the mass of
this system is our observable, that is loosely related to the top mass. We
considered its distributions both at the particle level, and by assuming that
experimental inaccuracies can be summarized by a simple smearing with a
resolution function, a Gaussian with a width of 15~GeV, which is the typical
resolution achieved on the top mass by the LHC collaborations. This
observable is an oversimplified version of the mass observables that are used
in direct top-mass measurements, that are the methods that lead to the most
precise mass determinations.

We have found a very consistent picture in the comparison of our three
generators when they are interfaced to \PythiaEightPtwo{}, and thus we begin
by summarizing our results for this case.  We first recall \emph{what we
  expect} from such comparison.  When comparing the \hvq{} and the
\ttNLOdec{} generators, we should remember that the latter has certainly
better accuracy in the description of spin correlations, since it implements
them correctly both at the leading and at the NLO level. However, we do not
expect spin correlations to play an important role in the reconstructed top
mass. As a further point, the \ttNLOdec{} generator implements NLO
corrections in decay. In the \hvq{} generator, the decay is handled by the
shower, where, by default, \PythiaEightPtwo{} includes matrix-element
corrections (MEC). These differ formally from a full NLO correction only by a
normalization factor, that amounts to the NLO correction to the top
width. Thus, as long as the MEC are switched on, we do not expect large
differences between \hvq{} and \ttNLOdec{}.  As far as the comparison between
\ttNLOdec{} and \bbfourl{}, we expect the difference to be given by NLO
off-shell effects, and by interference of radiation in production and decay,
since these effects are not implemented in \ttNLOdec{}. This comparison is
particularly interesting, since the interference between production and decay
can be considered as a ``perturbative precursor'' of colour reconnection
effects.

The results of these comparisons can be summarized as follows:
\begin{itemize}
\item The \ttNLOdec{} and the \bbfourl{} generators yield very similar
  results for most of the observables that we have considered, implying that
  NLO off-shell effects and interference between production and decay are
  modest.
\item
  As far as \mwbjmax{} (the peak of the reconstructed mass distribution) is
  concerned, the \ttNLOdec{} and the \hvq{} generators yield very similar
  results, confirming the fact that the MEC implementation in
  \PythiaEightPtwo{} has an effect very similar to the \POWHEG{}
  implementation of NLO corrections in decay in the \ttNLOdec{}. We have also
  observed that, if we switch off the MEC, the agreement between the two
  generators is spoiled.  More quantitatively, we find that the spread in the
  peak of the reconstructed mass at the particle level among the three NLO+PS
  generators is never above 30~MeV.  On the other hand, if resolution effects
  are accounted for with our smearing procedure, we find that the \hvq{}
  result is 147~MeV smaller, and the \ttbnlodec{} result 140~MeV larger than
  the \bbfourl{} one. These values are safely below currently quoted errors
  for the top-mass measurements with direct methods.

  If we switch off the MEC in \PythiaEightPtwo{}, we find that the peak position
  at the particle level in the \hvq{} case is displaced by 61~MeV, while, if
  smearing effects are included, the shift is of $916$~MeV, a rather large
  value, that can however be disregarded as being due to the poor accuracy of
  the collinear approximation in $b$ radiation when MEC corrections are off.
\item
  The jet-energy peak seems to be more sensitive to the modeling of radiation
  from the $b$ quark. In fact, while the \ttNLOdec{} and the \bbfourl{}
  results are quite consistent with each other, with the peak positions
  differing by less than 200~MeV, the \hvq{} result differs from them by more
  than 500~MeV. This would correspond to a difference in the extracted mass
  of the top quark roughly equal to twice that amount.  On the other hand, if
  the MEC in \hvq{} are switched off, the shift in the $b$-jet energy peak is
  more than 1.9~GeV. This leads us to conclude that the impact of modeling of
  $b$ radiation on the $b$-jet peak is much stronger than in the
  reconstructed top mass peak.  We stress, however, that the difference
  between \hvq{} (with MEC on) and the other two generators is safely below
  the errors quoted in current measurements~\cite{CMS-PAS-TOP-15-002}.

\item
  For the leptonic observables, we generally see a reasonable agreement
  between the different generators. The largest differences are found in the
  \hvq{} case, for the $\pt(\ell^+\ell^-)$ and $m(\ell^+\ell^-)$, larger than
  500~MeV with respect to the other two.  In Ref.~\cite{Frixione:2014ala} it
  was noticed that these observables had larger errors due to a stronger
  sensitivity to radiative corrections, and to spin-correlation effects, that
  are modelled incorrectly by \hvq{}.
\end{itemize}
Several sources of possible uncertainties have been explored in order to
check the reliability of these conclusions. First of all, two different
matching procedures for interfacing the \ttNLOdec{} and \bbfourl{} generators
to \PythiaEightPtwo{} have been implemented. For example, for the
reconstructed mass peak, we have checked that switching between them leads to
differences below 20~MeV for both generators. The effect of scales,
$\alpha_s$ and PDF uncertainties have also been examined, and were found to
yield very modest variations in the reconstructed mass peak. It was found, in
particular, that scale variations lead to a negligible peak displacement
(below 7~MeV) in the \ttNLOdec{} and \hvq{} case, while the effect is of
${}^{+86}_{-53}$~MeV for \bbfourl{}. The lack of scale dependence in the
\hvq{} and \ttNLOdec{} is easily understood as being due to the fact that the
peak shape is obtained by smearing an on-shell distribution with a
Breit-Wigner form, that does not depend upon any scale, and it suggests that,
in order to get realistic scale-variation errors, the most accurate
\bbfourl{} generator should be used.  We have also computed results at the
shower level, excluding the effects of hadronization and multi-parton
interactions, in order to see if the consistent picture found at the hadron
level is also supported by the parton-level results, and we have found that
this is indeed the case.

We have thus seen that the overall picture of the comparison of our three
NLO+PS generators within the framework of the \PythiaEightPtwo{} shower is quite
simple and consistent.  For the most precise observable, i.e.~the peak of the
reconstructed mass distribution, it leads to the conclusions that the use of
the most accurate generator may lead to a shift in the measured mass of at
most 150~MeV, which is well below the present uncertainties quoted by the
experimental collaborations.

Our study with \HerwigSevenPone{} instead reveals several problems. We can
summarize our findings as follows:
\begin{itemize}
\item The results obtained with \HerwigSevenPone{} differ substantially from
  those obtained with \PythiaEightPtwo{}. In particular, the peak of the
  reconstructed mass distribution at the particle level is shifted by -66 and
  -39~MeV in the \bbfourl{} and \ttNLOdec{} cases, and by +235 MeV in the
  \hvq{} case. When the experimental resolution is accounted for, using our
  smearing procedure, the shift raises to -1091 and -1179~MeV in the
  \bbfourl{} and \ttNLOdec{} cases, and to -251 MeV in the \hvq{} case.
\item The results obtained within the \HerwigSevenPone{} framework display large
  differences between the \hvq{} generator with respect to \bbfourl{} and
  \ttNLOdec{} ones. In particular, while the \ttNLOdec{} result exceeds the
  \bbfourl{} one only by about 50~MeV in both the particle level and smeared
  cases, \hvq{} exceeds \bbfourl{} by 311~MeV at particle level, and by
  693~MeV after smearing.
\end{itemize}
These results are quite alarming. The shifts reach values that are
considerably larger than current experimental uncertainties.

In the \hvq{} case, which is the NLO+PS generator currently used for top-mass
studies by the experimental collaborations, the difference in the mass-peak
position between \HerwigSevenPone{} and \PythiaEightPtwo{}, for the smeared
distribution, is -251~MeV, uncomfortably large but still below current
errors.  One would then be tempted to conclude that the large shifts may be
linked to some problems concerning the new generators. However, we also
notice that the same difference is +235~MeV when no smearing is applied, so
it is about as large in magnitude but with the opposite sign. This indicates
that the shape of the reconstructed mass distribution is considerably
different in the two shower models.  Lastly, if we use the internal \POWHEG{}
implementation of top decay (rather than the MEC) in \HerwigSevenPone{}, the
difference with respect to \PythiaEightPtwo{} raises to 607~MeV. Thus, we
conclude that in the \hvq{} case the smaller difference between
\HerwigSevenPone{} and \PythiaEightPtwo{} is accidental, and is subject to
considerable variations depending upon the settings.

Also in this case we checked whether the MEC yield an improved agreement
between the \hvq{} and the other two generators, as was observed for
\PythiaEightPtwo{}.  We find that, by switching off MEC, the
\hvq{}+\HerwigSevenPone{} result decreases by 307~MeV at particle level, and
by 1371~MeV in the smeared case. These effects are qualitatively similar to
what was observed in \PythiaEightPtwo{}. However, in the present case, when
MEC are switched off, the \hvq{} result exceeds the \bbfourl{} one by a
negligible amount at the particle level, and is lower than the \bbfourl{} one
by 678~MeV in the smeared case.

The discrepancy between \hvq{} and the other two generators is mitigated if,
instead of the MEC procedure, the internal \POWHEG{} option of
\HerwigSevenPone{} for top decay is used.  In this case, the discrepancy
between \hvq{} and \bbfourl{} is reduced to 244~MeV with no smearing, and to
337~MeV with smearing. We thus see that the consistency of the three NLO+PS
generators interfaced to \HerwigSevenPone{} is not optimal as in
\PythiaEightPtwo{}. It is however acceptable if the internal \POWHEG{}
feature is used rather than MEC in \HerwigSevenPone{}.

We have performed several studies to determine the origin of the difference
between \PythiaEightPtwo{} and \HerwigSevenPone{}, and to check whether it
could be attributed to some problem in our matching procedure. They can be
summarized as follows:
\begin{itemize}
\item
  We have shown that the difference is mostly due to the shower model, since
  it is already largely present at the parton level.
\item
  We have considered the $R$ dependence of the \HerwigSevenPone{} result. It
  differs from the one in \PythiaEightPtwo{}, leading to the hope that both
  generators may not represent the same set of data well, and tuning them may
  reduce their differences.  However, we have also noticed that the
  difference in slope is much smaller than the difference in size.
\item
  We have already mentioned that we have also compared results by making use
  of the internal \POWHEG{} implementation of top decay in
  \HerwigSevenPone{}, rather than using MEC. We have found non-negligible
  differences in this case.
\item
  We have implemented alternative veto procedure in the matching of
  \HerwigSevenPone{} with the NLO+PS generators. We found differences of the
  order of 200~MeV, not large enough to cover the discrepancy with
  \PythiaEightPtwo{}.
\item
  When interfacing \POWHEG{} generators to angular-ordered showers, in order
  to maintain the double-logarithmic accuracy of the shower, one should
  introduce the so called ``truncated showers''~\cite{Nason:2004rx}. One
  could then worry that the lack of truncated showers is at the origin of the
  discrepancies that we found.  Fortunately, \HerwigSevenPone{} offers some
  optional settings that are equivalent to the introduction of truncated
  showers. We found that these options lead to a shift of only 200~MeV in the
  peak position.
\end{itemize}
In summary, we found no indication that the discrepancy with
\PythiaEightPtwo{} is due to the specific matching procedure and general
settings that we have used in \HerwigSevenPone{}.

When comparing \HerwigSevenPone{} and \PythiaEightPtwo{} in the computation
of the $b$-jet energy peak, we have found even larger differences: when using
\bbfourl{} and \ttbnlodec{}, the shifts are of the order of 2~GeV, while for
\hvq{} the shift is around 1~GeV. They correspond to differences in the
extracted mass of around 4~GeV in the first two cases, and 2~GeV in the last
one.  This is not surprising, in view of the stronger sensitivity of the
$b$-jet peak to the shower model.

Finally, when considering leptonic observables, we find again large
differences between \HerwigSevenPone{} and \PythiaEightPtwo{}. Most
differences already arise at the shower level. Notice that this is in
contrast with the naive view that leptonic observables should be less
dependent upon QCD radiation effects and jet modeling. The comparison between
\HerwigSevenPone{} and \PythiaEightPtwo{} for leptonic observables can by
appreciated by looking at Fig.~\ref{fig:leptObs}, representing the value of
the extracted top mass from a sample generated with \bbfourl{} interfaced to
\PythiaEightPtwo{}.

\section{Conclusions}
\label{sec:Conc}
We focus our conclusions on the results obtained for the reconstructed mass
peak, since the issues that we have found there apply to the direct top mass
measurements, that are the most precise.  The experimental collaborations
extensively use the \hvq{} generator for this kind of analyses, and since new
generators of higher accuracy, the \ttbnlodec{} and the \bbfourl{} ones, have
become available, we have addressed the question of whether the physics
effects not included in \hvq{} may lead to inaccuracies in the top-mass
determination. The answer to this question is quite simple and clear when our
generators are interfaced to \PythiaEightPtwo{}. The differences that we find
are large enough to justify the use of the most accurate generators, but not
large enough to drastically overturn the conclusions of current measurements.
Notice that, since the \hvq{} generator does not include NLO corrections in
decays, we might have expected a very different modeling of the $b$-jet in
\hvq{} with respect to the other two generators, leading to important shifts
in the extracted top mass value. It turns out, however, that the
\PythiaEightPtwo{} handling of top decay in \hvq{}, improved with the
matrix-element corrections, does in practice achieve NLO accuracy up to an
irrelevant normalization factor.

This nicely consistent picture does not hold anymore if we use
\HerwigSevenPone{} as shower generator. In particular, it seems that the MEC
implemented in \HerwigSevenPone{} do not have the same effect as the handling of
radiation in decay of our modern NLO+PS generators, leading to values of the
extracted top mass that can differ up to about 700~MeV. Furthermore,
interfacing our most accurate NLO+PS generator (the \bbfourl{} one)
to \HerwigSevenPone{} leads to an extracted top mass of up to 1.2~GeV
smaller with respect to the corresponding result with \PythiaEightPtwo{}.

At this point we have two options:
\begin{itemize}
\item
  Dismiss the \HerwigSevenPone{}
  results, on the ground that its MEC handling of top decay does not match
  our modern generators.
\item
  Consider the \HerwigSevenPone{} result as a variation to be included
  as theoretical error.
\end{itemize}
We believe that the first option is not soundly motivated. In fact, the
implementation of MEC in \PythiaEightPtwo{} is also \emph{technically} very close
to what \POWHEG{} does. The hardest radiation is essentially generated in the
same way, and in both cases the subsequent radiation is generated with a
lower transverse momentum. Thus the good agreement between the two is not
surprising. The case of \HerwigSevenPone{} is completely different, since in
angular-ordered showers the hardest radiation is not necessarily the
first~\cite{Seymour:1994df}. It is thus quite possible that the differences
we found when \HerwigSevenPone{} handles the decay with MEC, with respect to the
case when \POWHEG{} does, are due to the fact that the two procedures,
although \emph{formally} equivalent (i.e.~both leading to NLO accuracy) are
\emph{technically} different. In this last case, their difference should be
attributed to uncontrolled higher-order effects, and should thus be
considered as a theoretical uncertainty.

A further question that this work raises is whether we should consider the
variation between the \PythiaEightPtwo{} and the \HerwigSevenPone{} programs
as an error that should be added to current top-mass measurements. By doing
so, current errors, that are of the order of 500-600~MeV, would become larger
than 1~GeV. We believe that our crude modeling of the measurement process
does not allow us to draw this conclusion. The analysis procedures used in
direct measurements are much more complex, and involve adequate tuning of the
MC parameters and jet-energy calibration using hadronic $W$ decays in the
same top events. It is not unlikely that these procedures could lead to an
increased consistency between the \PythiaEightPtwo{} and \HerwigSevenPone{}
results. However, in view of what we have found in our study, it is difficult
to trust the theoretical errors currently given in the top quark mass
determination if alternative NLO+PS and shower generators combinations are
not considered.

\section*{Acknowledgments}
We would like to thank Peter Richardson for important suggestions on the
settings of \HerwigSevenPone{} and on our \POWHEG{}-\HerwigSevenPone{}
interfaces, and Stefan Prestel for help with the \PythiaEightPtwo{} code and
for suggestions on the implementation of our \POWHEG{}-\PythiaEightPtwo{}
interface.  We also like to thank Simon Pl\"atzer and Stephen Webster for
help and suggestions on our \HerwigSevenPone{} input files, and for
illustrating us aspects of the \HerwigSevenPone{} code. Finally we thanks
Roberto Franceschini, Alexander Mitov and Stefano Pozzorini for suggestions
on the manuscript, and Mike Seymour, Peter Skands, T\"orbjorn Sj\"ostrand,
Michelangelo Mangano, Andreas Papaefstathiou, and Andrzej Siodmok for useful
discussions.

We acknowledge the CINECA and the Regione Lombardia award under the LISA
initiative 2016-2018, for the availability of high performance computing
resources and support.
C.O.~wishes to thank Maurizio Cremonesi for the help in running the code on
the CINECA resources, under the project LISA PWHG-RES.  The research of
T.J.~was supported by the Swiss National Science Foundation~(SNF) under
contracts BSCGI0-157722 and CRSII2-160814.

\appendix

\section{The treatment of remnants}
\label{app:remnants}
In \POWHEG{} it is possible to separate the real cross section, in a given
singular region $\alpha$, into two contributions
\begin{equation}
R^\alpha=R^\alpha_s+R^\alpha_f\,,
\end{equation}
where $R^{\alpha}_f$ does not contain any singularities, while $R^{\alpha}_s
$ is singular.  Only $R^{\alpha}_s$ is exponentiated in the Sudakov form
factor and used for the computation of $\tilde{B}$, while the leftover
$R^{\alpha}_f$, dubbed the remnant contribution, is finite upon phase space
integration~\cite{Nason:2004rx}.

In all our three NLO generators it is possible to achieve this separation for
initial-state radiation~(ISR) emissions by setting the parameter {\tt
  hdamp}\footnote{We used an {\tt hdamp} value equal to the input top-quark
  mass, i.e.~the {\tt qmass} parameter for the \hvq{} generator, {\tt tmass}
  for \bbfourl{} and \ttbnlodec{}.  } in the {\tt powheg.input} file.
Denoting with $\aprod$ the production region, $R^{\aprod}_s$ and
$R^{\aprod}_f$ are defined as
\begin{align}
R^{\aprod}_s = & \frac{{\tt hdamp}^2}{{\tt
    hdamp}^2+(\pt^{\aprod})^2}R^{\aprod}\,, \\ \;\;\; R^{\aprod}_f = &
\frac{(\pt^{\aprod})^2}{{\tt hdamp}^2+(\pt^{\aprod})^2}R^{\aprod}\,,
\end{align}
where $\pt^{\aprod}$ is the transverse momentum of the emitted parton
relative to the beam axis.  The {\tt scalup} variable contained in the Les
Houches event, that is used by the parton shower program to veto emissions
harder than the \POWHEG{} one, is set equal to $\pt^{\aprod}$.

Since remnant events are non-singular, the associated radiation has
transverse momenta of the order of the partonic center-of-mass energy.
We can thus define {\tt scalup} as
\begin{equation}
{\tt scalup} = \frac{\hat{s}}{2}\,.
\end{equation}
We have checked that, by using as {\tt scalup} the default \POWHEG{}
scale (i.e.~the transverse momentum of the radiated parton) the
$\mwbj{}$ and the $\Ebjmax$ values are very close to the ones we have
presented in this paper. This is consistent with the expectation that
these observables should be relatively insensitive to radiation in
production, that in our case is always treated as ISR.  The same holds
for the leptonic observable $m(\ell^+\ell^-)$.  For the remaining
ones, a higher sensitivity to ISR effects is not excluded, and in fact
the differences of the first Mellin moments reported in
Tab.~\ref{tab:Olep-summary-py} with the corresponding ones obtained
with the default {\tt scalup} value, for the \hvq{} generator showered
with \PythiaEightPtwo{}, are given by
\begin{equation}
  \begin{split}
  \delta \langle\pt(\ell^+) \rangle & = 125\pm 46~{\rm MeV}\,,
  \\ \delta \langle\pt(\ell^++\ell^-) \rangle & = 298\pm 54~{\rm
    MeV}\,, \\ \delta \langle E(\ell^+ \ell^-) \rangle & = 214\pm
  149~{\rm MeV}\,, \\ \delta \langle\pt(\ell^+)+\pt(\ell^-) \rangle &
  = 219\pm 87~{\rm MeV}\,.
  \end{split}
\end{equation}
In comparison with Tab.~\ref{tab:Olep-summary-py}, we see that these
variations are of the same order or smaller than those arising from scale and
PDF uncertainties.

In the \bbfourl{} code, when ISR remnants are generated, no radiation in decay
is produced.\footnote{This behaviour may be changed in the future.} Thus, in
this case, radiation off the resonances is fully handled by the parton
shower, without the use of a veto algorithm to limit the $\pt$ of the
radiated partons.

The \ttbnlodec{} generator does instead implement radiation in decay also for
remnants, and thus in this case vetoing is performed as for the standard events.

The absence of emissions from the $t$ and $\bar{t}$ resonances in remnant
events for the \bbfourl{} generator, in contrast with the \ttbnlodec{} one,
is probably the reason why the former generator displays a slightly larger
sensitivity to matrix-element corrections~(see Tabs.~\ref{tab:mwbj_MEC},
\ref{tab:Ebj_MEC} and~\ref{tab:leptobs_MEC}).

To summarize:
\begin{itemize}
\item \hvq{}: Emissions in decay are never vetoed. For remnant events the
  {\tt scalup} value used to limit radiation in production is set to
  $\sqrt{\hat{s}}/2$.
\item \ttbnlodec{}: Emissions in decay are always vetoed. For remnant events
  the {\tt scalup} value is set to $\sqrt{\hat{s}}/2$.
\item \bbfourl{}: Emissions in decay are always vetoed except if the event is
  a remnant, in which case they are never vetoed. For remnant events the {\tt
    scalup} value is set to $\sqrt{\hat{s}}/2$.
\end{itemize}

\section{Fitting procedure}
\label{app:fit}
We always adopt the same fitting procedures in order to find the maximum of a
distribution. Calling $Y(x)$ the histogram of our distribution, and
$y(x,\{a\})$ our fitting functional form, where $\{a\}$ represent the fitting
parameters, we proceed as follows:
\begin{itemize}
  \item We find the bin with the highest value, and assign its center to the
    variable $x_{\rm max}$.
  \item We find all surrounding bins whose value is not less than  $Y(x_{\rm max})/2$.
    We assign to the variable $\Delta$ the
    range covered by these bins divided by two.    
  \item
    We minimize the $\chi^2$ computed from the difference of the integral of
    $y(x,\{a\})$ in each bin, divided by the bin size, with respect to $Y(x)$,
    choosing as a range all bins that overlap with the segment
    $[x_{\rm max}-\Delta, x_{\rm max}+\Delta]$.
  \item
    From the fitted function we extract the maximum position and assign it to
    $x_{\rm max}$.
  \item If the reduced $\chi^2$ of the fit is less than 2, we keep this
    result.  If not, we replace $\Delta\to 0.95\times \Delta$ and repeat the
    operation until this condition is met.
\end{itemize}

\section{{\tt PowhegHooks.h}}
In \PythiaEightPtwo{} the transverse-momentum definition used in the veto
algorithm for radiation in production is different from the \POWHEG{} one.
In order to deal with this issue, the authors of \PythiaEightPtwo{} implemented
a veto employing the \POWHEG{} transverse momentum definition, by
constructing a {\tt UserHooks} subclass in the {\tt PowhegHooks.h} file,
which is currently part of the \PythiaEightPtwo{} distribution.

The \PythiaEightPtwo{} manual suggests to use the {\tt PowhegHooks} class whenever
showering a \POWHEG{} style matched NLO+PS process. In
order to implement the features of the {\tt PowhegHooks} class in our
generators, avoiding at the same time conflicts with the {\tt PowhegHooksBB4L}
one (that performs vetoing also for resonance decays), we added them
to the {\tt PowhegHooksBB4L} class, where they are activated by setting
\verb!POWHEG:veto = 1!.

All the results presented in this paper were obtained using
\verb!POWHEG:veto = 0!. However, we have also investigated
the sensitivity of our results to this setting, by showering all our
samples with \verb!POWHEG:veto = 1!.
The differences with respect to the \verb!POWHEG:veto = 0! setting
are listed in Tab.~\ref{tab:delta_pwgveto}
\begin{table}
  \centering
  { \begin{tabular}{|c|c|c|c|}
\cline{1-4}
\multicolumn{4}{|c|}{ \phantom{\Big|} {\tt PowhegHooks} ${}{-}$ no {\tt PowhegHooks} [MeV]} \\
\cline{1-4}
\phantom{\Big|} observable & \bbfourl & \ttbnlodec & \hvq \\
\cline{1-4}
 $\phantom{\Big|} \mwbjmax$ no smearing                                & $          35 \pm           6 $& $          18 \pm           5 $& $          17 \pm           5 $\\ \cline{1-4}
 $\phantom{\Big|} \mwbjmax$ smearing                                   & $          77 \pm           2 $& $          78 \pm           2 $& $          71 \pm           2 $\\ \cline{1-4}
 $\phantom{\Big|} \Ebjmax$                                             & $           4 \pm         115 $& $         130 \pm          87 $& $         157 \pm          91 $\\ \cline{1-4}
 $\phantom{\Big|} \langle \pT(\ell^+)\rangle $                         & $          57 \pm          70 $& $          74 \pm          47 $& $          50 \pm          46 $\\ \cline{1-4}
 $\phantom{\Big|} \langle \pT(\ell^+\ell^-)\rangle $                   & $         166 \pm          84 $& $         173 \pm          56 $& $         150 \pm          54 $\\ \cline{1-4}
 $\phantom{\Big|} \langle m(\ell^+\ell^-)\rangle $                     & $          25 \pm         140 $& $          16 \pm          91 $& $         -18 \pm          90 $\\ \cline{1-4}
 $\phantom{\Big|} \langle E(\ell^+\ell^-)\rangle $                     & $         145 \pm         230 $& $         143 \pm         152 $& $         123 \pm         149 $\\ \cline{1-4}
 $\phantom{\Big|} \langle \pT(\ell^+)+\pT(\ell^-)\rangle $             & $         123 \pm         135 $& $         144 \pm          89 $& $         107 \pm          87 $\\ \cline{1-4}
 \end{tabular}
}
  \caption{Differences between the predictions obtained using the {\tt
      POWHEG:veto = 1} and the {\tt POWHEG:veto = 0} settings for the three
    generators interfaced with \PythiaEightPtwo. }
  \label{tab:delta_pwgveto}
\end{table}
 for all the generators under study.  The shifts obtained are not large and
 mostly compatible among the different generators.

\section{Truncated showers}
\label{app:TS}
We briefly remind the need for truncated showers in the specific example of a
\POWHEG{} implementation of top decay interfaced to an angular-ordered parton
shower.

If top decay is treated without NLO corrections, the parton shower will
generate radiation from the $b$ quark with an unrestricted initial angle. The
hardest radiation will take place along the shower after an arbitrary number
of soft radiations at larger angles.

If \POWHEG{} style NLO corrections are included, the hardest radiation,
consisting in the emission of a gluon, will be generated first by \POWHEG{},
and the parton shower will build angular-ordered jets starting from the
$b$ quark and the \POWHEG{} gluon. The $b$ quark will be assigned an initial
angle for showering equal to $\theta_{bg}$, the angle between the $b$ and
the \POWHEG{} gluon. The soft radiation emitted by the $b$ quark at angles
larger than $\theta_{bg}$ will thus be missing. In order to remedy to this
problem, it was proposed in Ref.~\cite{Nason:2004rx} to let both the
$b$ quark and the gluon radiate with initial angle $\theta_{bg}$ and with a
$\pt$ veto set to the gluon relative transverse momentum, and to add a
$b$ quark $\pt$ vetoed shower starting with the angle that would have been
assigned if the gluon had not be radiated (i.e.~an unrestricted angle), and
stopping at the $\theta_{bg}$ angle.\footnote{A simple example is also
  illustrated in Sec.~7.2 of Ref.~\cite{Nason:2004rx}.}

The veto technique introduced in Ref.~\cite{Schofield:2011zi}, and activated
in \HerwigSevenPone{} with the settings of eq.~(\ref{eq:TSsettings}), performs a
fully equivalent task. In fact, with these settings, the initial angle for
radiation from a gluon is taken as the maximum angle between the gluon and
its two colour partners, that, in our case, leads to unrestricted radiation
from the gluon, i.e.~$\theta_{gg} \lesssim 1$.  However, the colour factor
$C_A$ associated with this radiation is reduced by a factor of two if
$\theta_{gg}>\theta_{bg}$, while it is restored to $C_A$ for smaller
angles. Since $C_A/2 \approx C_F$ in the large $N_c$ limit, we see that this
is equivalent to the inclusion of a vetoed truncated shower from the
$b$ quark down to the angle~$\theta_{bg}$.


\providecommand{\href}[2]{#2}\begingroup\raggedright\endgroup


\end{document}